\documentclass[aps,prd,twocolumn,superscriptaddress,nofootinbib,showpacs]{revtex4-2}
\usepackage{amsmath}
\usepackage{amssymb}
\usepackage{graphicx}
\usepackage{xcolor,color}
\usepackage{url}
\usepackage{bm}
\usepackage{mathrsfs}
\usepackage[utf8]{inputenc}
\usepackage{hyperref}
\usepackage{enumerate}
\usepackage{amsthm}
\usepackage{bbm}
\usepackage[normalem]{ulem}
\usepackage{upgreek}
\usepackage{tensor}
\usepackage{siunitx}
\usepackage{orcidlink}
\usepackage[caption=false]{subfig}
\usepackage{float}
\usepackage{colortbl}

\definecolor{colour1}{HTML}{0571b0} 
\definecolor{colour2}{HTML}{92c5de} 
\definecolor{colour3}{HTML}{f4a582} 
\definecolor{colour4}{HTML}{ca0020} 
\definecolor{colour5}{HTML}{fe4a49} 
\definecolor{colour6}{HTML}{2d3092} 

\definecolor{colour99}{HTML}{0000FF} 

\hypersetup{colorlinks=true, linkcolor=colour99, citecolor=colour99,
filecolor=colour99, urlcolor=colour99}

\theoremstyle{definition}

\newtheorem{open}{Open Problem}

\usepackage{color,xcolor}
\usepackage{ulem}
\usepackage{float}
\usepackage{subfig}
\usepackage{cleveref}
\usepackage{multirow}
\usepackage{booktabs}
\usepackage{booktabs}
\usepackage{threeparttable}

\usepackage[export]{adjustbox}
\usepackage{times}
\usepackage{commath}
\usepackage{lipsum}

\crefname{equation}{Eq.}{Eq.}
\Crefname{equation}{Eqs.}{Eqs.}

\crefname{section}{Sec.}{Sec.}
\Crefname{section}{Secs.}{Secs.}

\crefname{table}{Table}{Table}
\Crefname{table}{Tables}{Tables}
\crefname{figure}{Fig.}{Fig.}
\Crefname{figure}{Figs.}{Figs.}

\crefname{appendix}{Appendix}{Appendix}

\crefname{open}{Open Problem }{Open Problem}
\crefname{open}{Open Problems }{Open Problems}

\usepackage{threeparttable}
\graphicspath{{Figures/}}

\begin{document}

\title{Complete quasinormal modes of Type-D black holes}

\author{Changkai Chen\,\orcidlink{0000-0002-4023-0682}}
\affiliation{School of Physics and Astronomy, Beijing Normal University, Beijing, 100875, China}
\affiliation{Department of Physics, Key Laboratory of Low Dimensional Quantum Structures and Quantum Control of Ministry of Education, Synergetic Innovation Center for Quantum Effects and Applications, Hunan Normal University, Changsha, 410081, Hunan, China}
\affiliation{Institute for Frontiers in Astronomy and Astrophysics, Beijing Normal University, Beijing, 102206, China}

\author{{Jiliang} {Jing}\,\orcidlink{0000-0002-2803-7900} } \email[Corresponding author: ]{jljing@hunnu.edu.cn}
\affiliation{Department of Physics, Key Laboratory of Low Dimensional Quantum Structures and Quantum Control of Ministry of Education, Synergetic Innovation Center for Quantum Effects and Applications, Hunan Normal University, Changsha, 410081, Hunan, China}

\author{{Zhoujian} {Cao}\,\orcidlink{0000-0002-1932-7295}, } \email[Corresponding author: ]{zjcao@amt.ac.cn}
\affiliation{School of Physics and Astronomy, Beijing Normal University, Beijing, 100875, China}
\affiliation{Institute for Frontiers in Astronomy and Astrophysics, Beijing Normal University, Beijing, 102206, China}
\affiliation{School of Fundamental Physics and Mathematical Sciences, Hangzhou Institute for Advanced Study, UCAS, Hangzhou, 310024, China}

\author{{Mengjie} {Wang} }
\affiliation{Department of Physics, Key Laboratory of Low Dimensional Quantum Structures and Quantum Control of Ministry of Education, Synergetic Innovation Center for Quantum Effects and Applications, Hunan Normal University, Changsha, 410081, Hunan, China}

\date{\today}
\begin{abstract}
Quasinormal mode (QNM) spectra of black holes exhibit two open problems [Conf. Proc. C {\textbf{0405132}}, 145 (2004); CQG {\textbf{26}}, 163001 (2009)]: (i) the discontinuity in highly damped QNMs between Schwarzschild and Kerr solutions as $a \to 0$,
and (ii) the unexplained spectral proximity between QNMs and algebraically special (AS) frequencies, particularly the anomalous multiplet splitting for Kerr $\ell=2$, $m \geq 0$ modes.
We develop a novel method to compute complete QNM spectra for Type-D black holes, solving both problems and establishing a mathematical framework for boundary value problems of dissipative systems.
Using analytic continuation of radial eigenvalue equations, our method eliminates the dependence on auxiliary parameters in the connection formulas for confluent Heun solutions.
This breakthrough overcomes the long-standing challenge of calculating QNMs that cross or lie on the negative imaginary axis (NIA).
For Schwarzschild and Kerr spacetimes ($0 \leq {n} \leq 41$, $2 \leq \ell \leq 16$), we present complete spectra validated through scattering amplitudes with errors $<10^{-10}$.
The results provide definitive solutions to both open problems:
(i) The inability of conventional methods to compute QNMs crossing or residing on the NIA leads to apparent discontinuities in Kerr spectra as $a \to 0$.
(ii) When a QNM exactly coincides with the AS frequency, an additional QNM (unconventional mode) appears nearby.
For the Kerr case with $\ell=2$, overtone sequences from both unconventional and AS modes exhibit precisely $2\ell+1$ branches, without multiplets or supersymmetry breaking at AS frequencies.
Our method confirms Leung's conjecture of high-$\ell$ unconventional mode deviations from the NIA through the first $\ell=3$ calculation.
Moreover, this paradigm surpasses the state-of-the-art Cook's continued fraction and isomonodromic methods in computational efficiency.

\end{abstract}
\maketitle

\section{Introduction}

\subsection{Test GR Using QNMs}
The coalescence of binary black holes (BBHs) provides a unique gravitational laboratory to probe strong-field gravity through quasinormal modes (QNMs) \textemdash exponentially damped oscillations imprinted in the ringdown phase of gravitational wave (GW) signals \cite{Kokkotas:1999bd,Nollert:1999ji,Berti:2009kk,Konoplya:2011qq,Chung:2024ira,Berti:2025hly}.
These complex frequencies, determined solely by the BH's intrinsic parameters and modal indices (azimuthal $m$, orbital $\ell$, overtone ${n}$) rather than initial perturbations, establish QNMs as spacetime fingerprints \cite{Echeverria:1989hg,Dreyer:2003bv,Berti:2016lat}.
As dictated by general relativity (GR) and the no-hair theorems \cite{Carter:1971zc, Robinson:1975bv}, astrophysical BHs are fully characterized by their mass and spin, with QNM spectral features encoding these properties via damped sinusoidal eigenstates \cite{Regge:1957td,Press:1971wr,Echeverria:1989hg,Dreyer:2003bv}.
This universality makes QNM spectroscopy a cornerstone for testing GR \cite{Dreyer:2003bv, Berti:2005ys, Cardoso:2016ryw, Berti:2018vdi, Isi:2019aib, Isi:2021iql, Capano:2021etf}, as deviations from predicted frequencies could signal new physics.
Recent detections of over 100 BH mergers \cite{KAGRA:2021vkt} have enabled precision tests of spacetime dynamics, yet fundamental tensions persist between GR and quantum gravity unification \cite{Calmet:2004mp,Canetti:2012zc}, dark matter anomalies \cite{Sofue:2000jx,Bertone:2016nfn}, and cosmic acceleration \cite{SupernovaSearchTeam:1998fmf,SupernovaCosmologyProject:1998vns}.
Modified gravity theories addressing these issues often predict BH solutions with additional ``hairs" or spacetime modifications \cite{Kanti:1995vq, Jacobson:2007veq, Horava:2009uw,  Yunes:2011we, Yagi:2012ya, Wagle:2021tam, Blazquez-Salcedo:2017txk, Cardoso:2022whc}, imprinting characteristic shifts in QNM spectra.
By mapping gravitational-wave observables to beyond-GR couplings through parametric QNM analyses \cite{Dreyer:2003bv, Berti:2005ys, Berti:2018vdi,Franchini:2023eda}, we transform ringdown signals into a powerful diagnostic tool for exotic spacetime structures and high-energy physics phenomena inaccessible to terrestrial experiments.

\subsection{QNM Precision and High-Overtone Modes}

The pursuit of \textit{high-precision} QNM calculations is not merely a numerical challenge but a fundamental requirement for GW astronomy.
As even marginal variations in BH parameters (e.g., mass shifts of $\sim1\%$) can induce fractional changes in QNM frequencies at the sub-percent level, resolving these subtle spectral shifts demands computational accuracy surpassing $10^{-4}$ relative error \cite{Konoplya:2011qq}.
Consequently, significant efforts have been devoted in recent years to developing precise numerical and semi-analytical approaches for solving the QNM eigenvalue problem.
Traditional methods like the WKB approximation \cite{Iyer:1986np,Iyer:1986nq,Kokkotas:1988fm,Seidel:1989bp}, while effective for low-overtone modes, accumulate systematic errors when probing higher overtones, which necessitating advanced spectral techniques such as continued fraction method (CFM) \cite{Leaver:1985ax}.
From an astrophysical perspective, GW measurements of QNMs may allow us to measure BH masses and spins with unprecedented accuracy, offering a direct observational probe of the no-hair theorem in GR.
Although there is scientific debate about the hypothesized link between BH oscillation frequencies and quantum-scale properties, this conjecture has had a significant impact. It has driven methodological advances, enabling us to calculate previously unexplored regimes of the QNM spectrum.
These developments highlight persistent mathematical challenges in QNM analysis.
As emphasized by Berti et al. \cite{Berti:2003jh,Berti:2004md,Berti:2009kk}, enhancing mathematical rigor remains imperative for boundary value problems involving wave dissipation. Systematic development of asymptotic expansion techniques\textemdash replacing current ad hoc formulations\textemdash and improved spectral algorithms for highly damped (high-overtone) QNMs continue to represent critical unsolved problems in the field.

The QNM spectrum for gravitational perturbations exhibits a significantly more complex structure compared with other perturbation fields.
In 1975, Chandrasekhar and Detweiler performed the first numerical computation of weakly damped modes in gravitational QNM spectra for Schwarzschild BHs \cite{Chandrasekhar:1975zza}.
It was not until 1985 that Leaver \cite{Leaver:1985ax,Leaver:1986gd,Leaver:1986vnb} developed the most accurate method, using the CFM to calculate gravitational QNM spectra for both Schwarzschild and Kerr BHs.
Among them, the Schwarzschild QNMs include highly damped modes ( $n=59$), while the Kerr QNMs only contain weakly damped modes.
It is found that for the two Schwarzschild QNMs from Leaver's results \cite{Leaver:1985ax},
\begin{align}
\omega &= \quad 0.0000000 - 1.999000i,\quad {\kern 9pt}\ell  = 2, n =8;  \\
\omega &= \,-0.0001295 - 10.0078265i,\quad \ell  = 3, n =40.
\end{align}
These two modes  show proximity to algebraically special (AS) modes ${\omega_{\ell=2} ^{{\rm{AS}}}} = -2i$ and ${\omega_{\ell=3} ^{{\rm{AS}}}} = -10i$, which are also known as total transmission modes (TTMs).
The AS mode ${\omega_{\ell} ^{{\rm{AS}}}}$ of the Schwarzschild BH can be expressed as \cite{Couch:1973zc},
\begin{equation}\label{eq:ASmode}
{\omega_{\ell} ^{{\rm{AS}}}} = \pm i\frac{{(\ell  - 1)\ell (\ell  + 1)(\ell  + 2)}}{{12}}\end{equation}
In fact, the AS frequency is so called precisely because of the requirement that the perturbations should be ``special'': they should contain only ingoing waves (the perturbed Weyl scalar $\psi_0\neq 0$) or only outgoing waves ($\psi_4\neq 0$) throughout the entire spacetime.
Among them, purely ingoing waves are also called TTM-right (${\rm TTM}_{\rm R}$), and purely outgoing waves are called TTM-left (${\rm TTM}_{\rm L}$).
This boundary condition differs fundamentally from those of QNMs, which require pure ingoing waves at the horizon and pure outgoing waves at infinity.
A critical question arises: Why do the Schwarzschild QNMs approach AS modes? Is this merely an algebraic coincidence, or does it reflect deeper physical significance?

The AS modes for Schwarzschild BHs were first derived by Couch and Newman in 1973 \cite{Couch:1973zc}.
In the same year, Wald \cite{Wald:1973wwa} established the necessary and sufficient condition for AS modes in Kerr BHs, which is that the Starobinsky-Teukolsky constant vanishes.
In 1984, Chandrasekhar \cite{Chandrasekhar:1984mgh} further derived the AS modes for Reissner-Nordstr\"{o}m (RN) BHs and discussed the mathematical connections between these three BH types: Schwarzschild, Kerr, and RN. Additionally, he provided numerical results for Kerr AS modes.
To date, AS modes for other perturbation fields have not been identified.
It must be emphasized that purely imaginary modes are \emph{not} AS modes. Only modes satisfying TTM boundary conditions qualify as AS modes.
Only some AS modes may exhibit the form of purely imaginary modes.
For gravitational QNMs, the weakly damped (highly damped) regime is defined, somewhat arbitrarily, as corresponding to imaginary parts smaller (larger) than the AS mode \cite{Berti:2009kk}.

In 1990, Leaver \cite{Leaver:1990zz} employed the CFM to compute weakly damped modes in the QNM spectrum of RN BHs.
When the imaginary part of QNMs becomes large, the convergence of Leaver's CFM \cite{Leaver:1985ax} deteriorates significantly.
This necessitates increasing the depth of continued fractions for high-overtone modes, resulting in computational time explosion.
To address this issue, Nollert (1993) circumvented the slow convergence of CFM and successfully computed Schwarzschild QNMs with overtone numbers up to $n=2000$ \cite{Nollert:1993zz}.
For decades, the highly damped regime of QNM spectra remained largely unexplored beyond the Schwarzschild case.
Physically, the dominance of slowly damped modes in BH responses initially diminished motivation to study large-$|\omega_I|$ regimes.
Technically, conventional numerical methods, designed to distinguish outgoing/ingoing waves, fail catastrophically beyond critical damping thresholds, as exponentially decaying solutions become overwhelmed by numerical noise.
This impasse persisted until the 21st century.
In 2003, Berti's team \cite{Berti:2003jh} pioneered highly damped QNM studies for Kerr BHs using cross-checked implementations of the CFM,
though the validity of their results became questionable for $\ell=2\,(n>50)$ and $\ell=3\,(n>41)$ overtones.
For RN BHs, their CFM achieved calculations up to $n\sim10^5$ overtones, though challenges persisted for highly damped modes and large charge parameters \cite{Berti:2003zu,Berti:2004md}.
A pivotal advancement emerged in 2014 when Cook \cite{Cook:2014cta}, working within the framework of the confluent Heun equation for completeness, developed an exceptionally efficient spectral technique for the angular Teukolsky equation.
Though only two gravitational mode datasets \cite{cook_2024_14024959} were published ($0 \leq n \leq 15$, $2 \leq \ell \leq 16$; and $0 \leq n \leq 32$, $2 \leq \ell \leq 4$), code tests \cite{cookgb_kerrmodes_2023} revealed $\ell=2$ overtones up to $n=200$.
Cook's method (improved CFM) achieves high-efficiency computation of Kerr QNMs while maintaining high precision.
This method demonstrates remarkable applicability, and is capable of handling various extreme parameters (including near-extreme spins $a/M \rightarrow 1-10^{-10}$ and highly damped modes $n\sim 200$).
Although numerous improved CFM algorithms and alternative numerical methods exist \cite{Berti:2009kk,Berti:2025hly}, to our knowledge none surpass Cook's method in performance.

\subsection{ Incomplete QNM Spectra and Controversies}
The QNM spectrum exhibits extraordinary complexity, with BH perturbation theory leading us\textemdash in Chandrasekhar's vivid metaphor\textemdash ``into a realm of the rococo: splendorous, joyful, and immensely ornate" \cite{Chandrasekhar:1985kt}.
Nollert, Berti, and Cook have apparently perfectly mapped the QNM spectra for Schwarzschild \cite{Nollert:1993zz}, RN \cite{Berti:2003zu}, and Kerr \cite{Cook:2014cta} BHs respectively.
However, these ornate and complicated spectra ultimately reveal incompleteness and imperfections.
These deficiencies arise because there is an essential singularity at $\omega = 0$ and a branch cut extending from this singularity to $-i\infty$ \cite{Berti:2006wq,Ching:1994bd,Jensen:1985in}.
In other words, the negative imaginary axis (NIA) is a branch cut ${\mathscr{B}^{\,\,\omega}_{-i\infty, 0}}=(-i\infty, 0)$ for the QNM spectrum.
If the complete QNM spectrum  crosses or falls on this branch cut and the QNMs are directly solved without handling this branch cut,  the resulting QNM spectrum would lack these modes.

\textbf{ \textit{For Schwarzschild BHs}}, in 2003,
Leung et al. \cite{Leung:2003ix} were the first to address the branch cut, employing a fitting procedure to discover the first set of unconventional modes \cite{Leung:2003eq},
\begin{equation}\label{eq:UnconventionalMode1}
   \omega_{\pm}=\mp 0.01350 - 1.99835i ,\quad \ell  = 2.
\end{equation}
These modes are located on the unphysical sheet beyond the branch cut (hence referred to as ``unconventional").
QNMs possess two eigenfrequencies, traditionally classified in the literature \cite{Berti:2003jh,Berti:2009kk,Cook:2014cta} as: (i) positive frequencies with ${\rm Re}(\omega) > 0$, and (ii) negative frequencies with ${\rm Re}(\omega) < 0$.
However, the unconventional modes exhibit precisely the opposite behavior: the positive frequency $\omega_+$ has a negative real part, whereas the negative frequency $\omega_-$ has a positive real part.

Subsequently, in the same year, Berti et al. also derived a similar result using the extrapolation method \cite{Berti:2003jh},
\begin{equation}\label{eq:UnconventionalMode2}
   \omega_+ \simeq - 0.016240 - 1.998282i, \quad \ell = 2.
\end{equation}

In 2010, Fiziev provided a high-precision result ($\sim 10^{-10}$) using the epsilon method \cite{Fiziev:2010yy,Fiziev:2011mm},
\begin{equation}\label{eq:UnconventionalMode3}
\omega_{\pm} = \mp 0.015324503 - 1.998411845i, \quad \ell = 2.
\end{equation}
While higher-order unconventional modes with $\ell>2$ should theoretically exist, conventional numerical approaches fail to resolve them due to insufficient precision \cite{Leung:2003eq}. The methodology developed in this work successfully overcomes this limitation and provides reliable calculations for these modes.
Although the unconventional modes of the Schwarzschild BHs can be calculated for the QNMs that cross the branch cut, these three methods \cite{Leung:2003eq,Berti:2003jh,Fiziev:2010yy} and other numerical methods \cite{Leaver:1985ax,Konoplya:2011qq} are all unable to calculate the QNMs that lie on the branch cut.
In other words, for each $\ell$, there should exist a purely imaginary mode located precisely on the NIA.
While most numerical results are only close to (but do not exactly coincide with) the purely imaginary modes, our results demonstrate perfect agreement with the expected purely imaginary spectrum.

\textbf{ \textit{Regarding the controversy of Schwarzschild QNMs}},
the academic controversy \cite{Berti:2009kk,MaassenvandenBrink:2000iwh,Berti:2004md,Onozawa:1996ux} can be summarized as follows: ``Which modes (QNM, ${\rm TTM}_{\rm L}$, and ${\rm TTM}_{\rm R}$) actually exist at AS frequencies \eqref{eq:ASmode}?"
Since Schwarzschild AS mode lies on the branch cut, numerical methods (e.g., Leaver's CFM) oscillate and fail to converge along the NIA \cite{Cook:2016fge}, making it hard to verify which mode type exists. Regarding this issue, scholars have proposed the following different conjectures and explanations:

\begin{enumerate}
  \item  Leung and Maassen van den Brink found \cite{Leung:1999fr} that the Regge--Wheeler equation (RWE) and the Zerilli equation require distinct treatment at the AS frequency \eqref{eq:ASmode}, since the supersymmetry transformation leading to the proof of isospectrality is singular there. The RWE exhibits \textit{no modes} at the Schwarzschild AS frequency while the Zerilli equation supports \textit{both a QNM and a ${\rm TTM}_{\rm L}$} at $\omega_\ell^{\rm AS}$.
      Consequently, using slow-rotation expansions of the perturbation equations, Maassen van den Brink concluded \cite{MaassenvandenBrink:2000iwh} that at $\omega_\ell^{\rm AS}$, \textit{no ${\rm TTM}_{\rm R}$ exists} with only \textit{${\rm TTM}_{\rm L}$ and QNMs present}.

  \item Onozawa \cite{Onozawa:1996ux} employed the CFM to analyze the numerical behavior of TTMs and QNMs at the AS frequency \eqref{eq:ASmode}. The results demonstrate that positive- and negative- frequency modes cancel out at ${a}=0$, leaving no QNMs at $\omega_{\ell}^{\rm{AS}}$.
      Similarly, Cook \cite{Cook:2014cta} concluded the absence of QNMs at $\omega_{\ell}^{\rm{AS}}$.

  \item Although Fiziev calculated high-precision unconventional modes \cite{Fiziev:2010yy,Fiziev:2011mm,Fiziev:2019ewy}, there are no modes at $\omega_{\ell}^{\rm{AS}}$ in the Schwarzschild QNM spectrum ($n \leq 24$) with 16-digits precision that he provided \cite{CardosoWeb,CoGWeb}.

  \item Generally, most authors \cite{Chandrasekhar:1984mgh,Leaver:1985ax,Andersson:1994tt,Onozawa:1996ux,Cook:2014cta} agree that the AS frequency \eqref{eq:ASmode} of Schwarzschild geometry represents TTMs, where $s = -2$ corresponds to ${\rm TTM}_{\rm L}$ and $s = 2$ to ${\rm TTM}_{\rm R}$.

\end{enumerate}
The unconventional modes and the AS modes contribute to the imperfection of the Schwarzschild QNM spectrum.
However, most scholars appear skeptical about the existence of unconventional modes.
Following the discovery of unconventional modes, neither major QNM datasets \cite{BertiWeb_SchBH,KonoplyaWeb,BHPToolkit,Stein:2019mop,Fortuna:2020obg,slashdotfield2023} nor the subsequent reviews \cite{Berti:2009kk,Konoplya:2011qq,Berti:2025hly} have included or plotted these modes.
Only Fiziev provided detailed datasets \cite{Fiziev:2010yy,Fiziev:2011mm,Fiziev:2019ewy}, which were not incorporated into Berti's QNM website \cite{BertiWeb} but were included on other websites \cite{CardosoWeb,CoGWeb}.
The Schwarzschild QNMs serve as initial values for computing QNM spectra of other general relativistic BHs.
As a result, precisely because most researchers have overlooked unconventional modes, the numerically obtained QNMs for Kerr and RN BHs exhibit particularly anomalous and mysterious features.

\textbf{ \textit{For Kerr BHs}}, if certain overtone QNMs lie on or cross the NIA as the rotation parameter $a$ varies, no existing literature has addressed how to compute such modes. This leads to two phenomena:
\begin{enumerate}\label{item1}
  \item Partial regions of the spin parameter $a$ corresponding to a certain QNM sequence vanish.

  For instance, Cook's high-precision numerical results \cite{Cook:2016ngj,Cook:2016fge,cook_2024_14024959} demonstrate that for the AS mode overtones ($n=8$ and $m \geq 0$), QNMs are missing for the parameter range where $a$ varies from 0 to a certain critical value (The QNM corresponding to this critical value is  close to the NIA).
  Furthermore, in the case of $(\ell,m,n)=(2,-2,13)$, QNMs are absent for partial intermediate values of the rotation parameter $a$.
  \item While each $\ell=2$ overtone should theoretically possess exactly five QNM sequences \footnote{The index $m$ has $2\ell + 1$ possible values. When the spin parameter $a$ varies, each overtone's QNMs split into $2\ell + 1$ distinct branches. This splitting is identified as the Zeeman effect for the QNMs.}, some overtone spectra exhibit more than five sequences.

      For instance, Cook's high-precision numerical results \cite{Cook:2016ngj,Cook:2016fge,cook_2024_14024959} show that the overtone ($n=8$) spectrum originating from $\omega_{\ell}^{\rm{AS}}$ contains eight QNM sequences.
\end{enumerate}
While Cook \cite{Cook:2016ngj,Cook:2016fge} successfully computed high-precision QNMs approaching the NIA for Kerr BHs\footnote{The error in Cook's purely imaginary modes is below $10^{-10}$, allowing these modes to be approximately treated as lying on the NIA.}, modes truncated by the NIA remain computationally inaccessible.

\textbf{ \textit{Regarding the controversy of Kerr QNMs}},
the academic controversy \cite{Berti:2009kk,MaassenvandenBrink:2000iwh,Berti:2004md,Onozawa:1996ux} can be summarized as follows: ``What causes the missing partial QNMs and additional QNM sequences in Kerr BHs?"
Regarding this controversy, scholars have proposed the following different explanations and numerical results:
\begin{enumerate}
  \item Missing partial QNMs with $a$ varies from 0 to a certain critical value
  \begin{description}
    \item[1.1] In 1997, Onozawa \cite{Onozawa:1996ux} first investigated Kerr QNMs for $n=8$ at ${\omega_{\ell=2} ^{{\rm{AS}}}}$, and provided higher-precision numerical results for TTMs than Chandrasekhar's results \cite{Chandrasekhar:1984mgh}.
        Neither his plotted figures \cite{Onozawa:1996ux} contained extractable information about QNMs of  $n=8$ near ${\omega_{\ell=2} ^{{\rm{AS}}}}$, nor did he provide any datasets. Only the initial $a$-value QNMs for $n=8$ were discernible at some distance from the NIA, not particularly close.

    \item[1.2]  While studying the limit $a\rightarrow 0$ in detail, Maassen van den Brink conjectured \cite{MaassenvandenBrink:2000iwh} that Schwarzschild AS modes with $\ell=2$ possess a set of ``standard'' Kerr QNMs that split with increasing $a$.
        He further derived analytical expressions for this set of QNMs under the small $a$ approximation,
        \begin{align}\label{eq:MaassenEq}
           2M\omega=&-4i-\frac{33078176}{700009}ma +\frac{3492608}{41177}i a^2 \nonumber\\
                    &+{\cal O}(ma^2) +{\cal O}(a^4),\quad \quad\quad\quad\ell =2,
        \end{align}
       and more formulas with $\ell>2$ can be seen in Ref. \cite{MaassenvandenBrink:2000iwh}.
        Numerical studies by Berti et al. \cite{Berti:2003jh} found QNMs approaching the AS frequency $\omega_\ell^{\rm AS}$, but these numerical results were {\it not} in agreement with this prediction \eqref{eq:MaassenEq}.

    \item[1.3] Initially, when Cook proposed improvements \cite{Cook:2014cta} to the CFM (focusing solely on angular Teukolsky equation computations without addressing the branch cut), his numerical results led him to conclude: ``As the spin parameter $a$ increases from zero, the QNMs  first appear on the physical sheet at the NIA at some finite  value of $a$." Subsequent extensive and precise numerical tests by Cook \cite{Cook:2016fge,Cook:2016ngj} further supported the existence of QNMs whose initial values do not originate from the zero-spin case $a=0$.
        In other words, the QNM spectrum of Kerr BHs remains non-degenerate in the Schwarzschild limit ($a \rightarrow 0$).
  \end{description}

  \item Additional QNM sequences

  \begin{description}
    \item[2.1] Through numerical calculations for $n=8$ at ${\omega_{\ell=2} ^{{\rm{AS}}}}$, Berti et al. \cite{Berti:2003jh} first discovered in 2003 that the $m=2$ case exhibits two distinct QNM sequences, and similarly, the $m=1$ case also shows two corresponding QNM sequences.
        They also mentioned the discovery of an additional QNM sequence near ${\omega_{\ell=3} ^{{\rm{AS}}}}$.
        However, they conjectured that these additional modes might likely be ``spurious" modes due to numerical inaccuracies.

   \item[2.2] In the same year, Maassen van den Brink predicted \cite{Leung:2003eq} the existence of Schwarzschild unconventional modes \eqref{eq:UnconventionalMode1}.
       If this prediction is true, an {\it additional} QNM multiplet should emerge near ${\omega_{\ell} ^{{\rm{AS}}}}$ as $a$ increases. This multiplet {\it ``may well be due to $\omega_\pm$ splitting (since spherical symmetry is broken) and moving through the negative imaginary axis as $a$ is tuned''} \cite{Leung:2003eq}.
       However, Berti's numerical results \cite{Berti:2003jh,Berti:2003zu} indeed reveal the emergence of such multiplets, but these do not seem to behave exactly as predicted in Ref. \cite{Leung:2003eq}.
       These additional QNM multiplets all lack QNMs with spin parameter $a$ ranging from 0 to some critical value. Whether their Schwarzschild limits correspond to AS modes or unconventional modes remains an open question.
       Berti et al. \cite{Berti:2003jh,Berti:2003zu,Berti:2004md,Berti:2009kk} refer to them as "overtone multiplets," even though these additional modes violate the Zeeman splitting of QNM spectra.

  \item[2.3] It was not until 2014 that Cook \cite{Cook:2014cta} provided higher-precision QNM spectra for $\ell=2$ and $n=8$. Then in 2016, additional QNM sequences were discovered beyond just $m=1,2$ cases - notably, the $m=0$ case was also found to correspond to two QNM sequences \cite{Cook:2016fge,Cook:2016ngj}. Furthermore, for other high-overtone QNMs (highly damped modes), even more complicated additional QNM sequences were identified. Moreover, Cook \cite{Cook:2014cta, Cook:2016ngj,Cook:2016fge} also refers to these additional QNM sequences arising from anomalous Zeeman splitting as ``overtone multiplets".

  \end{description}

\end{enumerate}
The controversy regarding Kerr's QNM spectrum can be fundamentally resolved by providing the QNMs located on the unphysical sheet. This work accomplishes precisely that.

\textbf{ \textit{For RN BHs}}, Andersson and Onozawa employed two distinct numerical approaches \cite{Leaver:1990zz,Andersson1993} to investigate weakly damped modes \cite{Andersson:1996xw}. Subsequently, Berti et al. examined highly damped modes \cite{Berti:2003zu,Berti:2004md}, identifying a QNM frequency very close to, but {\it not exactly equal to}, the AS frequency of RN BHs \cite{Chandrasekhar:1984mgh}.
Their numerical results revealed that as the overtone number increases, oscillatory behavior in the frequency's real part emerges earlier with varying charge $Q$. For larger $n$, these oscillations accelerate while the CFM convergence slows, requiring longer computational times. Consequently, tracking roots numerically becomes increasingly challenging near extremal charge values as the imaginary part grows, explaining why data for very large $n$ cannot cover the full range of $Q$. The study of highly damped modes in RN BHs presents fewer controversies compared to other cases, with the primary challenge lying in computing complete QNM spectra across all $Q$ values for high overtones.

Although the controversies surrounding QNM spectra have been documented in the literature by Berti et al. \cite{Berti:2003jh,Berti:2003zu,Berti:2004md,Berti:2009kk}, we have systematically reclassified and updated the progress of Cook's results \cite{Cook:2014cta,Cook:2016fge,Cook:2016ngj,cook_2024_14024959} to facilitate reader comprehension.
In the review \cite{Berti:2004md}, Berti et al. presented five open problems regarding the overwhelmingly rich and complicated QNM spectra.
These controversies can be precisely summarized as two of these open problems, specifically:

  \begin{open}\label{open1}
   A general lesson from the study of charged and rotating BHs seems to be: the spectrum of the limit is not the limit of the spectrum.
  This is a well-known fact in spectral theory, but it is quite striking in the BH context.
  ``Ordinary'' BH QNMs suggest that there should be some continuity between the Schwarzschild and Kerr (RN) solutions as $a\to 0$ ($Q\to 0$).
  At a closer look, the asymptotic spectrum in the limit $|\omega_I|\to \infty$ ($n\to \infty$) seems to violate this continuity requirement.
  What is the mathematical and (more importantly) physical motivation of this discontinuity?
  \end{open}

  \begin{open}\label{open2} The AS mode is a mystery on its own.
  Numerical methods show the presence of a QNM {\it close to} (but not quite {\it at}) the algebraically special frequencies determined by Chandrasekhar \cite{Chandrasekhar:1984mgh} \textemdash  this is true both in the Schwarzschild and in the extremal RN cases.
  In the Kerr case, a {\it doublet} of modes with $m\geq0$\footnote{Cook's results \cite{Cook:2014cta} demonstrate the existence of multiplet structures in the QNM sequences for modes with $m \geq 0$.} comes out of the Schwarzschild AS frequency, but no mode splitting is observed when $m<0$. Analytical work on the ``supersymmetry breaking'' occurring at the Schwarzschild AS frequency only partially clarifies the situation \cite{MaassenvandenBrink:2000iwh}. We definitely need a better understanding of the meaning of QNMs and TTMs in this situation.
  \end{open}
These two long-standing open problems, which have seen limited progress for years, are the primary focus of our study.

\subsection{BH Perturbation and Heun Class Equation}
Using the Newman-Penrose null-tetrad formalism \cite{Newman:1961qr}, the first-order curvature perturbation equation for Kerr BHs can be derived \cite{Teukolsky:1973ha}. The resulting equation, known as the Teukolsky equation, governs the wave equation for the null-tetrad components of the Weyl tensor.
The Teukolsky formalism has seen significant advancements, with the emergence of new theoretical frameworks and computational techniques, and it is no longer limited to Type-D BHs in GR.
Li and Chen developed a novel Teukolsky formalism to derive perturbation equations for rotating BHs in modified gravity theories \cite{Li:2022pcy,Li:2023ulk,Wagle:2023fwl,Weller:2024qvo,Li:2025fci}.
They used this formalism to study QNMs of BHs in dynamic Chern-Simons gravity \cite{Wagle:2023fwl,Li:2025fci}.
Moreover, Pound's team established a second-order Teukolsky formalism for BH perturbations \cite{Spiers:2023mor,Spiers:2023cip}, enabling calculations of second-order gravitational self-force \cite{Pound:2012nt,Pound:2019lzj,Wardell:2021fyy,Warburton:2021kwk,vandeMeent:2023ols,Miller:2023ers,Bourg:2024vre} and  Schwarzschild quadratic QNMs \cite{Bourg:2024jme,Bourg:2025lpd} in GR.

It is known, but  historically somewhat overlooked in the physics literature, that wave equations (RWEs and Teukolsky Equations) of BH perturbations are examples of the Heun class equation \cite{Ronveaux:1995,slavyanov2000special}.
There are two main reasons: First, the efficient numerical computation of Heun class functions presents significant challenges; Second, it is difficult to derive connection formulas between different singular points for Heun class functions, especially between irregular and regular singular points of confluent Heun functions.
However, these problems now have some efficient solutions.
The three major mathematical softwares ( \textit{MATLAB},  \textit{Mathematica}, and  \textit{Maple}) can all perform numerical computations for Heun class functions.
For the connection formulas of Heun class functions, Bonelli et al. derived them using Liouville conformal field theory \cite{Bonelli:2022ten,Iossa:2023gvh,Bonelli:2021uvf}.
Furthermore, in our previous work \cite{Chen:2023ese}, inspired by the Mano-Suzuki-Takasugi (MST) method \cite{Mano_1996,Mano1996RWE,Sasaki:2003xr,Fujita_2004,Fujita_2005}, we also derived connection formulas for confluent Heun functions.
Fiziev provided exact solutions \cite{Fiziev:2005ki,Fiziev:2009kh} to the source-free perturbation equations (RW or Teukolsky equations) in terms of the confluent Heun function, and for the first time applied confluent Heun functions to compute Schwarzschild QNMs \cite{Fiziev:2005ki}.
Following Fiziev's work, numerous researchers have employed Heun class functions to study source-free perturbation equations in various spacetimes, such as QNMs \cite{GurtasDogan:2019kde} of Dirac field in $2+1$ dimensional GW background \cite{Zhang:2014kla}, QNMs \cite{Hatsuda:2020sbn} of Kerr-de Sitter BHs, QNMs \cite{Motohashi:2021zyv} of Kerr-Newman–de Sitter BHs, QNMs \cite{Noda:2022zgk} of the massive scalar field in Kerr-AdS$_5$ BHs, and gravitational QNMs \cite{Chen:2024rov} of uniformly accelerated BHs (the spinning C-metric).
However, except for Fiziev's highly damped ($n\leq24$) Schwarzschild QNMs \cite{CardosoWeb}, these studies exclusively calculated only the fundamental modes.

In this work, we combine the analytic continuation method with confluent Heun solutions to construct the eigenvalue equation for QNMs. This enables the computation of complete QNM spectra for Type-D BHs, thereby solving these two \cref{open1,open2}.
The remainder of the paper is arranged as follows.
In \cref{sec:Exact-Solution-TE}, we present exact solutions to both angular and radial Teukolsky equations for Type-D BHs using confluent Heun functions, along with detailed discussions of boundary conditions for QNMs and TTMs. The QNM eigenvalue equations are then developed through analytic continuation, thus solving branch cut issues in QNM spectra.
Based on our previous work \cite{Chen:2023ese}, we introduce the HeunC-MST method for computing asymptotic amplitudes to validate QNM results.
\cref{sec:HeunC-TTM} extends this methodology to solve and verify TTMs.
Our methodology is systematically applied to two key scenarios: Schwarzschild QNMs in \cref{sec:SchBH-QNM} and Kerr QNMs in \cref{sec:QNM-Kerr}.
The conclusions are presented in \cref{sec:Conclusion}.
In this paper, we use geometrized units $c=G=1$, and $M=1$.
\\
\\\textbf{ \textit{QNM Datasets:}} Compared to two Cook's datasets \cite{cook_2024_14024959}, which only include gravitational perturbations with $s=-2$:
$$0 \le \bar{n} \le 15,\quad 2 \le \ell \le 16;\quad 0 \le \bar{n} \le 32,\quad 2 \le \ell \le 4.$$
where $\bar{n}$ denotes the overtone number in Cook's incomplete QNM spectrum.

This work provides a significantly more comprehensive QNM dataset \cite{ChenQNM} covering:
$$s=\{-2,-1,0\},\quad 0 \le \hat{n} \le 41,\quad 2 \le \ell \le 16.$$
where $\hat{n}$ denotes the overtone number in the complete QNM spectrum.


\section{HeunC Method for Solving and Verifying QNMs}\label{sec:Exact-Solution-TE}

\subsection{The Teukolsky Equations in Type-D BHs}
Using the Newman-Penrose formalism, Teukolsky \cite{Teukolsky:1973ha} showed that several types of perturbation fields in Type-D BHs are governed by a single master equation
\begin{equation}\label{eq:Teuk-Master-Eq}
  {\cal O}\psi  = 4\pi \Sigma T,
\end{equation}
where $T$ represents the source term and $\psi$ is a scalar of spin weight $s$.  Scalar fields are represented by $s=0$,
neutrino fields by $s=\pm\tfrac{1}{2}$, electromagnetic fields by $s=\pm1$, and gravitational fields by
$s=\pm2$.
Assuming the vacuum case ($T=0$), \cref{eq:Teuk-Master-Eq} can be separated into radial and angular parts if we use
\begin{equation}\label{eq:separationEq-Teuko}
  \psi(t,r,\theta,\phi) = e^{-i\omega{t}} e^{im\phi}S_{\ell m}({\Theta }){R_{\ell m }}(r).
\end{equation}

With $\Theta\equiv\cos\theta\in[-1,1]$, the function $S_{\ell m}({\Theta })$ is the angular function satisfying the general form of angular Teukolsky equations (ATEs) in Type-D BHs:
\begin{equation}\label{eq:GFoATE}
  \left[ {\frac{d}{{d\Theta }}\left( {{\nabla}\frac{d}{{d\Theta }}} \right) + U(\Theta )} \right]S_{\ell m} = 0.
\end{equation}
And the function ${R_{\ell m }}(r)$ is the radial function satisfying the general form of radial Teukolsky equations (RTEs) in Type-D BHs:
\begin{equation}\label{eq:GFoRTE}
 \left[ {\Delta^{ - s + 1}\frac{d}{{dr}}\left( {\Delta^{s + 1}\frac{d}{{dr}}} \right) + V(r)} \right]{R_{\ell m }} = 0,
\end{equation}
where ${\nabla}$ and $\Delta$ are power series in $\Theta$ and $r$, respectively.
Solving $\Delta=0$ yields the horizon solutions of the Type-D BH.
When there exist two BH horizons, both \Cref{eq:GFoATE,eq:GFoRTE} can be transformed into the confluent Heun equation (CHE). For the case of three or four BH horizons, they can be transformed into the general Heun equation.
This paper specifically focuses on the Teukolsky equation for Type-D BHs, which include two horizons. This implies,
\begin{subequations}\label{eq:Delta2}
\begin{align}
{\Delta} &= ( {r - {r_ - }} )( {r - {r_ + }} ),\\
{\nabla} &= (1 - \Theta )(1 + \Theta ),
\end{align}
\end{subequations}
where $r_- $ is the inner horizon and $r_+ $ is the outer (event) horizon.

\subsection{Confluent Heun Equations}

The CHE is derived from the general Heun equation through a confluence process \cite{Ronveaux:1995,slavyanov2000unified,olver2010nist}.
The CHE has many standard forms, and we adopt the mathematical form used in the software \textit{Maple}.
This form can be written as
\begin{align}\label{eq:HeunC-Eq}
&\mathbb{H} '' - \frac{{ { - {x^2}\alpha  + \left( { - 2 - \beta  - \gamma  + \alpha } \right)x + 1 + \beta } }}{{x\left( {x - 1} \right)}}\mathbb{H}'\nonumber\\
 &  - \big{[} \left( {\left( { - 2 - \beta  - \gamma } \right)\alpha  - 2\delta } \right)x
 + \left( {\beta  + 1} \right)\alpha
\nonumber\\
 &    + \left( { - \gamma  - 1} \right)\beta  - \gamma  - 2\eta \big{]}\frac{\mathbb{H} }{{2x\left( {x - 1} \right)}} = 0.
\end{align}
where $\mathbb{H} (x) = {\rm{HeunC}}(\alpha, \beta,\gamma,\delta,\eta;x) $.
The parameters $(\alpha ,\beta ,\gamma ,\delta ,\eta)$ introduced in this form \eqref{eq:HeunC-Eq} reveal various symmetries of the confluent Heun function, which cannot be represented by other standard forms.
\begin{enumerate}
  \item The symmetry of the parameter $\alpha$ in CHE \eqref{eq:HeunC-Eq} can be given as follows,
\begin{align}\label{eq:symmetry_HeunC_alpha}
  {{\rm HeunC}} (\alpha ,\beta ,\gamma ,\delta ,\eta;x) = {{\rm e}^{ - \alpha x}}  {{\rm HeunC}}( - \alpha ,\beta ,\gamma ,\delta,&\eta; \,x), \nonumber \\
   \beta \notin \mathbb{Z}.&
\end{align}
where $\mathbb{Z}$ is the set of integers.

  \item  The symmetry of the parameter $\gamma$ in CHE \eqref{eq:HeunC-Eq} can be given as follows,
  \begin{align}\label{eq:symmetry_HeunC_gamma}
  {\rm{HeunC}}( {\alpha ,\beta ,\gamma ,\delta ,\eta;x}) ={( {1 - x})^{ - \gamma }}{\rm{HeunC}}( \alpha ,\beta , - &\gamma ,\delta,\eta;x ), \nonumber \\
  \beta \notin \mathbb{Z}.&
\end{align}
   \item The symmetry of the parameter $\beta$ is only displayed in the Teukolsky equations. It can be seen that the parameters $(\alpha,\beta,\gamma)$ in the general solution \eqref{eq:GSol_ATE} of the ATE \eqref{eq:GFoATE} and the general solution \eqref{eq:GSol_RTE} of the RTE \eqref{eq:GFoRTE} are interchangeable.
       The symmetry properties of the general solution $\mathbb{S} _{\ell m}$ and ${\mathbb{R} _{\ell m }}$ are as follows,
\begin{subequations}\label{eq:symetry_ATE}
  \begin{align}
&{\mathbb{S} _{\ell m}}(\alpha ,\beta ,\gamma ;x) = {\mathbb{S} _{\ell m}}( - \alpha ,\beta ,\gamma ;x),\\
&{\mathbb{S} _{\ell m}}(\alpha ,\beta ,\gamma ;x) = {\mathbb{S} _{\ell m}}(\alpha ,\beta , - \gamma ;x),\\
&{\mathbb{S} _{\ell m}}(\alpha ,\beta ,\gamma ;x) = {\mathbb{S} _{\ell m}}(\alpha , - \beta ,\gamma ;x),\,\,\,{\rm if} \,\,\,{D_1} = {D_2}.
  \end{align}
\end{subequations}
and
\begin{subequations}\label{eq:symetry_RTE}
  \begin{align}
&{\mathbb{R} _{\ell m }}(\alpha ,\beta ,\gamma ;x) = {\mathbb{R} _{\ell m }}( - \alpha ,\beta ,\gamma ;x),\\
&{\mathbb{R} _{\ell m }}(\alpha ,\beta ,\gamma ;x) = {\mathbb{R} _{\ell m }}(\alpha ,\beta , - \gamma ;x),\\
&{\mathbb{R} _{\ell m }}(\alpha ,\beta ,\gamma ;x) = {\mathbb{R} _{\ell m }}(\alpha , - \beta ,\gamma ;x),\,\,\,{\rm if} \,\,\, {C_1} = {C_2}.
 \end{align}
\end{subequations}
where $\{D_1,D_2\}$ and $\{C_1,C_2\}$ are the constants in the general solutions \eqref{eq:GSol_ATE} and \eqref{eq:GSol_RTE}, respectively.

\end{enumerate}

\subsection{HeunC Solutions of ATEs and RTEs}
The first step in converting the Teukolsky equations \eqref{eq:GFoATE} and \eqref{eq:GFoRTE} into the CHEs is to perform a M\"{o}bius (isomorphic) transformation, which is a linear fractional transformation,
\begin{equation}\label{eq:mtran0}
x = \frac{{  \hat{a}y +\hat{b}}}{\hat{c}y+\hat{d}},
\end{equation}
where $y={\Theta }$ for ATEs \eqref{eq:GFoATE} and $y={r }$ for RTEs \eqref{eq:GFoRTE}.
The quantities $\{ \hat a, \hat b, \hat c, \hat d \}$ represent constant parameters.

Next, we can use the S-homotopic transformation to construct the exact solutions for ATEs and RTEs.
The exact solution of the ATE \eqref{eq:GFoATE} can be expressed as
\begin{equation}\label{eq:Sm-Sol}
  S_{\ell m}(\Theta ) = S_\theta (\alpha, \beta,\gamma;x){\mathbb{H}}(x),
\end{equation}
and the S-homotopic transformation in $\theta$ direction,
\begin{equation}\label{eq:S-homotopic-theta-d2}
S_\theta (\alpha, \beta,\gamma;x) = {( - x)^{\tfrac{1}{2}\beta }}{(1 - x)^{\tfrac{1}{2}\gamma }}{{\mathop{\rm e}\nolimits} ^{\tfrac{1}{2}\alpha x}}.
\end{equation}
The exact solution of the RTE \eqref{eq:GFoRTE} can be expressed as
\begin{equation}\label{eq:Rm-Sol}
  {R_{\ell m }}(r) = S_r(\alpha, \beta,\gamma;x){\mathbb{H}}(x),
\end{equation}
and the S-homotopic transformation in $r$ direction,
\begin{equation}\label{eq:S-homotopic-r}
 S_{r}(\alpha, \beta,\gamma;x) = {\left( { - x} \right)^{ \tfrac{1}{2}(\beta-s) }}{\left( {1 - x} \right)^{\tfrac{1}{2}(\gamma-s) }}{{\rm{e}}^{\tfrac{1}{2}\alpha x}}.
\end{equation}
For conciseness, $S_{r}(\alpha, \beta,\gamma;x)$ is abbreviated as $S_{r}(\alpha, \beta,\gamma)$ in subsequent expressions.

\subsubsection{Eigenvalues of ATEs}
The general solution ${{\mathbb{S}}_{\ell m}}$ of the ATE \eqref{eq:GFoATE} is derived from the linear combination of two linearly independent particular solutions,
\begin{align}\label{eq:GSol_ATE}
{{\mathbb{S}}_{\ell m}} &= {D_1}S_\theta (\alpha, \beta,\gamma;x){\rm{HeunC}}(\alpha ,\beta ,\gamma ,\delta ,\eta;x) \nonumber \\
&+ {D_2}S_\theta (\alpha, -\beta,\gamma;x){\rm{HeunC}}(\alpha , - \beta ,\gamma ,\delta ,\eta;x).
\end{align}
where $D_1$ and $D_2$ are constants. Because the ATE contains two regular singularities at $\Theta =+1$ and $-1$, the parameters included in the general solution \eqref{eq:GSol_ATE} for the singularity at $\Theta =+1$ are denoted with a subscript $+1$, and those for the singularity at $\Theta =-1$ with a subscript $-1$.

Firstly, the M\"{o}bius transformation around singularity $\Theta =+1$ can be expressed as
\begin{equation}\label{eq:M_trans_ATE_+1}
 x:= x_{+1} = \tfrac{1}{2}(1 - \Theta ).
\end{equation}
Substituting \Cref{eq:Sm-Sol,eq:M_trans_ATE_+1,eq:S-homotopic-theta-d2} into the ATE \eqref{eq:GFoATE}, we will obtain the CHE \eqref{eq:HeunC-Eq}. The two local solutions of the CHE \eqref{eq:HeunC-Eq} can be represented as
\begin{widetext}
\begin{subequations}\label{eq:GSol_RTE-ATE-1}
\begin{align}
&{\rm HeunC}(\alpha_{+1} ,\beta_{+1} ,\gamma_{+1} ,\delta_{+1} ,\eta_{+1} ;x_{+1}) ,\\
&(x_{+1})^{-\beta_{+1}}{\rm HeunC}(\alpha_{+1} ,-\beta_{+1} ,\gamma_{+1} ,\delta_{+1} ,\eta_{+1} ;x_{+1}).
\end{align}
\end{subequations}
Therefore, two linearly independent solutions ${S_{\ell m}}({x_{ + 1}})$ of the ATE \eqref{eq:GFoATE} around $\Theta=+1$ can be represented as
\begin{subequations}
\begin{align}
&{S_\theta }({\alpha _{ + 1}},{\beta _{ + 1}},{\gamma _{ + 1}};{x_{ +1}}){\rm{HeunC}}({\alpha _{ + 1}},{\beta _{ + 1}},{\gamma _{ + 1}},{\delta _{ + 1}},{\eta _{ + 1}};{x_{ + 1}}),\label{eq:ATE_sol_A1}\\
&{S_\theta }({\alpha _{ + 1}}, - {\beta _{ + 1}},{\gamma _{ + 1}};{x_{+1}}){\rm{HeunC}}({\alpha _{ + 1}}, - {\beta _{ + 1}},{\gamma _{ + 1}},{\delta _{ + 1}},{\eta _{ + 1}};{x_{ + 1}}).\label{eq:ATE_sol_A2}
\end{align}
\end{subequations}

Similarly, M\"{o}bius transformation around singularity $\Theta =-1$ can be expressed as
\begin{equation}
 x:= x_{-1} = \tfrac{1}{2}(1 + \Theta ).
\end{equation}
The two local solutions of the ATE \eqref{eq:GFoATE} around the regular singularity $\Theta=-1$ can also be represented as
\begin{subequations}\label{eq:GSol_RTE-ATE-2}
\begin{align}
&{\rm HeunC}(\alpha_{-1} ,\beta_{-1} ,\gamma_{-1} ,\delta_{-1} ,\eta_{-1} ; x_{-1}) ,\\
&(x_{-1})^{-\beta_{-1}}{\rm HeunC}(\alpha_{-1} ,-\beta_{-1} ,\gamma_{-1} ,\delta_{-1} ,\eta_{-1} ; x_{-1}).
\end{align}
\end{subequations}

Therefore, two linearly independent solutions ${S_{\ell m}}({x_{ - 1}})$ of the ATE \eqref{eq:GFoATE} around $\Theta=-1$ can be represented as
\begin{subequations}
\begin{align}
&{S_\theta }({\alpha _{ - 1}},{\beta _{ - 1}},{\gamma _{ - 1}};{x_{ - 1}}){\rm{HeunC}}({\alpha _{ - 1}},{\beta _{ - 1}},{\gamma _{ - 1}},{\delta _{ - 1}},{\eta _{ - 1}};{x_{ - 1}}),  \label{eq:ATE_sol_B1} \\
&{S_\theta }({\alpha _{ - 1}}, - {\beta _{ - 1}},{\gamma _{ - 1}};{x_{ - 1}}){\rm{HeunC}}({\alpha _{ - 1}}, - {\beta _{ - 1}},{\gamma _{ - 1}},{\delta _{ - 1}},{\eta _{ - 1}};{x_{ - 1}}).\label{eq:ATE_sol_B2}
\end{align}
\end{subequations}

Using these solutions of the ATE \eqref{eq:GFoATE}, we can calculate the spin-weighted spheroidal harmonics and the angular separation constants ${}_s A_{\ell m}(a\omega)$.
Solutions will be regular at both poles\footnote{Boundary conditions for the ATE \eqref{eq:GFoATE} are that $S_{\ell m}$ be finite at both poles $\Theta=+1$ and $\Theta=-1$, where the indices are $\pm \tfrac{1}{2}\beta$ and $\pm \tfrac{1}{2}\gamma$, respectively. }, if and only if ${S_{\ell m}}({x_{ - 1}}) = {\rm constant}\,\times {S_{\ell m}}({x_{ + 1}})$, or, equivalently, if the Wronskian vanishes \cite{Fiziev:2009kh}.
To avoid the multivalued nature caused by parameters $\beta$ and $\gamma$, the absolute values of $\beta$ and $\gamma$ must be taken when calculating the Wronskian. Choose one solution from \cref{eq:GSol_RTE-ATE-1} and one solution from \cref{eq:GSol_RTE-ATE-2}, and use them to constitute the Wronskian.
\begin{equation}
  {W_\theta } = {S_{\ell m}}({x_{ - 1}})\frac{{\rm{d}}}{{{\rm{d}}\Theta }}{S_{\ell m}}({x_{ + 1}}) - {S_{\ell m}}({x_{ + 1}})\frac{{\rm{d}}}{{{\rm{d}}\Theta }}{S_{\ell m}}({x_{ - 1}}).
\end{equation}
To obtain a concise expression of the Wronskian, we select these two solutions.
\begin{align}
 &{S_{\ell m}}({x_{ + 1}}) = {S_\theta }({\alpha _{ + 1}},\left| {{\beta _{ + 1}}} \right|,\left| {{\gamma _{ + 1}}} \right|;{x_{ - 1}}){\rm{HeunC}}({\alpha _{ + 1}},\left| {{\beta _{ + 1}}} \right|,\left| {{\gamma _{ + 1}}} \right|,{\delta _{ + 1}},{\eta _{ + 1}};{x_{ + 1}}),\\
   &{S_{\ell m}}({x_{ - 1}}) = {S_\theta }({\alpha _{ - 1}},\left| {{\beta _{ - 1}}} \right|,\left| {{\gamma _{ - 1}}} \right|;{x_{ - 1}}){\rm{HeunC}}({\alpha _{ - 1}},\left| {{\beta _{ - 1}}} \right|,\left| {{\gamma _{ - 1}}} \right|,{\delta _{ - 1}},{\eta _{ - 1}};{x_{ - 1}}).
   \label{eq:ATE_W2}
\end{align}
Select $\Theta=0$ and by eliminating the common factors, the Wronskian can be written as
\begin{align}\label{eq:ATE_Wronskian_0}
 {W_\theta } &= {\rm{HeunC}}({\alpha _{ - 1}},\left| {{\beta _{ - 1}}} \right|,\left| {{\gamma _{ - 1}}} \right|,{\delta _{ - 1}},{\eta _{ - 1}};\tfrac{1}{2}){\rm{HeunCPrime}}({\alpha _{ + 1}},\left| {{\beta _{ + 1}}} \right|,\left| {{\gamma _{ + 1}}} \right|,{\delta _{ + 1}},{\eta _{ + 1}};\tfrac{1}{2}) \nonumber \\
 &+ {\rm{HeunC}}({\alpha _{ + 1}},\left| {{\beta _{ + 1}}} \right|,\left| {{\gamma _{ + 1}}} \right|,{\delta _{ + 1}},{\eta _{ + 1}};\tfrac{1}{2})\left[ {{\rm{HeunCPrime}}(  {\alpha _{ - 1}},\left| {{\beta _{ - 1}}} \right|,\left| {{\gamma _{ - 1}}} \right|,{\delta _{ - 1}},{\eta _{ - 1}};\tfrac{1}{2})} \right. \nonumber \\
&\left. { + \left( {\tfrac{{{\alpha _{ + 1}}}}{2} + \tfrac{{{\alpha _{ - 1}}}}{2} + \left| {{\beta _{ + 1}}} \right| + \left| {{\beta _{ - 1}}} \right| - \left| {{\gamma _{ + 1}}} \right| - \left| {{\gamma _{ - 1}}} \right|} \right){\rm{HeunC}}({\alpha _{ + 1}},\left| {{\beta _{ + 1}}} \right|,\left| {{\gamma _{ + 1}}} \right|,{\delta _{ + 1}},{\eta _{ + 1}};\tfrac{1}{2})} \right].
\end{align}
Using the symmetry \eqref{eq:symetry_ATE} of ${\mathbb{S} _{\ell m}}$ and selecting appropriate HeunC parameters (${\alpha _{ + 1}} =  - {\alpha _{ - 1}},\left| {{\beta _{ + 1}}} \right| =  - \left| {{\beta _{ - 1}}} \right|,\left| {{\gamma _{ + 1}}} \right| =  - \left| {{\gamma _{ - 1}}} \right|$), the factor $(\frac{{{\alpha _{ + 1}}}}{2} + \frac{{{\alpha _{ - 1}}}}{2} + \left| {{\beta _{ + 1}}} \right| + \left| {{\beta _{ - 1}}} \right| - \left| {{\gamma _{ + 1}}} \right| - \left| {{\gamma _{ - 1}}} \right|)$ can be eliminated. Lastly, the Wronskian can be simplified as
\begin{align}\label{eq:ATE_Wronskian}
 {W_\theta } &={\rm{HeunC}}(  {\alpha _{ - 1}},\left| {{\beta _{ - 1}}} \right|,\left| {{\gamma _{ - 1}}} \right|,{\delta _{ - 1}},{\eta _{ - 1}};\tfrac{1}{2}){\rm{HeunCPrime}}({\alpha _{ + 1}},\left| {{\beta _{ + 1}}} \right|,\left| {{\gamma _{ + 1}}} \right|,{\delta _{ + 1}},{\eta _{ + 1}};\tfrac{1}{2}) \nonumber \\
 &+{\rm{HeunC}}({\alpha _{ + 1}},\left| {{\beta _{ + 1}}} \right|,\left| {{\gamma _{ + 1}}} \right|,{\delta _{ + 1}},{\eta _{ + 1}};\tfrac{1}{2}){\rm{HeunCPrime}}(  {\alpha _{ - 1}},\left| {{\beta _{ - 1}}} \right|,\left| {{\gamma _{ - 1}}} \right|,{\delta _{ - 1}},{\eta _{ - 1}};\tfrac{1}{2}).
\end{align}
To solve $ {W_\theta }=0$, we can obtain the angular separation constants ${}_s A_{\ell m}(a\omega)$ of the ATE \eqref{eq:GFoATE}.
\end{widetext}

\subsubsection{QNM and TTM Conditions of RTEs}\label{sec:BC_RTE}
The asymptotic solutions for the RTE \eqref{eq:GFoRTE} can be expressed as
\begin{equation}
{R_{\ell m}} \to \left\{ \begin{array}{l}
Z_{{\rm{in}}}^{\rm{H}}R_{{\rm{in}}}^{\rm{H}} + Z_{{\rm{up}}}^{\rm{H}}R_{{\rm{up}}}^{\rm{H}},\quad r \to {r_ + },\\
Z_{{\rm{in}}}^\infty R_{{\rm{in}}}^\infty  + Z_{{\rm{up}}}^\infty R_{{\rm{up}}}^\infty ,\quad r \to \infty.
\end{array} \right.
\end{equation}
where $Z_{{\rm{in}}}^{\rm{H}}$ and $Z_{{\rm{up}}}^{\rm{H}}$ are complex amplitudes of the ingoing wave $R_{{\rm{in}}}^{\rm{H}}$ and the outgoing wave $R_{{\rm{up}}}^{\rm{H}}$ near the horizon, respectively.  $Z_{{\rm{in}}}^\infty$ and $Z_{{\rm{up}}}^\infty$ are complex amplitudes of the ingoing wave $R_{{\rm{in}}}^\infty$ and the outgoing wave $R_{{\rm{up}}}^\infty$ near infinity, respectively.

We summarize the boundary conditions of three types of complex frequencies \cite{Berti:2009kk}:

\noindent 1)  The standard QNM is defined by purely ingoing waves at the horizon and purely outgoing waves at infinity.
The corresponding boundary conditions \cite{Cook:2014cta} for the QNMs can be expressed as
\begin{equation}\label{eq:QNM_Conditon}
\left\{ \begin{array}{lcl}
\mathop {\lim }\limits_{r \to {r_ + }} {R_{\ell m}} \to Z_{{\rm{in}}}^{\rm{H}}R_{{\rm{in}}}^{\rm{H}}, \\
\mathop {\lim }\limits_{r \to \infty } {R_{\ell m}} \to Z_{{\rm{up}}}^\infty R_{{\rm{up}}}^\infty   .
\end{array} \right.
\end{equation}
\noindent 2)  ${\rm{TTM}_{\rm{L}}}$ of gravitational perturbations for the Weyl scalar $\psi_4$ is defined by purely outgoing waves at the horizon and infinity.
The corresponding boundary conditions for the ${\rm{TTM}_{\rm{L}}}$ can be expressed as
\begin{equation}\label{eq:TTM_L_BC}
\left\{ {\begin{array}{*{20}{lcl}}
{\mathop {\lim }\limits_{r \to {r_ + }} {R_{\ell m}} \to Z_{{\rm{up}}}^{\rm{H}}R_{{\rm{up}}}^{\rm{H}}},\\
{\mathop {\lim }\limits_{r \to \infty } {R_{\ell m}} \to Z_{{\rm{up}}}^\infty R_{{\rm{up}}}^\infty }.
\end{array}} \right.
\end{equation}

\noindent 3)  ${\rm{TTM}_{\rm{R}}}$ of gravitational perturbations for the Weyl scalar $\psi_0$ is defined by purely ingoing waves at the horizon and infinity.
The corresponding boundary conditions for the ${\rm{TTM}_{\rm{R}}}$ can be expressed as
\begin{equation}\label{eq:TTM_R_BC}
\left\{ {\begin{array}{*{20}{lcl}}
{\mathop {\lim }\limits_{r \to {r_ + }} {R_{\ell m}} \to Z_{{\rm{in}}}^{\rm{H}}R_{{\rm{in}}}^{\rm{H}}},\\
{\mathop {\lim }\limits_{r \to \infty } {R_{\ell m}} \to Z_{{\rm{in}}}^\infty R_{{\rm{in}}}^\infty }.
\end{array}} \right.\end{equation}
For Kerr BHs, the boundary (necessary and sufficient) conditions of the TTMs are, in fact, equivalent to the vanishing of the Starobinsky-Teukolsky constant (STC) \cite{Wald:1973wwa,Chandrasekhar:1984mgh}. The STC can be written as \cite{Teukolsky:1974yv,Chandrasekhar:1984mgh,Cook:2018ses,Cook:2022kbb}
\begin{align}\label{eq:Starobinsky_const}
  {|{\bf{C}}_{\rm ST}|}^2=\lambdabar^2(\lambdabar&+2)^2+ 8\lambdabar{a}\omega\left(6({a}\omega+m) -5\lambdabar({a}\omega-m)\right)\nonumber\\
      &+ 144\omega^2\left(1+{a}^2({a}\omega-m)^2\right)=0,
\end{align}
where
\begin{subequations}\label{eq:Starobinsky_const}
  \begin{align}
    &\lambdabar= {{\kern 1pt} _s}{A_{\ell m}}(a\omega )+ a^2\omega^2 - 2ma\omega,  \quad\quad & s=-2,  \\
    &\lambdabar={{\kern 1pt} _s}{A_{\ell m}}(a\omega )+ a^2\omega^2 - 2ma\omega+2s, \quad &s=+2.
  \end{align}
\end{subequations}

For $\psi \sim{e^{ - i\omega t}}$, when solving the TTMs using \cref{eq:Starobinsky_const}, the negative spin-weight case ($s = -2$) corresponds to ${\rm{TTM}_{\rm{L}}}$ solutions, whereas the positive spin-weight case ($s =2$) corresponds to ${\rm{TTM}_{\rm{R}}}$ solutions.

\subsubsection{HeunC Solutions of RTEs}\label{sec:HeunC_Sol_RTE}

The general solution ${{\mathbb{R}}_{\ell m }}$ of the RTE \eqref{eq:GFoRTE} is derived from the linear combination of two linearly independent particular solutions around the singularity $x=0$,
\begin{align}\label{eq:GSol_RTE}
&{{\mathbb{R}}_{\ell m }}={C_1}S_r (\alpha_r, \beta_r,\gamma_r){\rm{HeunC}}(\alpha_r,\beta_r,\gamma_r,\delta_r,\eta_r;x_r) \nonumber\\
&+ {C_2}S_r(\alpha_r, -\beta_r,\gamma_r){\rm{HeunC}}(\alpha_r, - \beta_r,\gamma_r,\delta_r,\eta_r;x_r) .
\end{align}
where $C_1$ and $C_2$ are constants that can be determined based on different boundary conditions.

The radial S-homotopic transformation is given in \cref{eq:S-homotopic-r}.
Additionally, the M\"{o}bius transformation around singularity $x_r = 0$ can be expressed as
\begin{equation}\label{eq:mtran}
x_r =- \frac{{  r - r_+}}{{r_ + } - {r_ - }}.
\end{equation}

The HeunC parameters $(\alpha_r,\beta_r,\gamma_r,\delta_r,\eta_r;x_r)$ for the singularity $x_r = 0$ are denoted with a subscript $r$.
Although the RTE \eqref{eq:GFoRTE} has three singularities ($x_r = 0,1,\infty $), this paper focuses more on the local solutions around the singularity $x_r=0$. Because the singularity $x_r=0$ corresponds to the event horizon, this facilitates the discussion of the QNM conditions.
A more detailed derivation of the general solution \eqref{eq:GSol_RTE} of the RTE \eqref{eq:GFoRTE} can be seen in our previous work \cite{Chen:2023ese,Chen:2023lsa}.

The selection of the general solution \eqref{eq:GSol_RTE} is advantageous in constructing the solutions that satisfy the pure ingoing wave condition at the horizon, and the pure outgoing wave condition at infinity,
\begin{align}
    & R_{\ell m }^{{\rm{in}}}\to \left\{ \begin{array}{*{20}{l}}
{B_{\ell m}^{{\rm{trans}}}R_{{\rm{in}}}^{\rm{H}},}&{r \to {r_ + },}\\
{B_{\ell m }^{{\rm{ref}}}R_{{\rm{up}}}^\infty  + B_{\ell m }^{{\rm{inc}}}R_{{\rm{in}}}^\infty ,}&{r \to  + \infty ,}
\end{array} \right.\label{eq:boundary1}\\
 & R_{\ell m }^{{\rm{up}}} \to \left\{
\begin{array}{*{20}{l}}
{C_{\ell m}^{{\rm{up}}} R_{{\rm{in}}}^{\rm{H}} + C_{\ell m}^{{\rm{ref}}}R_{{\rm{up}}}^{\rm{H}},}&{r \to {r_ + },}\\
{C_{\ell m}^{{\rm{trans}}}R_{{\rm{up}}}^{\rm{H}},}&{r \to  + \infty .}
\end{array}\right.\label{eq:boundary2}
\end{align}
Using the asymptotic property of the confluent Heun function,
\begin{equation}\label{eq:Asy-HC-Hor}
   \mathop {\lim }\limits_{x_r \to 0} {\rm{ HeunC}}(\alpha_r,\beta_r,\gamma_r,\delta_r,\eta_r;x_r) = 1,
\end{equation}
and the boundary conditions \eqref{eq:boundary1} of the ingoing wave, we can construct the ingoing wave solution $R_{\ell m }^{{\rm{in}}}$.
$R_{\ell m }^{{\rm{in}}}$ at the horizon has purely ingoing property, which requires that $C_2=0$ \footnote{Actually, it is only after knowing the metric of the BH that we can identify the part in the general solution as $R_{\ell m }^{{\rm{in}}}$.}. Thus, the ingoing wave solution $R_{\ell m }^{{\rm{in}}}$ is given by
\begin{equation}\label{eq:uHor}
R_{\ell m }^{{\rm{in}}} = S_r (\alpha_r, \beta_r,\gamma_r){\rm{HeunC}}(\alpha_r,\beta_r,\gamma_r,\delta_r,\eta_r;x_r).
\end{equation}
Based on the symmetry \eqref{eq:symetry_RTE} of the general solution, we conclude that $R_{\ell m}^{{\rm{in}}}(\alpha_r ) = R_{\ell m}^{{\rm{in}}}( - \alpha_r )$.

The boundary conditions of the QNMs require that $R_{\ell m }^{{\rm{in}}} $ has no ingoing waves at infinity or $R_{\ell m }^{{\rm{up}}} $ has no outgoing waves at the event horizon, i.e., ${{B_{\ell m }^{{\rm{inc}}}}}=0$ or $C_{\ell m }^{{\rm{up}}}=0$.
If we use $C_{\ell m }^{{\rm{up}}}=0$ as the boundary condition for QNMs, we then need to solve for $C_1$ and $C_2$ in $R_{\ell m }^{{\rm{up}}}$, which is a complicated and tedious calculation. Therefore, using ${{B_{\ell m }^{{\rm{inc}}}}}=0$ as the boundary condition for QNMs is beneficial for solving the QNMs.

If the analytical expression for ${{B_{\ell m }^{{\rm{inc}}}}}$ is known, we can directly solve the QNMs.
Initially, Japanese scholars derived the asymptotic amplitudes (${{B_{\ell m }^{{\rm{inc}}}}}$ and ${{B_{\ell m }^{{\rm{ref}}}}}$) at infinity for $R_{\ell m }^{{\rm{in}}} $ by using the Mano-Suzuki-Takasugi (MST) method \cite{Mano1996RWE,Mano_1996,Fujita_2004,Fujita_2005}. The MST method is a power-series method based on special functions.
In the previous work \cite{Chen:2023ese}, we combined the MST method with the confluent Heun function, deriving all asymptotic amplitudes by incorporating the HeunC parameter as input.
The analytical expression for ${{B_{\ell m }^{{\rm{inc}}}}}$ introduces a parameter $\nu$, which represents the renormalized angular momentum and appears simultaneously with the QNM frequencies $\omega$ in the transcendental equation \eqref{eq:recurrence-fnv}.
Without an additional constraint equation, it is impossible to solve $\nu$ and $\omega$.
Even if a constraint equation for $\nu$ is imposed, using the MST method to solve the QNMs is not substantially different from Leaver's continued fraction method. Moreover, this method is more complicated, as the MST method involves two-sided infinite series and the calculation of the parameter $\nu$, while the Leaver method is a single-sided infinite series.
For further discussions on this parameter $\nu$, please refer to \Cref{sec:renormalized_AM}.

The asymptotic amplitudes (${{B_{\ell m }^{{\rm{inc}}}}}$ and ${{B_{\ell m }^{{\rm{ref}}}}}$) of $R_{\ell m }^{{\rm{in}}} $ are essentially related to a connection problem of the confluent Heun function. Besides the MST method, there exists an alternative approach.
In 2022, Bonelli et al. derived connection formulas for Heun-class functions using the Liouville conformal field theory \cite{Bonelli:2022ten}.
However, similar to the MST method, this connection formula also introduces additional parameters; for the MST method, it is the renormalized angular momentum $\nu$, while for Bonelli et al.'s connection formula, it involves the parameter $a(m_i, u, L)$ of the Nekrasov-Shatashvili function. The relationship \cite{Bautista:2023sdf} between the two is $a = \tfrac{1}{2} - \nu$.
Both methods employ the power series method of special functions, and their comparison can be found in Ref. \cite{Bautista:2023sdf}.
Furthermore, for QNM calculations using these two connection formulas, the QNM results derived from the Nekrasov-Shatashvili technology are documented in Ref. \cite{Aminov:2020yma,Bonelli:2021uvf}, whereas those obtained through the MST method are reported in Ref. \cite{Casals:2019vdb}.

\subsection{HeunC Method for Solving QNMs}\label{sec:HeunC-QNM}
Fiziev proposed another form of the asymptotic amplitude ${{B_{\ell m }^{{\rm{inc}}}}}$ to solve the QNMs \cite{Fiziev:2005ki}.
This form involves analytically extending the boundary conditions \eqref{eq:boundary1} for $R_{\ell m }^{{\rm{in}}} $ at infinity into the complex plane. By selecting a specific phase angle, the outgoing wave part ($R_{{\rm{up}}}^\infty$) is eliminated. Then, ${{B_{\ell m }^{{\rm{inc}}}}}$ can be expressed by $\mathop {\lim }\limits_{r \to \infty } \frac{{R_{\ell m }^{{\rm{in}}}}}{R_\infty ^{\rm{in}}}$, thereby constructing the QNMs equation.
However, his method has certain limitations:
\begin{enumerate}
  \item It cannot calculate algebraically special (AS) modes \cite{Fiziev:2011mm}.
  \item It cannot calculate high-overtone modes \cite{Staicova:2014ioa} of Kerr BHs.
  \item Its computational efficiency is too low \cite{Fiziev:2010yy}.
\end{enumerate}
Fiziev used the HeunC function from the software \textit{Maple}, and the numerical algorithm of the HeunC function provided by \textit{Maple} has slow convergence and limited parameter range. Therefore, Fiziev proposed a new algorithm to improve the computational efficiency and performance of the HeunC function \cite{Fiziev:2010yy}.
However, compared with our method (see \Cref{tab:Sch_QNM,tab:Sch_QNM_err}), the computational time of Fiziev's results is still too long.
For example, when calculating the first twelve QNMs of Schwarzschild BHs with the same level of precision, our method takes less than 1 second without using parallel computing, and the time will be even shorter with parallel computing. In contrast, the computational time for Fiziev's method is 4959 seconds \cite{Fiziev:2010yy}.

\subsubsection{New HeunC Method}

Next, we will use the confluent Heun function to construct an alternative form of the asymptotic amplitude ${{B_{\ell m }^{{\rm{inc}}}}}$. Although it adopts the idea of analytical continuation by Fiziev, it is different from his form.

First, we consider $\psi \sim {e^{ - i\omega t}}$.
Because the QNM condition \eqref{eq:QNM_Conditon} is $B_{\ell m }^{{\rm{inc}}}=0$, \cref{eq:boundary1} can be written as
 \begin{equation}\label{eq:QNM-Binc-MFD}
 B_{\ell m }^{{\rm{inc}}} = \mathop {\lim }\limits_{r \to \infty }  \left( {\frac{{R_{\ell m }^{{\rm{in}}}}}{R_{\rm{in}}^\infty} - B_{\ell m }^{{\rm{ref}}}\frac{R_{\rm{up}}^\infty}{R_{\rm{in}}^\infty}} \right)=0.
\end{equation}
where $R_{\rm{in}}^\infty = g_1({r^{ - 1}}){{\rm{e}}^{-i\omega {r_*}}} $ and $R_{\rm{in}}^\infty= g_2({r^{ - 1}}){{\rm{e}}^{ i\omega {r_*}}}$.
For Kerr BHs, $g_1({r^{ - 1}})={r^{ - 1}}$ and $g_2({r^{ - 1}})={r^{ - 1 - 2s}}$ can be obtained by performing an asymptotic analysis of the RTE \eqref{eq:GFoRTE} at infinity \cite{Teukolsky:1973ha}.

We assume that
\begin{equation}\label{eq:QNM-assume}
   \mathop {\lim }\limits_{r \to \infty } \frac{{R_{{\rm{up}}}^\infty }}{{R_{{\rm{in}}}^\infty }} = \mathop {\lim }\limits_{r \to \infty } \frac{{{g_2}({r^{ - 1}})}}{{{g_1}({r^{ - 1}})}}{{\rm{e}}^{2i\omega {r_*}}} = 0,
\end{equation}
 a constraint condition can be established,
\begin{equation}\label{eq:QNM-Cond-MFD}
\arg(r) = 2{\rm k}\pi + \frac{\pi}{2} - \arg(\omega ), \quad {\rm{k}} \in \mathbb{Z}.
\end{equation}

Therefore, we obtain the HeunC equation of the QNMs,
 \begin{equation}\label{eq:QNM-HeunC-MFD}
  B_{\ell m }^{{\rm{inc}}} = \mathop {\lim }\limits_{r \to \infty } \frac{{R_{\ell m }^{{\rm{in}}}}}{R_\infty ^{\rm{in}}} = 0.
 \end{equation}
Before deriving the simplified form for \cref{eq:QNM-HeunC-MFD}, it is necessary to discriminate the positivity or negativity of the HeunC parameters $(\alpha_r ,\beta_r ,\gamma_r )$. This step is particularly important as it will make it clear to the reader how the HeunC parameters $(\alpha_r ,\beta_r ,\gamma_r )$ are selected.
Since any Type-D BH can degenerate into a Schwarzschild BH under the Schwarzschild limit (is represented by $\mathop {\lim }\limits_{{\rm{BH}} \to {\rm{Sch}}}$), the positivity or negativity of the HeunC parameters can be discriminated based on the Schwarzschild limit.
\begin{subequations}\label{eq:HeunC_parameter_pm}
  \begin{align}
&\mathop {\lim }\limits_{{\rm{BH}} \to {\rm{Sch}}} {\alpha _r}^+ = {\alpha _{{\rm{Sch}}}} = 2i\omega {r_ + },\\
&\mathop {\lim }\limits_{{\rm{BH}} \to {\rm{Sch}}} {\alpha _r}^- =  - {\alpha _{{\rm{Sch}}}} =  - 2i\omega {r_ + }\\
&\mathop {\lim }\limits_{{\rm{BH}} \to {\rm{Sch}}} {\beta _r}^+ =    {\beta _{{\rm{Sch}}}} = -s - 2i\omega {r_ + },\\
&\mathop {\lim }\limits_{{\rm{BH}} \to {\rm{Sch}}} {\beta _r}^- =  - {\beta _{{\rm{Sch}}}} =  s + 2i\omega {r_ + },\\
&\mathop {\lim }\limits_{{\rm{BH}} \to {\rm{Sch}}} {\gamma _r}^+ = {\gamma _{{\rm{Sch}}}} = s,\\
&\mathop {\lim }\limits_{{\rm{BH}} \to {\rm{Sch}}} {\gamma _r}^- =  - {\gamma _{{\rm{Sch}}}} =  - s.
\end{align}
\end{subequations}
Even outside the Schwarzschild limit, these coefficients with $+$ and $-$ superscripts are mutually opposite in sign, for instance: $\alpha_r^+ = -\alpha_r^-$.

In the previous work \cite{Chen:2023ese}, the asymptotic analytical expressions (connection formulas) of the confluent Heun function at infinity were presented,
\begin{align}\label{eq:hc-hor-inf}
&\mathop {\lim }\limits_{|x_r| \to \infty } {\rm{HeunC}}(\alpha_r,\beta_r,\gamma_r,\delta_r,\eta_r;x_r) \to \nonumber \\
  &D_ \odot ^{{\beta _r}}\;{x_r}^{ - \frac{{{\beta _r} + {\gamma _r} + 2}}{2} - \frac{{{\delta _r}}}{{{\alpha _r}}}}+ D_ \otimes ^{{\beta _r}}{{\rm{e}}^{ - {\alpha _r}{x_r}}}{x_r}^{ - \frac{{{\beta _r} + {\gamma _r} + 2}}{2} + \frac{{{\delta _r}}}{{{\alpha _r}}}},
\end{align}
where $D_ \odot ^{{\beta _r}}$ and $D_ \otimes ^{{\beta _r}}$ are the connection coefficients, and their analytical expressions are shown in Appendix \ref{app:AsymptoticFormula}.

Lastly, using \Cref{eq:uHor,eq:hc-hor-inf}, \cref{eq:QNM-HeunC-MFD} can be simplified as
\begin{subequations}\label{eq:HeunC_QNM}
  \begin{align}
  \tilde B_{\ell m }^{{\rm{inc}}} = \mathop {\lim }\limits_{|x_r| \to \infty }\frac{{{\rm{HeunC}}({\alpha _r},{\beta^+ _r},{\gamma^+_r},{\delta _r},{\eta _r},{x_r})}}{{{x_r}^{ - \frac{{{\beta^+ _r} + {\gamma^+_r} + 2}}{2} - \frac{{{\delta _r}}}{{{\alpha _r}}}}}} = 0&, \nonumber\\
   {\rm{if}}\,\,\alpha_r  = {\alpha^+_r }&; \\
 \tilde B_{\ell m }^{{\rm{inc}}} = \mathop {\lim }\limits_{|x_r| \to \infty }\frac{{{\rm{HeunC}}({\alpha _r},{\beta^+ _r},{\gamma^+_r},{\delta _r},{\eta _r},{x_r})}}{{{{\rm{e}}^{ - {\alpha _r}{x_r}}}{x_r}^{ - \frac{{{\beta^+ _r} + {\gamma^+_r} + 2}}{2} + \frac{{{\delta _r}}}{{{\alpha _r}}}}}} = 0&, \nonumber\\
{\rm{if}}\,\, \alpha_r  = {\alpha^- _r }&.
  \end{align}
\end{subequations}
where $\tilde B_{\ell m }^{{\rm{inc}}}$ represents $ B_{\ell m }^{{\rm{inc}}}$ obtained by the HeunC method.
It is reiterated that Fiziev's form \cite{Fiziev:2005ki} of $ B_{\ell m}^{\text{inc}} $ is different from our results.

{
However, in practical numerical computations, choosing the set of HeunC parameters associated with the complex conjugate QNM frequency occasionally yields higher precision.
These results are discussed in detail in Appendix \ref{app:CompDetail_A}.
Therefore, the corresponding constraint equations for the complex conjugate QNM frequencies, $\omega^*$, are:
\begin{subequations}
  \begin{align}
{\tilde B_{\ell m}^{{\rm{inc}}} = \mathop {\lim }\limits_{|{x_r}| \to \infty } \frac{{{\rm{HeunC}}({\alpha _r},\beta _r^- ,\gamma _r^+ ,{\delta _r},{\eta _r},{x_r})}}{{{{\rm{e}}^{ - {\alpha _r}{x_r}}}{x_r}^{ - \frac{{\beta _r^-  + \gamma _r^+  + 2}}{2} + \frac{{{\delta _r}}}{{{\alpha _r}}}}}} = 0}&, \nonumber\\
{{\rm{if}}{\kern 1pt} {\kern 1pt} {\alpha _r} = \alpha _r^+ }&;\\
{\tilde B_{\ell m}^{{\rm{inc}}} = \mathop {\lim }\limits_{|{x_r}| \to \infty } \frac{{{\rm{HeunC}}({\alpha _r},\beta _r^- ,\gamma _r^+ ,{\delta _r},{\eta _r},{x_r})}}{{{x_r}^{ - \frac{{\beta _r^-  + \gamma _r^+  + 2}}{2} - \frac{{{\delta _r}}}{{{\alpha _r}}}}}} = 0}&, \nonumber\\
{{\rm{if}}{\kern 1pt} {\kern 1pt} {\alpha _r} = \alpha _r^- }&.
  \end{align}
\end{subequations}
with a constraint condition
\begin{equation}
\arg(r) = 2{\rm{k}}\pi - \frac{\pi}{2} - \arg(\omega^* ), \quad{\rm{k}} \in \mathbb{Z}.
\end{equation}
}

\subsubsection{Branch-Cut for QNMs and HeunC Function}
For QNM frequencies in the complex plane, there exists a branch cut line known as the NIA \cite{Berti:2006wq,Ching:1994bd,Jensen:1985in}.
In this subsubsection, we develop a method inspired by Fiziev's epsilon method \cite{Fiziev:2010yy,Fiziev:2011mm} to handle the branch cut in QNM spectra, providing a mathematically rigorous reformulation.
As a multivalued function, the confluent Heun function ${\rm{HeunC}}(\alpha ,\beta ,\gamma ,\delta ,\eta;x)$ has branch cut ${\mathscr{B}^{\,\,x}_{1,\infty}}=(1,+\infty)$ in its $x$-domain.
Using the M\"{o}bius transformation \eqref{eq:mtran}, we can derive the branch cut of $r$ is the negative real axis ${\mathscr{B}^{\,\,r}_{1,\infty}}=(-\infty,0)$.
Subsequently, we observe that the phase difference between the two branch cuts ${\mathscr{B}^{\,\,\omega}_{-i\infty, 0}}$ and ${\mathscr{B}^{\,\,r}_{1,\infty}}$ is exactly $\frac{\pi}{2}$, which precisely satisfies the phase constraint condition \eqref{eq:QNM-Cond-MFD}.
Remarkably, this condition \eqref{eq:QNM-Cond-MFD} coincidentally constructs a mapping relation between $r$ and $\omega$.
Therefore, we can introduce a small quantity ($-1<\varepsilon < 1$) in the phase condition \eqref{eq:QNM-Cond-MFD} to make the branch cuts stay away from QNM solutions.
Similarly, it also avoids the jump discontinuity of the confluent Heun function.
The modified phase condition is given by
\begin{equation}\label{eq:QNM-Cond-MFD-new}
\arg(r) = 2k\pi + \frac{\pi}{2}(1+\varepsilon) - \arg(\omega), \quad k \in \mathbb{Z},
\end{equation}
where for $\varepsilon < 0$, the corresponding QNM solution is $\omega_{\rm I}$ (Class-I).
While for $\varepsilon > 0$, the solution yields $\omega_{\rm II}$ (Class-II).

{The physical significance of this construction merits emphasis.
Conventional methods such as Cook's CFM fail near the NIA because they work with real radial coordinates and asymptotic expansions at spatial infinity.
When $\omega$ approaches the NIA (where $\arg(\omega) \to -\pi/2$), the oscillatory wave behavior $e^{\pm i\omega r_*}$ degenerates into pure exponential growth/decay $e^{\pm |\omega| r_*}$, causing catastrophic numerical cancellation between ingoing and outgoing components—the fundamental reason Cook et al.~documented convergence failures for purely imaginary modes \cite{Cook:2014cta,Cook:2016fge,Cook:2016ngj} .}

{Our phase constraint \eqref{eq:QNM-Cond-MFD-new} systematically rotates the computational contour in the complex $r$-plane to avoid this problematic region, analogous to Wick rotation in quantum field theory \cite{Wick:1954eu,Gibbons:1976ue}.
The parameter $\varepsilon$ controls which Riemann sheet we access: $\varepsilon < 0$ yields Class-I solutions $\omega_{\rm I}$ predominantly with $\mathrm{Re}(\omega) > 0$, while $\varepsilon > 0$ yields Class-II solutions $\omega_{\rm II}$ predominantly with $\mathrm{Re}(\omega) < 0$.
In the Schwarzschild limit, these families exhibit the symmetry $\omega_{\rm II} = -\omega^*_{\rm I}$, but for rotating BHs they evolve as distinct spectral branches.
This classification differs from the traditional ``positive/negative frequency'' labeling \cite{Berti:2009kk,Cook:2016ngj}, which becomes ambiguous for modes crossing the NIA—precisely the modes missing from Cook's incomplete spectra.
This approach enables computation of the complete QNM spectrum including modes that cross or lie on the NIA, which have been inaccessible to all previous methods.}

\subsection{HeunC Method for verifying QNMs}\label{sec:verify-QNM}

By solving \Cref{eq:ATE_Wronskian,eq:HeunC_QNM}, the QNM frequencies and the angular separation constants of Type-D BHs can be determined.
Besides calculating the numerical errors of \Cref{eq:ATE_Wronskian,eq:HeunC_QNM}, what other methods can be used to verify their correctness?
As mentioned in \cref{sec:HeunC_Sol_RTE}, the MST method can provide the analytical form of the asymptotic amplitude $B_{\ell m }^{\rm inc}$, which can be used to evaluate the accuracy of the QNMs. If the numerical result of $B_{\ell m }^{\rm inc}$ at a certain frequency is close to the machine precision ($10^{-16}$), then this frequency is considered to be a QNM.
Berti et al. calculated the excitation factors of QNMs \cite{Berti:2006wq,Zhang:2013ksa}: They first utilized Leaver's continued fraction method to obtain $\omega$ and ${}_s A_{\ell m}(a\omega)$, and then employed the MST method to compute $B_{\ell m }^{\rm inc}$ and subsequently the excitation factors. In this process, their idea of verifying the accuracy of QNMs is consistent with ours.
In this section, we will use the MST method to validate the accuracy of the QNM frequencies and the angular separation constants.

\subsubsection{Asymptotic Amplitudes}
In previous work \cite{Chen:2023lsa}, we have already derived all the asymptotic amplitudes \footnote{
These amplitudes $( B_{\ell m }^{{\rm{inc}}}, B_{\ell m }^{{\rm{ref}}},C_{\ell m }^{{\rm{ref}}},C_{\ell m }^{{\rm{up}}})$ have been normalized: $B_{\ell m }^{{\rm{trans}}} = 1$ and $C_{\ell m }^{{\rm{trans}}} = 1$} of the Kerr BH using the relationship between the HeunC method and the MST method.
These amplitudes are crucial for studying the physical properties of BH spacetimes. For instance, they can be used to calculate the excitation of the QNMs \cite{Leaver:1986gd,Berti:2006wq,Zhang:2013ksa}, reflection/absorption rates (or greybody factors) \cite{Brito:2015oca}, high-order tail \cite{Casals:2015nja,Casals:2016soq,Casals:2019vdb}, the tidal response \cite{Ivanov:2022qqt,Saketh:2023bul,OSullivan:2014ywd,OSullivan:2015lni,Penna:2017luh} of BHs, the amplitude of GW waveforms \cite{LISAConsortiumWaveformWorkingGroup:2023arg}, and GW echoes from extremely compact objects \cite{Srivastava:2021uku}, among a range of BH physics issues.

Now, we extend these amplitudes to arbitrary Type-D BHs.
The analytic expressions of the non-normalized asymptotic amplitudes $\hat B_{\ell m }^{\rm inc}$ and $\hat B_{\ell m }^{\rm ref}$ are given in the following form:

\noindent 1)  For $\psi \sim{e^{ - i\omega t}}$ and Class-I solutions $\omega_{\text{I}}$,
\begin{subequations}\label{eq:RTE_Amplitudes_1}
\begin{align}
{\hat B_{\ell m}^{{\rm{inc}}}}&{ = {{( - 1)}^{ - \frac{{{\beta _r} + {\gamma _r} + 2}}{2} - \frac{{{\delta _r}}}{{{\alpha _r}}}}}{{\left( {{r_ + } - {r_ - }} \right)}^{s + 1 + \frac{{{\delta _r}}}{{{\alpha _r}}}}}} \nonumber\\
{}&{ \times {{\left( {{r_ + } + {r_ - }} \right)}^{ - 2iM\omega }}{{\rm{e}}^{i\omega {r_ + }}}D_ \odot ^{{\beta _r}},}\\
{\hat B_{\ell m}^{{\rm{ref}}}}&{ = {{( - 1)}^{ - \frac{{{\beta _r} + {\gamma _r} + 2}}{2} + \frac{{{\delta _r}}}{{{\alpha _r}}}}}{{\left( {{r_ + } - {r_ - }} \right)}^{s + 1 - \frac{{{\delta _r}}}{{{\alpha _r}}}}}}\nonumber\\
{}& \times {\left( {{r_ + } + {r_ - }} \right)^{2iM\omega }}{{\rm{e}}^{ - i\omega {r_ + }}}D_ \otimes ^{{\beta _r}},\\
&\quad\quad\quad\quad\text{with}\quad\quad {\alpha _r} = \alpha _r^+ ,\quad {\beta _r} = \beta _r^+ .
\end{align}
\end{subequations}

\noindent 2)  For $\psi \sim{e^{ - i\omega t}}$ and Class-II solutions $\omega_{\text{II}}$,
\begin{subequations}\label{eq:RTE_Amplitudes_2}
\begin{align}
{\hat B_{\ell m}^{{\rm{inc}}}}&{ = {{( - 1)}^{ - \frac{{{\beta _r} + {\gamma _r} + 2}}{2} + \frac{{{\delta _r}}}{{{\alpha _r}}}}}{{\left( {{r_ + } - {r_ - }} \right)}^{s + 1 - \frac{{{\delta _r}}}{{{\alpha _r}}}}}} \nonumber\\
{}& \times {\left( {{r_ + } + {r_ - }} \right)^{2iM\omega }}{{\rm{e}}^{ - i\omega {r_ + }}}D_ \otimes ^{{\beta _r}},\\
{\hat B_{\ell m}^{{\rm{ref}}}}&{ = {{( - 1)}^{ - \frac{{{\beta _r} + {\gamma _r} + 2}}{2} - \frac{{{\delta _r}}}{{{\alpha _r}}}}}{{\left( {{r_ + } - {r_ - }} \right)}^{s + 1 + \frac{{{\delta _r}}}{{{\alpha _r}}}}}}\nonumber\\
{}&{ \times {{\left( {{r_ + } + {r_ - }} \right)}^{ - 2iM\omega }}{{\rm{e}}^{i\omega {r_ + }}}D_ \odot ^{{\beta _r}},}\\
&\quad\quad\quad\quad\text{with}\quad\quad {\alpha _r} = \alpha _r^- ,\quad {\beta _r} = \beta _r^+ .
\end{align}
\end{subequations}

\noindent 3)  For $\psi \sim{e^{  i\omega^* t}}$ and the complex conjugation of $\omega_{\text{I}}$,
\begin{subequations}\label{eq:RTE_Amplitudes_3}
\begin{align}
{\hat B_{\ell m}^{{\rm{inc}}}}&{ = {{( - 1)}^{ - \frac{{{\beta _r} + {\gamma _r} + 2}}{2} + \frac{{{\delta _r}}}{{{\alpha _r}}}}}{{\left( {{r_ + } - {r_ - }} \right)}^{s + 1 - \frac{{{\delta _r}}}{{{\alpha _r}}}}}} \nonumber\\
{}& \times {\left( {{r_ + } + {r_ - }} \right)^{2iM\omega }}{{\rm{e}}^{ - i\omega {r_ + }}}D_ \otimes ^{{\beta _r}},\\
{\hat B_{\ell m}^{{\rm{ref}}}}&{ = {{( - 1)}^{ - \frac{{{\beta _r} + {\gamma _r} + 2}}{2} - \frac{{{\delta _r}}}{{{\alpha _r}}}}}{{\left( {{r_ + } - {r_ - }} \right)}^{s + 1 + \frac{{{\delta _r}}}{{{\alpha _r}}}}}}\nonumber\\
{}&{ \times {{\left( {{r_ + } + {r_ - }} \right)}^{ - 2iM\omega }}{{\rm{e}}^{i\omega {r_ + }}}D_ \odot ^{{\beta _r}},}\\
&\quad\quad\quad\quad\text{with}\quad\quad {\alpha _r} = \alpha _r^+ ,\quad {\beta _r} = \beta _r^- .
\end{align}
\end{subequations}

\noindent 4)  For $\psi \sim{e^{  i\omega^* t}}$ and the complex conjugation of $\omega_{\text{II}}$,
\begin{subequations}\label{eq:RTE_Amplitudes_4}
\begin{align}
{\hat B_{\ell m}^{{\rm{inc}}}}&{ = {{( - 1)}^{ - \frac{{{\beta _r} + {\gamma _r} + 2}}{2} - \frac{{{\delta _r}}}{{{\alpha _r}}}}}{{\left( {{r_ + } - {r_ - }} \right)}^{s + 1 + \frac{{{\delta _r}}}{{{\alpha _r}}}}}} \nonumber\\
{}&{ \times {{\left( {{r_ + } + {r_ - }} \right)}^{ - 2iM\omega }}{{\rm{e}}^{i\omega {r_ + }}}D_ \odot ^{{\beta _r}},}\\
{\hat B_{\ell m}^{{\rm{ref}}}}&{ = {{( - 1)}^{ - \frac{{{\beta _r} + {\gamma _r} + 2}}{2} + \frac{{{\delta _r}}}{{{\alpha _r}}}}}{{\left( {{r_ + } - {r_ - }} \right)}^{s + 1 - \frac{{{\delta _r}}}{{{\alpha _r}}}}}}\nonumber\\
{}& \times {\left( {{r_ + } + {r_ - }} \right)^{2iM\omega }}{{\rm{e}}^{ - i\omega {r_ + }}}D_ \otimes ^{{\beta _r}},\\
&\quad\quad\quad\quad\text{with}\quad\quad {\alpha _r} = \alpha _r^+ ,\quad {\beta _r} = \beta _r^- .
\end{align}
\end{subequations}
where either $\gamma_r^+$ or $\gamma_r^-$ can be chosen for $\gamma_r$.

For these asymptotic amplitudes, the connection coefficients $D_ \odot ^{{\beta _r}}$ and $D_ \otimes ^{{\beta _r}}$ utilize the matching technique derived from the MST method: where two series sums based on special functions are matched within the convergence interval. We refer to this method as the MST-HeunC method.
$(\hat B_{\ell m }^{\rm inc},\hat B_{\ell m }^{\rm ref})$ represent $( B_{\ell m }^{\rm inc}, B_{\ell m }^{\rm ref})$ obtained by the MST-HeunC method.
The general form of $ B_{\ell m}^{\text{trans}}$ is not provided here. Its specific form can only be determined when the spacetime metric is known.
For a detailed discussion on these amplitudes, please refer to Refs.\ \cite{Chen:2023ese,Chen:2023lsa}.

The MST-HeunC method differs from the MST method in that it introduces the HeunC parameter.
To calculate the asymptotic amplitudes of any Type-D BHs, the MST-HeunC method only requires inputting the corresponding HeunC parameter into \Cref{eq:RTE_Amplitudes_1,eq:RTE_Amplitudes_2,eq:RTE_Amplitudes_3,eq:RTE_Amplitudes_4} to obtain these amplitudes.
In contrast, the MST method necessitates the reconstruction of the ingoing wave and outgoing wave solutions based on the series form of special functions, followed by a discussion of the asymptotic behavior of these solutionsto calculate the asymptotic amplitudes.

\subsubsection{Renormalized Angular Momentum}\label{sec:renormalized_AM}
The connection coefficients $D_ \otimes ^{{\beta _r}}$ and $D_ \odot ^{{\beta _r}}$ are related to the expansion coefficient $f_{\rm n}^\nu$,
and $f_{\rm n}^\nu$ satisfies the following three-term recurrence relation,
\begin{equation}\label{eq:recurrence-fnv}
 {{\hat \alpha }_{\rm n}}f_{{\rm n} + 1}^\nu  + {{\hat \beta }_{\rm n}}f_{\rm n}^\nu  + {{\hat \gamma }_{\rm n}}f_{{\rm n} - 1}^\nu  = 0,
\end{equation}
where
\begin{subequations}
\begin{align}
{{\hat \alpha }_{\rm n}} &=   \frac{{\left( {2{\rm n} + 2\nu  + 2 - \beta_r + \gamma_r } \right)}}{{8\left( {{\rm n} + \nu  + 1} \right)\left( {2{\rm n} + 2\nu  + 3} \right)}}\\
&\times \left( {\alpha_r {\rm n} + \alpha_r \nu  + \alpha_r  -\delta_r } \right)\left( {2{\rm n} + 2\nu  + 2 - \gamma_r  - \beta_r} \right),\nonumber \\
{{\hat \beta}_{\rm n}} &= \eta_r  + \frac{\delta_r }{2} - \frac{{{\beta_r^2}}}{4} - \frac{{{\gamma_r ^2}}}{4} + \left( {{\rm n} + \nu } \right)\left( {{\rm n} + \nu  + 1} \right) \nonumber \\
&+ \frac{{\delta_r \left( {\gamma_r  + \beta_r} \right)\left( {\beta_r - \gamma_r } \right)}}{{8\left( {n + \nu } \right)\left( {n + \nu  + 1} \right)}},\\
{{\hat \gamma }_{\rm n}} &= -\frac{{\left( {2{\rm n} + 2\nu  + \beta_r - \gamma_r } \right)}}{{8\left( {{\rm n} + \nu } \right)\left( 2{\rm n} + 2\nu  - 1 \right)}} \nonumber \\
&\times \left( {\alpha_r {\rm n} + \alpha_r \nu  + \delta_r } \right)\left( {2{\rm n} + 2\nu  + \gamma_r  + \beta_r} \right).
\end{align}
\end{subequations}

The phase parameter $\nu$, also known as the renormalized angular momentum, can be obtained by solving an eigenvalue equation expressed as the sum of two infinite continued fractions.
\begin{equation}\label{eq:nu}
  {{\hat \beta }_0} = \frac{{{{\hat \alpha }_{ - 1}}{{\hat \gamma }_0}}}{{{{\hat \beta }_{ - 1}} - }}\frac{{{{\hat \alpha }_{ - 2}}{{\hat \gamma }_{ - 1}}}}{{{{\hat \beta }_{ - 2}} - }}\frac{{{{\hat \alpha }_{ - 3}}{{\hat \gamma }_{ - 2}}}}{{{{\hat \beta }_{ - 3}} - }}   \cdots
 + \frac{{{{\hat \alpha }_0}{{\hat \gamma }_1}}}{{{{\hat \beta }_1} - }}\frac{{{{\hat \alpha }_1}{{\hat \gamma }_2}}}{{{{\hat \beta }_2} - }}\frac{{{{\hat \alpha }_2}{{\hat \gamma }_3}}}{{{{\hat \beta }_3} - }} \cdots .
\end{equation}
Although $\nu$ has been introduced via Leaver's solutions \cite{Leaver:1986vnb} or the MST method \cite{Mano1996RWE,Mano_1996}, it is indeed a fundamental parameter. The parameter $\nu$ appears in other analyses of BH perturbations that do not use the MST formalism.
For instance, in subextremal Kerr, $\nu$ is related to the monodromy of the outgoing wave solution at $r=\infty$ \cite{Castro:2013lba}.
Furthermore, $\nu(\nu+1)$ is the eigenvalue of the Casimir operator for the $\mathfrak{sl}(2,\mathbb{R})$ factor within the algebra of the near-horizon geometry of extremal Kerr BHs \cite{Gralla:2015rpa}.
We summarize some properties and symmetries related to the MST formalism and $\nu$:
\begin{itemize}
\item The MST formalism is fundamentally invariant under $\nu \to -\nu - 1$.
The reason is that $\nu$ was introduced as a parameter in the RTE \eqref{eq:GFoRTE} through the combination ``$\nu(\nu+1)$''.
This leads to the symmetry,
\begin{equation}\label{eq:fn->f-n}
f_{\rm n}^{-\nu-1}=f_{-{\rm n}}^{\nu}.
\end{equation}

\item The periodicity of $\nu$ is reflected in that $\nu + \hat{k}$ is also a solution to \cref{eq:nu}, for any $\hat{k}\in \mathbb{Z}$.
The reason is that $\nu$ only appears in \cref{eq:nu} in the combination $\nu+{\rm n}$, where ${\rm n}\in \mathbb{Z}$.

\item It has been shown (analytically, but with an assumption supported numerically) in subextremal Kerr in~\cite{Fujita_2004,Fujita_2005,Sasaki:2003xr} and in extremal Kerr in~\cite{Casals:2019vdb} that,
    for $\omega$ real, $\nu$ is either real valued or else complex valued with a real part that is equal to a half-integer.

\item It follows from the above property that, for $\omega$ real, complex conjugation of $\nu$ can be achieved by applying the MST symmetries of $\nu \to -\nu - 1$ and the addition of an integer to $\nu$.

\item The series coefficients $(\hat\alpha_{\rm n},\hat\beta_{\rm n},\hat\gamma_{\rm n})$ are all invariant under $m\to -m$ and $\omega\to -\omega$, and, therefore, $\nu$ is also invariant.
\end{itemize}

Solving for $\nu$, one of the greatest challenges is the selection of initial values.
For a given QNM frequency, the initial value $\nu_0$ needs to be sufficiently close to the exact value of $\nu$.
Casals \cite{Casals:2018eev,Casals:2019vdb} provided approximate formulas for the frequencies corresponding to $\nu$ in both low-frequency and extremal spin cases.
For low frequencies, Casals derived the 3rd-order post-Minkowskian (PM) expansion, but in actual calculations, it is best to use the 2PM results as the initial value $\nu_0$. The expansion for $\nu$ can be written as
\begin{align}\label{eq:nu_2PM}
\nu&=\ell+\frac{1}{2 \ell+1}\bigg(-2-\frac{s^2}{\ell(\ell+1)}+\frac{\left[(\ell+1)^2-s^2\right]^2}{(2 \ell+1)(2 \ell+2)(2 \ell+3)}\nonumber\\
&-\frac{\left(\ell^2-s^2\right)^2}{(2 \ell-1) 2 \ell(2 \ell+1)}\bigg) ( 2M\omega)^2.
\end{align}

Fujita pointed out that when the real frequency exceeds a certain threshold, $\nu$ is either real or complex with a real part that is a half-integer \cite{Fujita_2004,Fujita_2005}.
Therefore, for the extremal spin, Casals derived the limit \cite{Casals:2018eev} of $\nu$ in the following form,
\begin{equation}\label{eq:nu_m}
  \mathop {\lim }\limits_{\omega  \to m/2} \nu  = {\nu _{c, \pm }}: =  - \frac{1}{2} \pm \sqrt {\frac{1}{4} +  {{\kern 1pt} _s}{K_{\ell m}} - 2{m^2}} ,
\end{equation}
where ${{\kern 1pt} _s}{K_{\ell m}} = {{\kern 1pt} _s}{A_{\ell m}} + {M^2}{\omega ^2} + s(s + 1)$.

The initial value $\nu_0$ is selected as follows: when $|\omega|$ is small, \cref{eq:nu_2PM} is selected as $\nu_0$, and when $|\omega|$ is large, \cref{eq:nu_m} is selected as $\nu_0$.
This initial value selection is not universally valid for all complex frequencies. The scheme fails for higher frequencies. Alternatively, one can directly compute $\nu$ using the \texttt{RenormalizedAngularMomentum} package in the BHPToolkit \cite{BHPToolkit}.
Recently, Barry Wardell has updated this package, which can be used to calculate the Teukolsky equation of complex frequencies and its amplitudes \cite{Wardell_Teukolsky_2025}.

\textbf{Note:} {
Both the MST method and the MST-HeunC method involve asymptotic amplitudes that contain numerous Gamma functions and Pochhammer symbols\footnote{Pochhammer symbols $(a)_{\rm n} = \frac{\Gamma(a+{\rm n})}{\Gamma(a)}$. The amplitudes $(\hat B_{\ell m }^{\rm inc},\hat B_{\ell m }^{\rm ref})$ contain the connection coefficients $(D_ \otimes ^{{\beta _r}},D_ \odot ^{{\beta _r}})$, and their expressions are shown in Appendix \ref{app:AsymptoticFormula}.}. This inadvertently imposes several restrictions on the input parameters, which is unfavorable for using the asymptotic amplitude of the MST-HeunC method to verify certain special QNM frequencies. This phenomenon will be discussed in detail in the numerical experiments of \cref{sec:AS-mode}.}

\subsubsection{Error Estimation}
To better assess the accuracy of QNM frequencies, we introduce three error quantities:
\begin{enumerate}
  \item With $ B_{\ell m}^{\text{trans}}=1$ normalized at the event horizon, the numerical error of the asymptotic amplitudes is
  \begin{equation}\label{eq:err1_QNM}
  \xi (\omega ) =  \left|\frac{{\hat B_{\ell m }^{{\rm{inc}}}(\omega )}}{{\hat B_{\ell m }^{{\rm{ref}}}(\omega )}} \right|.
\end{equation}

  \item $\vartheta (\omega )$ is defined as the numerical error of the QNM equation for a certain method.
  And the numerical error of the HeunC method \eqref{eq:HeunC_QNM} is
\begin{equation}\label{eq:err2_QNM}
  \vartheta (\omega ) = \left|\tilde B_{\ell m }^{{\rm{inc}}}(\omega ) \right|.
\end{equation}

  \item The relative error of the QNM frequencies is
\begin{equation}
  \mu (\omega ) = \left| {\frac{{{\omega _{{\rm{Exact}}}} - {\omega _{{\rm{x}}}}}}{{{\omega _{{\rm{Exact}}}}}}} \right|,
\end{equation}
where ${\omega_{{\rm{Exact}}}}$ represents the QNM frequencies of the exact solution, and ${\omega _{{\rm{x}}}}$ denotes the QNM frequencies of a certain method.
\end{enumerate}

\section{HeunC Method for Solving and Verifying TTMs}\label{sec:HeunC-TTM}
Similarly, the HeunC method for solving QNMs can be adapted to solve and verify the TTMs of gravitational perturbations. However, the TTMs only exist for specific spin-weight fields \cite{Chandrasekhar:1984mgh,Berti:2009kk,MaassenvandenBrink:2000iwh}. For $\psi \sim e^{-i\omega t}$, only the $\rm{TTM}_{\rm{L}}$ exists when $s=-2$, while only the $\rm{TTM}_{\rm{R}}$ exists when $s=2$.

\subsection{HeunC Method for Solving TTMs}\label{sec:TTMs-Eq}
The derivation of the eigenvalue equation for TTMs is similar to that for QNMs. We omit the derivation process and directly present the final result.
\subsubsection{${\rm{TTM}_{\rm{L}}}$ Frequencies}
When $\psi \sim {e^{ - i\omega t}}$, the outgoing wave solution $\tilde{\cal R}_{\ell m}^{{\rm{up}}}$ at the event horizon can be written as
\begin{equation}\tilde{\cal R}_{\ell m}^{{\rm{up}}} = {S_r}({\alpha _r},\beta _r^ - ,{\gamma _r}){\rm{HeunC}}({\alpha _r},\beta _r^ - ,{\gamma _r},{\delta _r},{\eta _r};{x_r}), \end{equation}
with the boundary condition
\begin{equation}\label{eq:BoundaryCondition-TTML}
{\tilde{\cal R}_{\ell m}^{{\rm{up}}}\to \left\{ {\begin{array}{*{20}{l}}
{\widetilde {\cal B}_{\ell m}^{{\rm{trans}}}R_{{\rm{up}}}^{\rm{H}},}&{r \to {r_ + },}\\
{\widetilde {\cal B}_{\ell m}^{{\rm{ref}}}R_{{\rm{up}}}^\infty  + \widetilde {\cal B}_{\ell m}^{{\rm{inc}}}R_{{\rm{in}}}^\infty ,}&{r \to  + \infty .}
\end{array}} \right.}\end{equation}

The ${\rm{TTM}_{\rm{L}}}$ requires that $\tilde{\cal R}_{\ell m}^{{\rm{up}}}$ have no ingoing wave at infinity, i.e., $\tilde{\cal B}_{\ell m}^{{\rm{inc}}} = 0$.
The eigenvalue equation for the ${\rm{TTM}_{\rm{L}}}$ frequencies can be written in the following form,
\begin{subequations}\label{eq:HeunC_TTM_L}
  \begin{align}
  \tilde {\cal B}_{\ell m }^{{\rm{inc}}} = \mathop {\lim }\limits_{|x_r| \to \infty }\frac{{{\rm{HeunC}}({\alpha _r},{\beta^- _r},{\gamma^+_r},{\delta _r},{\eta _r},{x_r})}}{{{x_r}^{ - \frac{{{\beta^- _r} + {\gamma^+_r} + 2}}{2} - \frac{{{\delta _r}}}{{{\alpha _r}}}}}} = 0&, \nonumber\\
   {\rm{if}}\,\,\alpha_r  = {\alpha^+_r }&; \\
 \tilde {\cal B}_{\ell m }^{{\rm{inc}}} = \mathop {\lim }\limits_{|x_r| \to \infty }\frac{{{\rm{HeunC}}({\alpha _r},{\beta^- _r},{\gamma^+_r},{\delta _r},{\eta _r},{x_r})}}{{{{\rm{e}}^{ - {\alpha _r}{x_r}}}{x_r}^{ - \frac{{{\beta^- _r} + {\gamma^+_r} + 2}}{2} + \frac{{{\delta _r}}}{{{\alpha _r}}}}}} = 0&, \nonumber\\
{\rm{if}}\,\, \alpha_r  = {\alpha^- _r }&;
  \end{align}
\end{subequations}
with a constraint condition,
\begin{equation}\label{eq:TTM_L-Cond-MFD}
\arg(r) = 2k\pi + \frac{\pi}{2}(1+\varepsilon) - \arg(\omega), \quad k \in \mathbb{Z}.
\end{equation}

For the case $\psi \sim e^{i\omega^* t}$, we similarly obtain the eigenvalue equation for the ${\rm{TTM}_{\rm{L}}}$ frequencies with complex conjugation.
\begin{subequations}\label{eq:HeunC_TTM_L-Conj}
  \begin{align}
{\tilde {\cal B}_{\ell m}^{{\rm{inc}}} = \mathop {\lim }\limits_{|{x_r}| \to \infty } \frac{{{\rm{HeunC}}({\alpha _r},\beta _r^ + ,\gamma _r^ + ,{\delta _r},{\eta _r},{x_r})}}{{{x_r}^{ - \frac{{\beta _r^ +  + \gamma _r^ +  + 2}}{2} - \frac{{{\delta _r}}}{{{\alpha _r}}}}}} = 0},&{} \nonumber \\
{{\rm{if}}{\kern 1pt} {\kern 1pt} {\alpha _r} = \alpha _r^-}&;\\
{\tilde {\cal B}_{\ell m}^{{\rm{inc}}} = \mathop {\lim }\limits_{|{x_r}| \to \infty } \frac{{{\rm{HeunC}}({\alpha _r},\beta _r^ + ,\gamma _r^ + ,{\delta _r},{\eta _r},{x_r})}}{{{{\rm{e}}^{ - {\alpha _r}{x_r}}}{x_r}^{ - \frac{{\beta _r^ +  + \gamma _r^ +  + 2}}{2} + \frac{{{\delta _r}}}{{{\alpha _r}}}}}} = 0},&{} \nonumber  \\
{{\rm{if}}{\kern 1pt} {\kern 1pt} {\alpha _r} = \alpha _r^ + }&;
\end{align}
\end{subequations}
with a constraint condition,
\begin{equation}\arg (r) = 2k\pi  - \frac{\pi }{2}(1 + \varepsilon ) - \arg ({\omega ^*}),\quad k \in \mathbb{Z} .\end{equation}

\subsubsection{${\rm{TTM}_{\rm{R}}}$ Frequencies}\label{TTM_R_Eq}

When $\psi \sim {e^{ - i\omega t}}$, the ingoing wave solution ${ R}_{\ell m}^{{\rm{in}}}$ at the event horizon can be written as
\begin{equation}R_{\ell m}^{{\rm{in}}} = {S_r}({\alpha _r},\beta _r^ + ,{\gamma _r}){\rm{HeunC}}({\alpha _r},\beta _r^ + ,{\gamma _r},{\delta _r},{\eta _r};{x_r}),\end{equation}
with the boundary condition
\begin{equation}R_{\ell m}^{{\rm{in}}} \to \left\{ {\begin{array}{*{20}{l}}
{B_{\ell m}^{{\rm{trans}}}R_{{\rm{in}}}^{\rm{H}},}&{r \to {r_ + },}\\
{B_{\ell m}^{{\rm{ref}}}R_{{\rm{up}}}^\infty  + B_{\ell m}^{{\rm{inc}}}R_{{\rm{in}}}^\infty ,}&{r \to  + \infty }.
\end{array}} \right.\end{equation}

The ${\rm{TTM}_{\rm{R}}}$ requires that $R_{\ell m}^{{\rm{in}}}$ have no outgoing wave at infinity, i.e., $B_{\ell m}^{{\rm{ref}}} = 0$.
The eigenvalue equation for the ${\rm{TTM}_{\rm{R}}}$ frequencies can be written in the following form,
\begin{subequations}\label{eq:HeunC_TTM_R}
  \begin{align}
{\tilde { B}_{\ell m}^{{\rm{ref}}} = \mathop {\lim }\limits_{|{x_r}| \to \infty } \frac{{{\rm{HeunC}}({\alpha _r},\beta _r^ + ,\gamma _r^ + ,{\delta _r},{\eta _r},{x_r})}}{{{{\rm{e}}^{ - {\alpha _r}{x_r}}}{x_r}^{ - \frac{{\beta _r^ +  + \gamma _r^ +  + 2}}{2} + \frac{{{\delta _r}}}{{{\alpha _r}}}}}} = 0},&{}\nonumber\\
{{\rm{if}}{\kern 1pt} {\kern 1pt} {\alpha _r} = \alpha _r^ + }&;\\
{\tilde B_{\ell m}^{{\rm{ref}}} = \mathop {\lim }\limits_{|{x_r}| \to \infty } \frac{{{\rm{HeunC}}({\alpha _r},\beta _r^ + ,\gamma _r^ + ,{\delta _r},{\eta _r},{x_r})}}{{{x_r}^{ - \frac{{\beta _r^ +  + \gamma _r^ +  + 2}}{2} - \frac{{{\delta _r}}}{{{\alpha _r}}}}}} = 0},&{}\nonumber\\
{{\rm{if}}{\kern 1pt} {\kern 1pt} {\alpha _r} = \alpha _r^ - }&;
  \end{align}
\end{subequations}
with a constraint condition
\begin{equation}\label{eq:TTM_R-Cond-MFD}
\arg (r) = 2k\pi  - \frac{\pi }{2}(1 + \varepsilon ) - \arg (\omega ),\quad k \in \mathbb{Z} .
\end{equation}

The ${\rm{TTM}_{\rm{R}}}$ uses the same ${ R}_{\ell m}^{{\rm{in}}}$ and boundary conditions (purely ingoing wave at the event horizon) as those for QNMs.
The difference is that ${\rm{TTM}_{\rm{R}}}$ corresponds to $B_{\ell m}^{{\rm{ref}}} = 0$ while QNMs correspond to $B_{\ell m}^{{\rm{inc}}} = 0$. This means no complex frequency can simultaneously be both a ${\rm{TTM}_{\rm{R}}}$ and a QNM.

For the case $\psi \sim e^{i\omega^* t}$, we similarly obtain the eigenvalue equation for the ${\rm{TTM}_{\rm{R}}}$ frequencies with complex conjugation.


\begin{subequations}\label{eq:HeunC_TTM_R-Conj}
  \begin{align}
{\tilde B_{\ell m}^{{\rm{ref}}} = \mathop {\lim }\limits_{|{x_r}| \to \infty } \frac{{{\rm{HeunC}}({\alpha _r},\beta _r^ - ,\gamma _r^ + ,{\delta _r},{\eta _r},{x_r})}}{{{{\rm{e}}^{ - {\alpha _r}{x_r}}}{x_r}^{ - \frac{{\beta _r^ -  + \gamma _r^ +  + 2}}{2} + \frac{{{\delta _r}}}{{{\alpha _r}}}}}} = 0},&{}\nonumber\\
{{\rm{if}}{\kern 1pt} {\kern 1pt} {\alpha _r} = \alpha _r^ - }&;\\
{\tilde B_{\ell m}^{{\rm{ref}}} = \mathop {\lim }\limits_{|{x_r}| \to \infty } \frac{{{\rm{HeunC}}({\alpha _r},\beta _r^ - ,\gamma _r^ + ,{\delta _r},{\eta _r},{x_r})}}{{{x_r}^{ - \frac{{\beta _r^ -  + \gamma _r^ +  + 2}}{2} - \frac{{{\delta _r}}}{{{\alpha _r}}}}}} = 0},&{} \nonumber\\
{{\rm{if}}{\kern 1pt} {\kern 1pt} {\alpha _r} = \alpha _r^ + }&;
\end{align}
\end{subequations}
with a constraint condition
\begin{equation}\arg (r) = 2k\pi  + \frac{\pi }{2}(1 + \varepsilon ) - \arg (\omega ),\quad k \in \mathbb{Z} .\end{equation}

The rationale for also considering the eigenvalue equations of complex-conjugate TTMs parallels the case for QNMs.
\begin{itemize}
  \item For ${\rm{TTM}_{\rm{L}}}$ frequencies, \cref{eq:HeunC_TTM_L-Conj} of the complex-conjugate TTMs yields higher-precision and more efficient results.
  \item For ${\rm{TTM}_{\rm{R}}}$ frequencies, \cref{eq:HeunC_TTM_R} similarly provides higher-precision and more efficient solutions.
\end{itemize}

\subsection{HeunC Method for Verifying TTMs}
The HeunC method can also be employed to construct asymptotic amplitudes for TTMs, thereby verifying the accuracy of the TTM spectrum. We summarize the asymptotic amplitudes for Class-I solutions $\omega_{\text{I}}$ as follows:

\noindent 1)  For $\psi \sim{e^{ - i\omega t}}$ and the ${\rm{TTM}_{\rm{L}}}$ equation \eqref{eq:HeunC_TTM_L},
\begin{subequations}\label{eq:TTM_L_Amplitudes_1}
\begin{align}
{\hat {\cal B}_{\ell m}^{{\rm{inc}}}}&{ = {{( - 1)}^{ - \frac{{{\beta _r} + {\gamma _r} + 2}}{2} - \frac{{{\delta _r}}}{{{\alpha _r}}}}}{{\left( {{r_ + } - {r_ - }} \right)}^{s + 1 + \frac{{{\delta _r}}}{{{\alpha _r}}}}}} \nonumber\\
{}&{ \times {{\left( {{r_ + } + {r_ - }} \right)}^{ - 2iM\omega }}{{\rm{e}}^{i\omega {r_ + }}}D_ \odot ^{{\beta _r}},}\\
{\hat {\cal B}_{\ell m}^{{\rm{ref}}}}&{ = {{( - 1)}^{ - \frac{{{\beta _r} + {\gamma _r} + 2}}{2} + \frac{{{\delta _r}}}{{{\alpha _r}}}}}{{\left( {{r_ + } - {r_ - }} \right)}^{s + 1 - \frac{{{\delta _r}}}{{{\alpha _r}}}}}}\nonumber\\
{}& \times {\left( {{r_ + } + {r_ - }} \right)^{2iM\omega }}{{\rm{e}}^{ - i\omega {r_ + }}}D_ \otimes ^{{\beta _r}},\\
&\quad\quad\quad\quad\text{with}\quad\quad {\alpha _r} = \alpha _r^+ ,\quad {\beta _r} = \beta _r^- .
\end{align}
\end{subequations}
where $(\hat {\cal B}_{\ell m }^{\rm inc},\hat {\cal B}_{\ell m }^{\rm ref})$ represents $( \tilde{ \cal B}_{\ell m }^{\rm inc}, \tilde{ \cal B}_{\ell m }^{\rm ref})$ obtained by the MST-HeunC method.

\noindent 2)  For $\psi \sim{e^{i\omega^* t}}$  and the ${\rm{TTM}_{\rm{L}}}$ equation \eqref{eq:HeunC_TTM_L-Conj},
\begin{subequations}\label{eq:TTM_L_Amplitudes_2}
\begin{align}
{\hat {\cal B}_{\ell m}^{{\rm{inc}}}}&{ = {{( - 1)}^{ - \frac{{{\beta _r} + {\gamma _r} + 2}}{2} + \frac{{{\delta _r}}}{{{\alpha _r}}}}}{{\left( {{r_ + } - {r_ - }} \right)}^{s + 1 - \frac{{{\delta _r}}}{{{\alpha _r}}}}}}\nonumber\\
{}& \times {\left( {{r_ + } + {r_ - }} \right)^{2iM\omega }}{{\rm{e}}^{ - i\omega {r_ + }}}D_ \otimes ^{{\beta _r}},\\
{\hat {\cal B}_{\ell m}^{{\rm{ref}}}}&{ = {{( - 1)}^{ - \frac{{{\beta _r} + {\gamma _r} + 2}}{2} - \frac{{{\delta _r}}}{{{\alpha _r}}}}}{{\left( {{r_ + } - {r_ - }} \right)}^{s + 1 + \frac{{{\delta _r}}}{{{\alpha _r}}}}}} \nonumber\\
{}&{ \times {{\left( {{r_ + } + {r_ - }} \right)}^{ - 2iM\omega }}{{\rm{e}}^{i\omega {r_ + }}}D_ \odot ^{{\beta _r}},}\\
&\quad\quad\quad\quad\text{with}\quad\quad {\alpha _r} = \alpha _r^+ ,\quad {\beta _r} = \beta _r^+ .
\end{align}
\end{subequations}

\noindent 3) For $\psi \sim{e^{ - i\omega t}}$  and the ${\rm{TTM}_{\rm{R}}}$ equation \eqref{eq:HeunC_TTM_R},
\begin{subequations}\label{eq:TTM_R_Amplitudes_1}
\begin{align}
{\hat B_{\ell m}^{{\rm{inc}}}}&{ = {{( - 1)}^{ - \frac{{{\beta _r} + {\gamma _r} + 2}}{2} - \frac{{{\delta _r}}}{{{\alpha _r}}}}}{{\left( {{r_ + } - {r_ - }} \right)}^{s + 1 + \frac{{{\delta _r}}}{{{\alpha _r}}}}}} \nonumber\\
{}&{ \times {{\left( {{r_ + } + {r_ - }} \right)}^{ - 2iM\omega }}{{\rm{e}}^{i\omega {r_ + }}}D_ \odot ^{{\beta _r}},}\\
{\hat B_{\ell m}^{{\rm{ref}}}}&{ = {{( - 1)}^{ - \frac{{{\beta _r} + {\gamma _r} + 2}}{2} + \frac{{{\delta _r}}}{{{\alpha _r}}}}}{{\left( {{r_ + } - {r_ - }} \right)}^{s + 1 - \frac{{{\delta _r}}}{{{\alpha _r}}}}}}\nonumber\\
{}& \times {\left( {{r_ + } + {r_ - }} \right)^{2iM\omega }}{{\rm{e}}^{ - i\omega {r_ + }}}D_ \otimes ^{{\beta _r}},\\
&\quad\quad\quad\quad\text{with}\quad\quad {\alpha _r} = \alpha _r^+ ,\quad {\beta _r} = \beta _r^+ .
\end{align}
\end{subequations}

\noindent 4)  For $\psi \sim{e^{i\omega^* t}}$  and the ${\rm{TTM}_{\rm{L}}}$ equation \eqref{eq:HeunC_TTM_R-Conj},
\begin{subequations}\label{eq:TTM_R_Amplitudes_2}
\begin{align}
{\hat B_{\ell m}^{{\rm{inc}}}}&{ = {{( - 1)}^{ - \frac{{{\beta _r} + {\gamma _r} + 2}}{2} + \frac{{{\delta _r}}}{{{\alpha _r}}}}}{{\left( {{r_ + } - {r_ - }} \right)}^{s + 1 - \frac{{{\delta _r}}}{{{\alpha _r}}}}}} \nonumber\\
{}& \times {\left( {{r_ + } + {r_ - }} \right)^{2iM\omega }}{{\rm{e}}^{ - i\omega {r_ + }}}D_ \otimes ^{{\beta _r}},\\
{\hat B_{\ell m}^{{\rm{ref}}}}&{ = {{( - 1)}^{ - \frac{{{\beta _r} + {\gamma _r} + 2}}{2} - \frac{{{\delta _r}}}{{{\alpha _r}}}}}{{\left( {{r_ + } - {r_ - }} \right)}^{s + 1 + \frac{{{\delta _r}}}{{{\alpha _r}}}}}}\nonumber\\
{}&{ \times {{\left( {{r_ + } + {r_ - }} \right)}^{ - 2iM\omega }}{{\rm{e}}^{i\omega {r_ + }}}D_ \odot ^{{\beta _r}},}\\
&\quad\quad\quad\quad\text{with}\quad\quad {\alpha _r} = \alpha _r^+ ,\quad {\beta _r} = \beta _r^- .
\end{align}
\end{subequations}
where either $\gamma_r^+$ or $\gamma_r^-$ can be chosen for $\gamma_r$.


\section{QNMs of Schwarzschild BHs}\label{sec:SchBH-QNM}
\begin{figure*}[htbp]
	\centering
\includegraphics[width=7in]{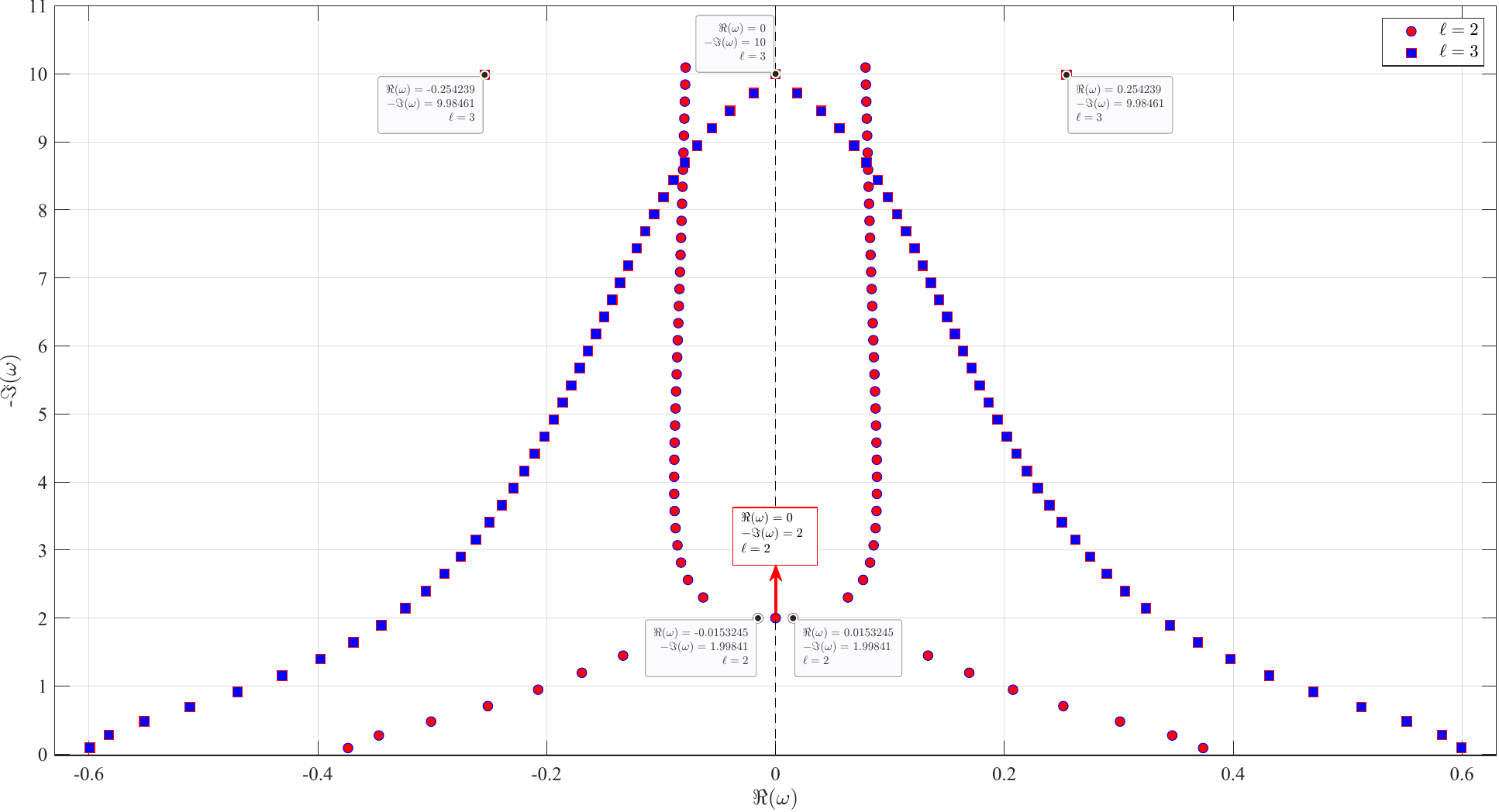}
\caption{ The first 42  QNMs for the gravitational perturbations of the Schwarzschild BH.}\label{fig:SchBH_L2L3}
\end{figure*}
\begin{table*}[htbp]
  \centering
  \caption{Numerical results of the first twelve QNMs $\omega_{\text{I}}$ of the Schwarzschild BH with $s=-2$ and $\ell=2$.}\label{tab:Sch_QNM}%
    \begin{threeparttable}
    \begin{tabular}{ccccc}
    \toprule
    overtone & HeunC-Chen$^\clubsuit$ & HeunC-Fiziev$^\diamondsuit$  & Berti-RTE-CFM$^\heartsuit$ & BHPToolkit-RWE-CFM$^\spadesuit$ \\
    \midrule
    0    & $\begin{array}{l}{\kern 7pt} 0.373671684418042 \\- 0.0889623156889363i \end{array}$ & $\begin{array}{l}{\kern 7pt} 0.37367168445 \\ - 0.0889623155 i \end{array}$ & $\begin{array}{l}{\kern 7pt} 0.3736716844181666 \\ - 0.0889623156892308 i \end{array}$ & $\begin{array}{l}{\kern 7pt} 0.37367168441804105 \\ - 0.08896231568893471 i \end{array}$ \\
\midrule
    1    & $\begin{array}{l}{\kern 7pt} 0.346710996879163 \\- 0.273914875291232i \end{array}$ & $\begin{array}{l}{\kern 7pt} 0.3467109969 \\ - 0.273914875 i \end{array}$ & $\begin{array}{l}{\kern 7pt} 0.3467109970795381 \\ - 0.27391487587943053 i \end{array}$ & $\begin{array}{l}{\kern 7pt} 0.3467109968791626 \\ - 0.27391487529123315 i \end{array}$ \\
\midrule
    2    & $\begin{array}{l}{\kern 7pt} 0.301053454612362 \\- 0.478276983223071i \end{array}$ & $\begin{array}{l}{\kern 7pt} 0.3010534546 \\ - 0.478276983 i \end{array}$ & $\begin{array}{l}{\kern 7pt} 0.30105362648761164 \\ - 0.47827715278020655 i \end{array}$ & $\begin{array}{l}{\kern 7pt} 0.3010534546123656 \\ - 0.4782769832230699 i \end{array}$ \\
\midrule
    3    & $\begin{array}{l}{\kern 7pt} 0.251504962185590 \\- 0.705148202433496i \end{array}$ & $\begin{array}{l}{\kern 7pt} 0.25150496205 \\ - 0.7051482025 i \end{array}$ & $\begin{array}{l}{\kern 7pt} 0.2515203561898016 \\ - 0.7051513755626464 i \end{array}$ & $\begin{array}{l}{\kern 7pt} 0.25150496218559065 \\ - 0.7051482024334933 i \end{array}$ \\
\midrule
    4    & $\begin{array}{l}{\kern 7pt} 0.207514579813078 \\- 0.946844890866350i \end{array}$ & $\begin{array}{l}{\kern 7pt} 0.2075145798 \\ - 0.946844891 i \end{array}$ & $\begin{array}{l}{\kern 7pt} 0.20773264759604257 \\ - 0.9468893264193199 i \end{array}$ & $\begin{array}{l}{\kern 7pt} 0.20751457981306362 \\ - 0.9468448908663512 i \end{array}$ \\
\midrule
    5    & $\begin{array}{l}{\kern 7pt} 0.169299403093033 \\- 1.19560805413585i \end{array}$ & $\begin{array}{l}{\kern 7pt} 0.1692994032 \\ - 1.195608054 i \end{array}$ & $\begin{array}{l}{\kern 7pt} 0.16994676478334064 \\ - 1.196626942722103 i \end{array}$  & $\begin{array}{l}{\kern 7pt} 0.16929940309304417 \\ - 1.1956080541358447 i \end{array}$ \\
\midrule
    6    & $\begin{array}{l}{\kern 7pt} 0.133252340245188 \\- 1.44791062616200i \end{array}$ & $\begin{array}{l}{\kern 7pt} 0.1332523405\\- 1.4479106265i \end{array}$ & Null & $\begin{array}{l}{\kern 7pt} 0.1332523402451876 \\ - 1.447910626162035 i \end{array}$ \\
\midrule
    7    & $\begin{array}{l}{\kern 7pt} 0.0928223336702147 \\- 1.70384117220615i \end{array}$ & $\begin{array}{l}{\kern 7pt} 0.0928223342 \\ - 1.70384112725 i \end{array}$ & Null & $\begin{array}{l}{\kern 7pt} 0.0928223336701992 \\ - 1.7038411722061295 i \end{array}$ \\
\midrule
    8    & $\begin{array}{l} -0.0153245047607925 \\- 1.99841184185869i \end{array}$ & $\begin{array}{l}- 0.015324503 \\ - 1.998411845 i \end{array}$ & Null & Null \\
\midrule
    9    & $\begin{array}{l}{\kern 7pt}0.00000000000000 \\- 2.00000000000000i \end{array}$ & Null & Null & ${\large \textcolor{red}{\maltese}}\,\begin{array}{l}{\kern 7pt}2.3075893869665945\times 10^{-16} \\- 1.9999999999998i \end{array}$ \\
\midrule
    10   & $\begin{array}{l}{\kern 7pt} 0.0632635051256196 \\- 2.30264476515842i \end{array}$ & $\begin{array}{l}{\kern 7pt} 0.063263509 \\ - 2.302644771 i \end{array}$ & Null & $\begin{array}{l}{\kern 7pt} 0.06326350512559752 \\ - 2.30264476515854 i \end{array}$ \\
\midrule
    11   & $\begin{array}{l}{\kern 7pt} 0.0765534628861350 \\- 2.56082661738148i \end{array}$ & $\begin{array}{l}{\kern 7pt} 0.0765534751 \\ - 2.560826636 i \end{array}$ & Null & $\begin{array}{l}{\kern 7pt} 0.07655346288598577 \\ - 2.560826617381502 i \end{array}$ \\\midrule
Time(s) & 0.6153 & 4959.000  & 12405.302 & 173.406 \\

\bottomrule
\end{tabular}%
\begin{tablenotes}
    \item[$\clubsuit$] The range of initial values is ${\rm Re} ({\omega _0}) \in [0,0.35],\, {\rm Im} ({\omega _0}) \in [0,2.5],$ and the step size is $ {\rm Re} (\square {\omega _0}) = 0.05+\tfrac{1}{2}\epsilon,{\rm Im} (\square {\omega _0}) = 0.25+ \tfrac{1}{2}\epsilon$. This grid has a total of 77 initial values.
    \item[$\diamondsuit$] In Ref. \cite{Fiziev:2010yy}, Fiziev provided these numerical results.
    \item[$\heartsuit$] On Berti's personal website \cite{BertiWeb}, he provided a code that uses the CFM of RTEs to solve the QNM spectrum of Kerr BHs.
    The range of initial values is ${\rm Re} ({\omega _0}) \in [0,0.35],\, {\rm Im} ({\omega _0}) \in [0,2.5],$ and the step size is 0.01.
    This grid has a total of 35571 initial values.
    \item[$\spadesuit$] The Black Hole Perturbation Toolkit (BHPToolkit) provides a \texttt{QuasiNormalModes} package for computing QNM spectra \cite{BHPToolkit}. For Schwarzschild BH QNMs, this package enhances the computational performance based on previous CFM studies \cite{Dolan:2009nk,Casals:2013mpa,Ansorg:2016ztf,Leaver:1985ax,BertiWeb,Nollert:1993zz} of RWEs.
    \item[${\textcolor{red}{\maltese}}$] In the BHPToolkit's code \cite{BHPToolkit}, this QNM value (from \texttt{s2l2.dat} in \cite{BertiWeb_SchBH}) of the AS mode is set artificially as the output and is not actually calculated using the CFM code.
    \end{tablenotes}
\end{threeparttable}

\end{table*}

\begin{table*}[htbp]
  \centering
  \caption{Numerical errors of the first twelve QNMs $\omega_{\text{I}}$ of the Schwarzschild BH with $s=-2$ and $\ell=2$. }
    \begin{tabular}{cccccc}
    \toprule
   overtone & Errors &HeunC-Chen & HeunC-Fiziev  & Berti-RTE-CFM & BHPToolkit-RWE-CFM \\
    \midrule
    0    & $\xi(\omega)$   & $4.507\times {10^{-13}}$ & $1.302\times {10^{-7}}$ & $2.176\times {10^{-10}}$ & $8.407\times {10^{-13}}$ \\
    \midrule
    1    & $\xi(\omega)$   & $4.644\times {10^{-13}}$ & $5.003\times {10^{-8}}$ & $1.065\times {10^{-7}}$ & $5.265\times {10^{-13}}$ \\
    \midrule
    2    & $\xi(\omega)$   & $1.797\times {10^{-13}}$ & $1.181\times {10^{-8}}$ & $1.280\times {10^{-5}}$ & $1.241\times {10^{-14}}$ \\
    \midrule
    3    & $\xi(\omega)$   & $1.321\times {10^{-14}}$ & $2.799\times {10^{-9}}$ & $2.913\times {10^{-4}}$ & $5.702\times {10^{-14}}$ \\
    \midrule
    4    & $\xi(\omega)$   & $9.305\times {10^{-14}}$ & $1.004\times {10^{-9}}$ & $1.660\times {10^{-3}}$ & $1.615\times {10^{-14}}$ \\
    \midrule
    5    & $\xi(\omega)$   & $3.212\times {10^{-14}}$ & $5.796\times {10^{-10}}$ & $4.030\times {10^{-3}}$ & $8.715\times {10^{-15}}$ \\
    \midrule
    6    & $\xi(\omega)$   & $5.293\times {10^{-14}}$ & $6.672\times {10^{-10}}$ & Null & $3.084\times {10^{-15}}$ \\
    \midrule
    7    & $\xi(\omega)$   & $1.273\times {10^{-14}}$ & $4.258\times {10^{-10}}$ & Null & $5.682\times {10^{-15}}$ \\
    \midrule
    8    & $\xi(\omega)$   & $5.795\times {10^{-14}}$ & $1.105\times {10^{-10}}$ & Null &  Null \\
    \midrule\multirow{2}[4]{*}{9}
         & $\xi(\omega)$   & MST method fails & Null & Null & MST method fails \\\cmidrule{2-6}
         &$\mu(\omega)$    & 0                          & Null & Null & $1.000\times {10^{-13}}$ \\
    \midrule
    10   & $\xi(\omega)$   & $3.687\times {10^{-14}}$ & $1.943\times {10^{-9}}$ & Null & $5.174\times {10^{-15}}$ \\
    \midrule
    11   & $\xi(\omega)$   & $4.141\times {10^{-14}}$ & $6.054\times {10^{-9}}$ & Null & $1.534\times {10^{-14}}$ \\
    \bottomrule
    \end{tabular}%
  \label{tab:Sch_QNM_err}%
\end{table*}
\begin{table*}[htbp]
  \centering
  \caption{The numerical error $\xi (\omega )$ (ratio of asymptotic amplitudes) of three complex frequencies for the different deviation $\varpi$.}
  \begin{threeparttable}
    \begin{tabular}{cccccccccc}
    \toprule
         & \multicolumn{4}{c}{$\omega_{\ell=2}^{\rm{AS}} = -2i$} &      & \multicolumn{4}{c}{$\omega_{\ell=3}^{\rm{AS}} = -10i$} \\
    \cmidrule{2-5}\cmidrule{7-10}    $\varpi$ & $10^{-10}$ & $10^{-15}$ & $10^{-20}$ & $10^{-25}$ &      & $10^{-10}$ & $10^{-15}$ & $10^{-20}$ & $10^{-25}$ \\
    \midrule
  ${\rm QNM}_{-2}$$^\clubsuit$  & $2.883\times10^{-12}$ & $2.883\times10^{-17}$ & $2.883\times10^{-22}$ & $2.883\times10^{-27}$ &      & $1.458\times10^{-13}$ & $1.458\times10^{-18}$ & $1.458\times10^{-23}$ & $1.458\times10^{-28}$ \\
    ${\rm QNM}_{+2}$  & $3.280$ & $3.280$ & $3.280$ & $3.280$ &      & $12962.923$ & $12962.923$ & $12962.923$ & $12962.923$ \\

 ${\rm TTM}_{\rm L}$$^\diamondsuit$ & $1.439\times10^{-11}$ & $1.439\times10^{-16}$ & $1.439\times10^{-21}$ & $1.439\times10^{-26}$ &      & $2.750\times10^{-14}$ & $2.750\times10^{-19}$ & $2.750\times10^{-24}$ & $2.750\times10^{-29}$ \\
    ${\rm TTM}_{\rm R}$$^\heartsuit$ & $0.305$ & $0.305$ & $0.305$ & $0.305$ &      & $7.714\times10^{-5}$ & $7.714\times10^{-5}$ & $7.714\times10^{-5}$ & $7.714\times10^{-5}$ \\
    \bottomrule
    \end{tabular}
 \begin{tablenotes}
    \item[$\clubsuit$] The numerical error of QNMs is defined by \cref{eq:err1_QNM}, which is $\xi (\omega ) =  \left|{{\hat B_{\ell m }^{{\rm{ref}}}(\omega )}}/{{\hat B_{\ell m }^{{\rm{inc}}}(\omega )}} \right|$. ${\rm QNM}_{-2}$ denotes the QNM of gravitational perturbations with spin weight $s = -2$; similarly, ${\rm QNM}_{+2}$ denotes that with $s = +2$.

    \item[$\diamondsuit$] The numerical error of ${\rm TTM}_{\rm L}$ is defined by $\xi (\omega ) =  \left|{{\hat{\cal B}_{\ell m }^{{\rm{inc}}}(\omega )}}/{{\hat {\cal B}_{\ell m }^{{\rm{ref}}}(\omega )}} \right|$.
    \item[$\heartsuit$] The numerical error of ${\rm TTM}_{\rm R}$ is defined by $\xi (\omega ) =  \left|{{\hat B_{\ell m }^{{\rm{ref}}}(\omega )}}/{{\hat B_{\ell m }^{{\rm{inc}}}(\omega )}} \right|$.
    \end{tablenotes}
\end{threeparttable}

  \label{tab:ASmode-Errors}%
\end{table*}%
\begin{table}[htbp]
  \centering,
  \caption{Computational times (single-core) of the first 42 QNMs $\omega_{\text{I}}$ of the Schwarzschild BH with the $L_\infty$-error of $\xi(\omega)$ set to $10^{-13}$.}
    \begin{tabular}{cccccc}
    \toprule
         & \multicolumn{2}{c}{HeunC-Chen} &      & \multicolumn{2}{c}{BHPTookit} \\
\cmidrule{2-3}\cmidrule{5-6}    $\ell=|s|$ & overtone & Time(s) &      & overtone & Time(s) \\
    \midrule
    2    & $[0,41]$ & 2.971 & & $[0,7],[9,34],[36,40]$ & 5183.430 \\
    1    & $[0,41]$ & 2.785 & & $[0,24],[26,28]$ & 5677.838 \\
    0    & $[0,41]$ & 4.469 & & $[0,39]$ & 5899.959 \\
    \bottomrule
    \end{tabular}%
  \label{tab:QNM_diff_s}%
\end{table}

Next, we apply our implementation of the HeunC method, denoted HeunC-Chen to distinguish it from Fiziev's approach, to compute the complex frequencies (QNMs and TTMs) of Type-D BHs. The corresponding dataset and code, maintained by the author (Changkai Chen), are publicly available at Ref. \cite{ChenQNM}. The accuracy of our simulations is evaluated using norm errors, and the specific numerical settings employed are detailed in Appendix \ref{app:CompDetail_B}.

The QNM frequencies and the angular separation constant of the Schwarzschild BH can serve as initial values for any Type-D BH. Therefore, before calculating the QNMs of other Type-D BHs, we should first provide a more complete and highly accurate QNM spectrum for the Schwarzschild BH. The ``more complete" here refers to the inclusion of weakly damped (low-overtone) modes, AS modes, and highly damped (high-overtone) modes.
The issue appears to have been effectively addressed by the CFM\footnote{
On Berti's personal website \cite{BertiWeb}, the QNM spectrum of the Schwarzschild BH is provided for overtone indices from 0 to 1001, though some duplicated results are observed.}. However, important modes of gravitational perturbations might be missing from CFM's results \cite{Leaver:1985ax,Nollert:1993zz,Berti:2009kk}, potentially leading to misconceptions when solving the QNM spectrum of rotating BHs.
To address this, the complete QNM spectrum of Schwarzschild BHs is first computed as initial values for subsequent calculations of rotating BH spectra.

For the Schwarzschild BHs, the potential of the ATE \eqref{eq:GFoATE} is
\begin{equation}\label{eq:U_ATE_Sch}
U(\Theta ) =  - \frac{{{{(m + s\Theta )}^2}}}{{1 - {\Theta ^2}}} + s + {\kern 1pt} {{\kern 1pt} _s}{A_{\ell m}},
\end{equation}
where the angular separation constant ${{\kern 1pt} _s}{A_{\ell m}}(a = 0)$ is known analytically,
\begin{equation}
   {{\kern 1pt} _s}{A_{\ell m}}(a = 0) = \ell (\ell  + 1) - s(s + 1).
\end{equation}

And the potential of the RTE \eqref{eq:GFoRTE} is
\begin{equation}\label{eq:V_RTE_Sch}
  V(r) = {\omega ^2}{r^4} - is\omega {r^2}\Delta' + \left( {4is\omega r - {\kern 1pt} {{\kern 1pt} _s}{A_{\ell m}} } \right)\Delta .
\end{equation}

The HeunC parameters of the ATE \eqref{eq:GFoATE} can be derived in the following form,
\begin{subequations}\label{eq:ATE_HCprameters_Sch}
 \begin{align}
&{{\alpha _{ - 1}} = 0,\quad\quad\quad {\alpha _{ + 1}} = 0,\quad }\\
&{{\beta _{ - 1}} = m - s,\quad {\beta _{ + 1}} = m + s,\quad }\\
&{{\gamma _{ - 1}} = m + s,\quad {\gamma _{ + 1}} = m - s,\quad }\\
&{{\delta _{ - 1}} = 0,\quad\quad\quad {\delta _{ + 1}} = 0,}\\
&{{\eta _{ - 1}} = {\eta _{ + 1}} = \tfrac{1}{2}{m^2} - \tfrac{1}{2}{s^2} - s - {{\kern 1pt} _s}{A_{\ell m}} .}
\end{align}
\end{subequations}

The HeunC parameters of the RTE \eqref{eq:GFoRTE} can be derived in the following form,
\begin{subequations}\label{eq:RTE_HCprameters_Sch}
 \begin{align}
&{\alpha^+_r} = 2i\omega {r_ + },\\
&{\beta^+_r} =   -s -2i\omega {r_ + },\\
&{\gamma^+_r} = s,\\
&{\delta _r} =  - 2is\omega {r_ + } - 2{\omega ^2}r_ + ^2,\\
&{\eta _r} = 2is\omega {r_ + } + 2{\omega ^2}r_ + ^2 - \tfrac{1}{2}{s^2} - s - {{\kern 1pt} _s}{A_{\ell m}} ,\\
&{r_ - } = 0,\quad {r_ + } = 2M.
\end{align}
\end{subequations}

We numerically solved the QNM spectrum for arbitrary perturbations of Schwarzschild BHs with indices $(\ell,\hat{n})$ where $2\leq\ell\leq16$ and $0\leq\hat{n}\leq41$.
At default machine precision, the QNM spectrum exhibits relative errors $\xi(\omega)$ and $ \vartheta (\omega )$ belonging to $[10^{-16}, 10^{-10}]$.
As mentioned before, previous gravitational QNM spectra lacked some modes. The following numerical tests focus on gravitational QNMs.
\Cref{tab:Sch_QNM,tab:Sch_QNM_err} present the numerical results and corresponding errors\footnote{The error $\xi (\omega )$ with different $\nu$ has some differences in its mantissa (or coefficient), but its exponent will be as much the same as possible, or not more than an order of magnitude different.} for the first 12 Class-I solutions $\omega_{\text{I}}$ of the Schwarzschild BH, computed using four different methods with $s=-2$ and $\ell=2$.
\cref{tab:QNM_diff_s} compares computational times for the first 42 Class-I solutions $\omega_{\text{I}}$ for different perturbation fields.
These numerical comparisons demonstrate that our method achieves significantly higher computational efficiency than the other three methods while maintaining comparable accuracy levels.
\cref{fig:SchBH_L2L3} displays the QNM spectrum of gravitational perturbations, showing the first 42 Class-I solutions $\omega_{\text{I}}$.
Additional QNM spectra are available in the dataset \cite{ChenQNM}.
Since purely imaginary modes divide the gravitational QNMs of the Schwarzschild BH into the lower QNM branch and the upper branch, we classify them into three categories: the weakly damped modes (lower QNM branch), purely imaginary modes, and the highly damped modes (upper branch).
Modes with $\hat{n} < 9$ are conventionally classified as weakly damped modes, while those with $\hat{n} > 9$ are highly damped modes.

\subsection{Weakly Damped and Unconventional Modes}
As shown in \cref{tab:Sch_QNM} and \cref{fig:SchBH_L2L3}, compared with previous results, our complete QNM spectrum includes two unconventional modes:
\begin{align}
\omega_{\rm{I}} = -0.0153245047607925 - 1.99841184185869&i, \nonumber \\ \ell = 2. \label{eq:SchNewModesL2}&\\
\omega_{\rm{I}} = -0.254239444199141 - 9.984607808258673&i, \nonumber \\ \ell = 3.\label{eq:SchNewModesL3}&
\end{align}

The unconventional mode \eqref{eq:SchNewModesL2} for $\ell=2$ achieves higher precision compared to previous results in \Cref{eq:UnconventionalMode1,eq:UnconventionalMode2,eq:UnconventionalMode3}.
The fitting procedure of Leung et al. has significant limitations \cite{Leung:2003eq};
It can only solve for the modes with $\ell = 2$ and is unable to solve for $\ell\geq3$ modes.
As they speculated, when $\ell \geq 3$, the corresponding unconventional modes are farther from the NIA compared to the $\ell = 2$ modes. Our results in \cref{fig:SchBH_L2L3} confirm that their speculation is correct.
So far, only our work has obtained the unconventional mode \eqref{eq:SchNewModesL3} for $\ell=3$.

\subsection{Purely Imaginary QNMs and TTMs}\label{sec:AS-mode}

To the best of our knowledge, no existing numerical results of purely imaginary QNMs have achieved perfect agreement with Chandrasekhar's exact solution \eqref{eq:ASmode}.
The most accurate numerical result was obtained by Mongwane et al. \cite{Mongwane:2024vao}, who achieved purely imaginary QNMs with numerical errors of $ 10^{-33} $ using the CFM within the Bondi-Sachs framework. Remarkably, our numerical results for the purely imaginary QNMs exhibit complete agreement with Chandrasekhar's exact solution \eqref{eq:ASmode}, with zero relative error.
At the AS mode \eqref{eq:ASmode}, the existence of modes (QNMs, ${\rm TTM}_{\rm L}$, or ${\rm TTM}_{\rm R}$) has long been controversial  \cite{MaassenvandenBrink:2000iwh,Berti:2009kk}. Generally, most authors \cite{Chandrasekhar:1984mgh,Leaver:1985ax,Onozawa:1996ux} agree that the AS mode \eqref{eq:ASmode} of Schwarzschild BHs represents the TTMs, where $s = -2$ corresponds to ${\rm TTM}_{\rm L}$ and $s = 2$ to ${\rm TTM}_{\rm R}$.

This paper provides analytical expressions\footnote{The QNM amplitudes are given by \Cref{eq:RTE_Amplitudes_1,eq:RTE_Amplitudes_2,eq:RTE_Amplitudes_3,eq:RTE_Amplitudes_4}, and the TTM amplitudes are given by \Cref{eq:TTM_L_Amplitudes_1,eq:TTM_L_Amplitudes_2,eq:TTM_R_Amplitudes_1,eq:TTM_R_Amplitudes_2}.} for asymptotic amplitudes at arbitrary complex frequencies, which can precisely determine the type of the AS mode $\omega_{\ell}^{\rm{AS}}$.
As shown in \cref{tab:Sch_QNM_err}, the MST-HeunC method fails to compute asymptotic amplitudes at default machine precision ($10^{-16}$). The underlying reason is that,
as discussed in \cref{sec:renormalized_AM}, these amplitudes involve numerous Gamma functions and Pochhammer symbols . The Gamma function is undefined at negative integers, which restricts complex frequencies. For instance, the amplitude \eqref{eq:RTE_Amplitudes_1} contains
\begin{equation}\Gamma \left( {\beta _r^ +  + 1} \right),\Gamma (\nu  + 1 \pm \frac{{{\delta _r}}}{{{\alpha _r}}}),\Gamma ( - \nu  \pm \frac{{{\delta _r}}}{{{\alpha _r}}}).\end{equation}
When $\omega_{\ell=2}^{\rm{AS}} = -2i$ and $\nu $ is an integer, these become,
\begin{equation}
\Gamma ( { - 6} ),\ \Gamma (\nu  + 1 \pm 6),\ \Gamma ( - \nu  \pm 6),
\end{equation}
all of which lie at poles of the Gamma function and are consequently undefined.

This benefits from the numerical algorithms for Gamma functions in mathematical software (\textit{MATLAB} or \textit{Mathematica}), and these computations of Gamma functions near their poles become feasible through increased floating-point precision.
We introduce a test frequency
\begin{equation}
\varpi_\ell^{{\rm{AS}}} = \omega_\ell^{{\rm{AS}}} - \varpi,
\end{equation}
where $\varpi$ denotes a small quantity representing the deviation between $\varpi_\ell^{{\rm{AS}}}$ and $\omega_\ell^{{\rm{AS}}}$.

With floating-point precision set to $10^{-300}$, \cref{tab:ASmode-Errors} provides the error $\xi (\omega )$ (amplitude ratio) for different values of $\varpi_\ell^{{\rm{AS}}}$.
The results demonstrate that as $\varpi$ decreases, the error $\xi (\omega )$ of $\varpi_\ell^{{\rm{AS}}}$ for both the ${\rm TTM}_{\rm L}$ and the ${\rm QNM}_{-2}$ approaches zero.
This confirms that $\omega_\ell^{{\rm{AS}}}$ simultaneously satisfies both ${\rm QNM}_{-2}$ and ${\rm TTM}_{\rm L}$ boundary conditions, proving that the AS modes $\omega_\ell^{{\rm{AS}}}$ of Schwarzschild BHs are both ${\rm QNM}_{-2}$ and ${\rm TTM}_{\rm L}$ frequencies.
We must reiterate that this conclusion represents an algebraic coincidence, where $\omega_\ell^{\rm AS}$ merely constitutes an intersection point of two distinct spectra.
Onozawa \cite{Onozawa:1996ux} previously noted from a scattering perspective that QNMs correspond to zero transmission amplitude at infinity ($B_{\ell m }^{{\rm{inc}}}=0$), and TTMs correspond to zero reflection amplitude at infinity ($B_{\ell m }^{{\rm{ref}}}=0$). He consequently argued that these modes cannot generally coincide.
His understanding of the TTM boundary conditions was incorrect. The correct boundary conditions \eqref{eq:TTM_L_BC} and \eqref{eq:TTM_R_BC} require purely outgoing waves or purely ingoing waves across the entire spatial domain.
Our present work provides analytic expressions for both QNM amplitudes and TTM amplitudes.
These expressions reveal crucial distinctions: when examining different modes, QNM amplitudes \eqref{eq:RTE_Amplitudes_1} belong to a different set than ${\rm TTM}_{\rm L}$ amplitudes \eqref{eq:TTM_L_Amplitudes_1}, while QNM amplitudes share the same set with ${\rm TTM}_{\rm R}$ amplitudes \eqref{eq:TTM_R_Amplitudes_1}.
Therefore, intersections can exist between QNMs and ${\rm TTM}_{\rm L}$, but are strictly impossible between QNMs and ${\rm TTM}_{\rm R}$.
The results in \cref{tab:ASmode-Errors} further confirm our conclusion in \cref{TTM_R_Eq}: \textit{ no complex frequency can be both a ${\rm{TTM}_{\rm{R}}}$ and a QNM}.

\subsection{Highly Damped Modes}

Using a variant of Leaver's CFM \cite{Leaver:1985ax}, Nollert \cite{Nollert:1993zz} first achieved reliable numerical computations of highly damped QNM frequencies in gravitational fields.
Based on these algorithms \cite{Dolan:2009nk,Casals:2013mpa,Ansorg:2016ztf,Leaver:1985ax,BertiWeb,Nollert:1993zz}, the BHPToolkit provides efficient code for computing QNMs of the Schwarzschild BH.
As shown in \cref{tab:QNM_diff_s}, our method demonstrates superior computational efficiency compared to the BHPToolkit when calculating high-overtone modes ($\hat{n} \in [0,41]$) for arbitrary perturbation fields.
Furthermore, unlike the BHPToolkit, our method successfully captures all high-overtone modes without any spectral omissions.

\section{QNMs of Kerr BHs}\label{sec:QNM-Kerr}

In this section, the QNM spectrum of the Kerr BH is presented, which is divided into two parts. In the first part, the QNMs of the massless perturbative field are calculated, and in the second part, the QNMs of the massive scalar field are calculated.
\subsection{Massless Kerr QNMs}

The QNM spectrum of Kerr BHs exhibits rich and complicated structures, particularly showing significant mode deficiencies in traditional gravitational QNMs.
The Kerr QNM spectrum exhibits relative errors $\xi(\omega)$ and $ \vartheta (\omega )$ belonging to $[10^{-16}, 10^{-12}]$ at default machine precision.
To our knowledge, the most advanced results for Kerr QNMs were achieved by Cook's improved CFM \cite{Cook:2014cta,Cook:2016fge,Cook:2016ngj,cook_2024_14024959}. While providing high-precision spectra for Kerr gravitational perturbations, Cook's solution still lacks modes crossing the NIA.
This section compares our complete QNM spectrum of the Class-I solution $\omega_{\text{I}}$ with Cook's results. Note that Cook's overtone indices are denoted by $\bar{n}$, distinct from our $\hat{n}$ notation.

For the Kerr BHs, the potential of the ATE \eqref{eq:GFoATE} is
\begin{equation}\label{eq:U_ATE_Kerr}
U(\Theta ) = {a^2}{\omega ^2}{\Theta ^2} - \frac{{{{(m + s\Theta )}^2}}}{{1 - {\Theta ^2}}} - 2a\omega s\Theta  + s + {\kern 1pt} {{\kern 1pt} _s}{A_{\ell m}},
\end{equation}
and the potential of the RTE \eqref{eq:GFoRTE} is
\begin{equation}\label{eq:V_RTE_Kerr}
 V(r) = {K^2} - isK\Delta ' + \left( {2isK' - \lambda } \right)\Delta,
\end{equation}
where $\lambda  = {\kern 1pt} {{\kern 1pt} _s}{A_{\ell m}}(a\omega ) + {a^2}{\omega ^2} - 2am\omega $ and $K = ({r^2} + {a^2})\omega - am$.

The HeunC parameters of the ATE \eqref{eq:GFoATE} can be derived in the following form,
\begin{subequations}\label{eq:ATE_HCprameters_Kerr}
 \begin{align}
&{{\alpha _{ - 1}} = 4a\omega ,\quad {\alpha _{ + 1}} =  - 4a\omega ,\quad }\\
&{{\beta _{ - 1}} = m - s,\quad {\beta _{ + 1}} = m + s,\quad }\\
&{{\gamma _{ - 1}} = m + s,\quad {\gamma _{ + 1}} = m - s,\quad }\\
&{{\delta _{ - 1}} = 4sa\omega ,\quad {\delta _{ + 1}} =  - 4sa\omega ,}\\
&{{\eta _{ - 1}} = \tfrac{1}{2}{m^2} - \tfrac{1}{2}{s^2} - s - 2ma\omega  - 2sa\omega  - \lambda ,}\\
&{{\eta _{ + 1}} = \tfrac{1}{2}{m^2} - \tfrac{1}{2}{s^2} - s - 2ma\omega  + 2sa\omega  - \lambda .}
\end{align}
\end{subequations}

The HeunC parameters of the RTE \eqref{eq:GFoRTE} can be derived in the following form,
\begin{subequations}\label{eq:RTE_HCparameters_Kerr}
 \begin{align}
&{\alpha^+ _r} = 2i\omega ({r_ + } - {r_ - }),\\
&{\beta^+ _r} = -s - \frac{{2i\omega (r_ + ^2 + {a^2}) - 2iam}}{{{r_ + } - {r_ - }}},\\
&{\gamma^+ _r} = s - \frac{{2i\omega (r_ - ^2 + {a^2}) - 2iam}}{{{r_ + } - {r_ - }}},\\
&{\delta _r} = 2is\omega ({r_ - } - {r_ + }) + 2{\omega ^2}\Delta, \\
&\eta_r  = 2is\omega {r_ + } - \tfrac{1}{2}{s^2} - s - \lambda \\
& -\frac{{2\left[ {\omega (r_ + ^2 + {a^2}) - am} \right]\left[ { ({a^2} + 2{r_ - }{r_ + } - r_ + ^2)\omega - am} \right]}}{{{{({r_ + } - {r_ - })}^2}}},\nonumber \\
&{r_ \pm } = M \pm \sqrt {{M^2} - {a^2}}.
\end{align}
\end{subequations}

We numerically compute the QNM spectrum for arbitrary perturbation fields of Kerr BHs, with the same three-index range as the Schwarzschild QNMs in \cref{sec:SchBH-QNM}.
The spin parameter typically covers $a \in [0,0.999]$, though some high-overtone QNMs for $m=0$ may not reach near-extremal spins at $a=0.999$.
Our dataset \cite{ChenQNM} provides the complete QNM spectra available.
The dataset uses $a$-spacing $\square a=0.001$ for most QNM sequences. For complicated structures (particularly the helical structures of $m=0$ modes or trajectories crossing the NIA), smaller spacing is used.
Following the Schwarzschild QNM analysis in \cref{sec:SchBH-QNM}, we particularly focus on gravitational QNMs in our numerical tests, as existing literature \cite{Berti:2003jh,Berti:2009kk,Cook:2014cta,Cook:2016fge} exhibits significant incompleteness in these modes.

\begin{figure*}[htbp]
	\centering
\subfloat[ Incomplete results from Cook\label{fig:l2n8-a} ]{\includegraphics[width=2.6in]{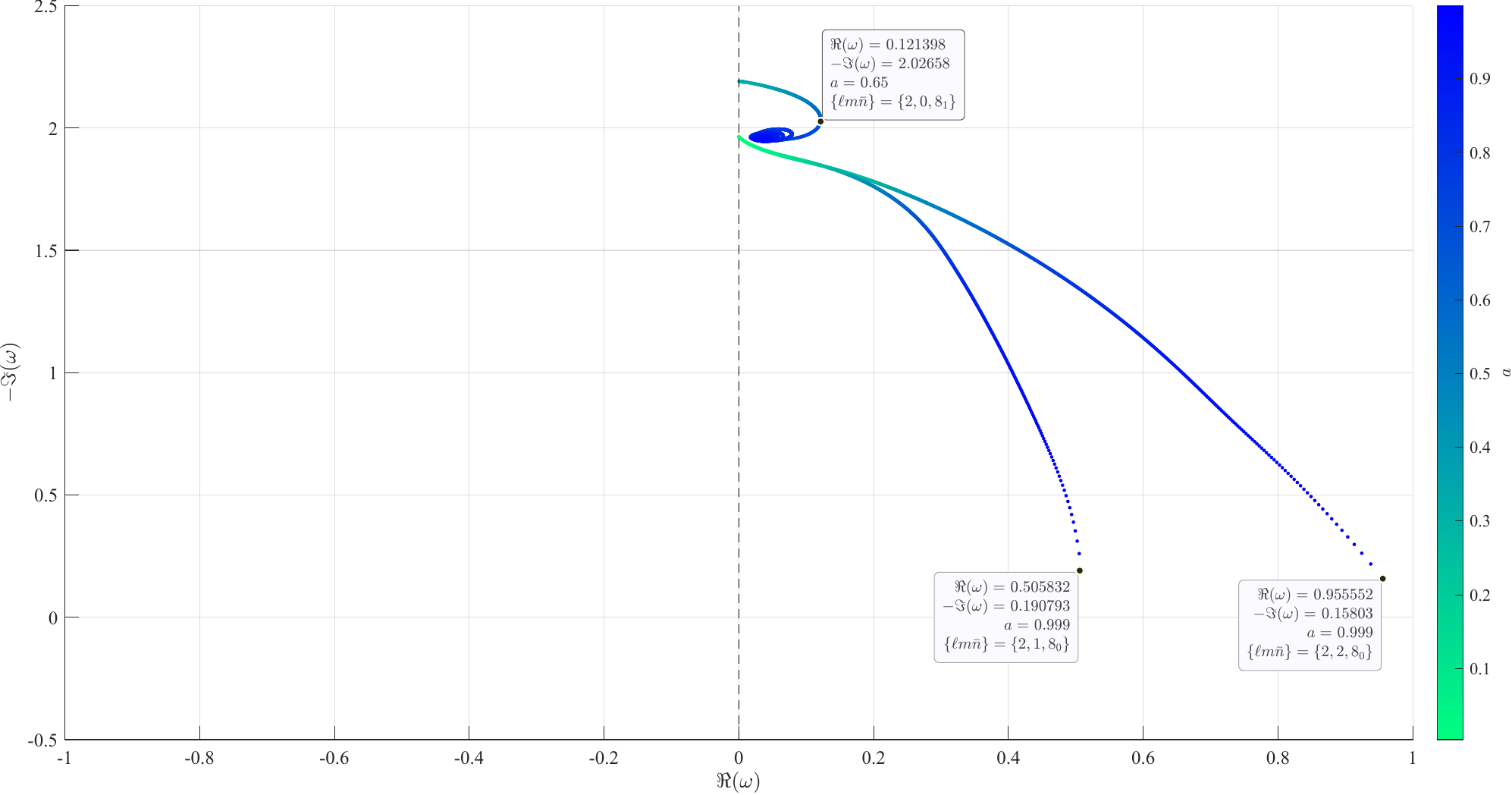}}\,\,
\subfloat[ HeunC-Chen method ]{\includegraphics[width=2.6in]{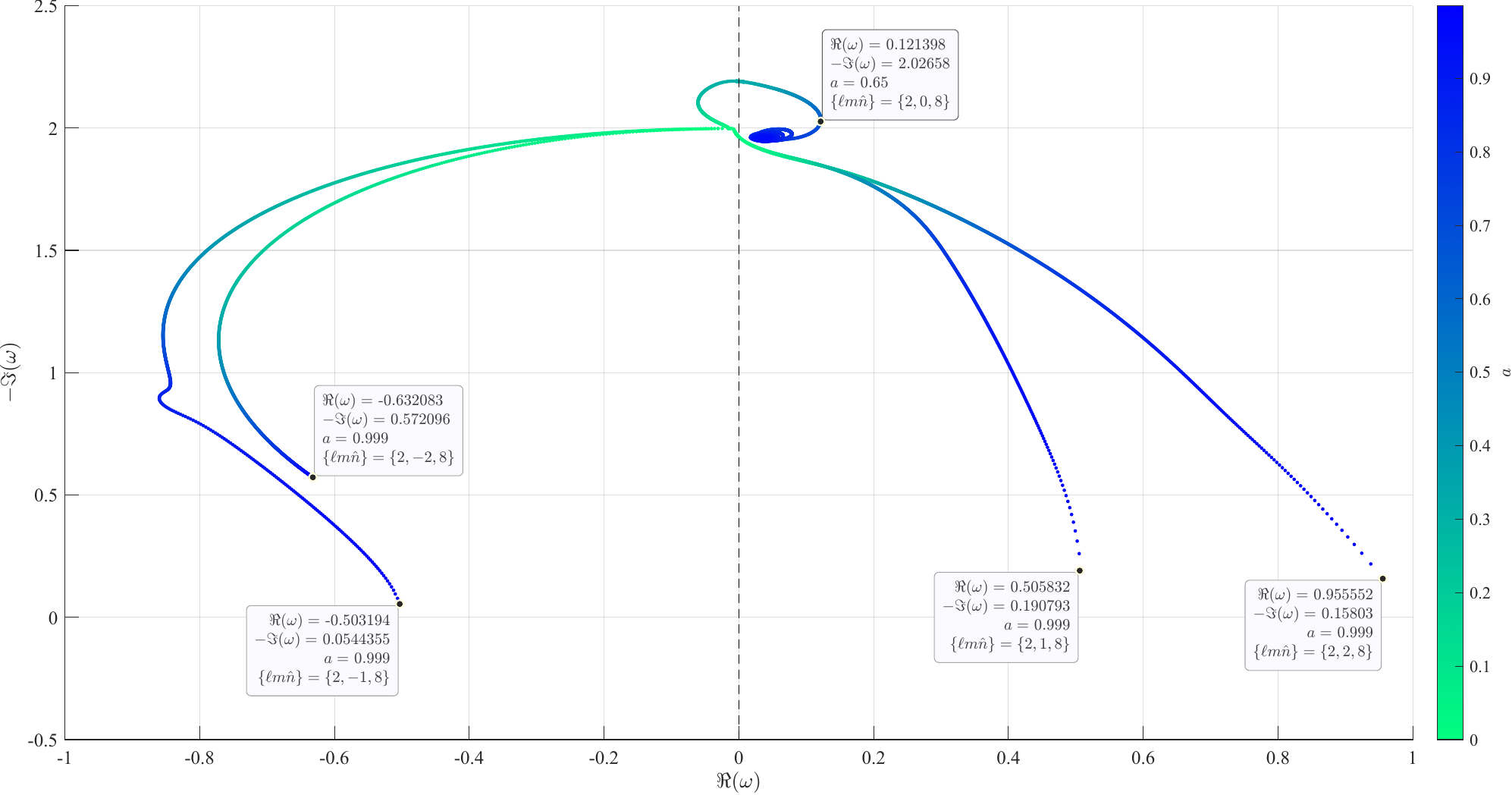}}\,\,
\subfloat[Close-up of Zeeman splitting \label{fig:l2n8-c}]{\includegraphics[width=1.7in]{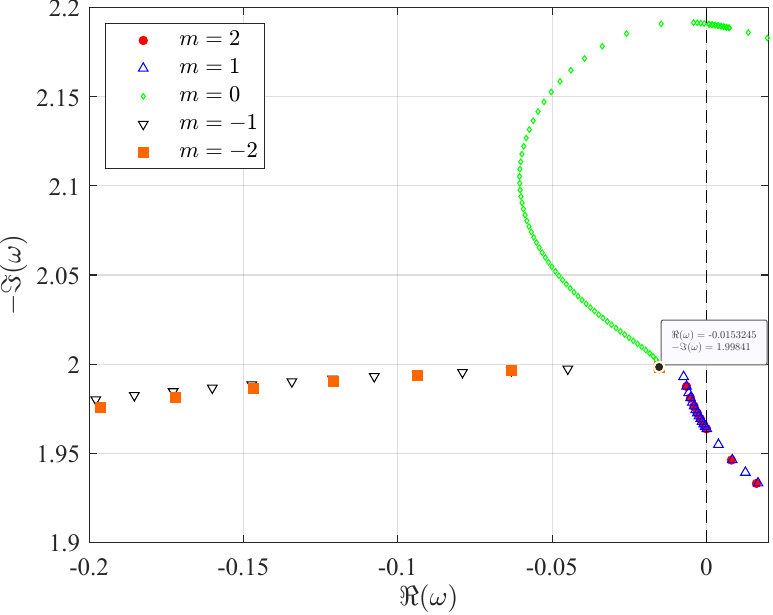}}
\caption{ QNM spectrum for $\ell= 2$ and $ \hat n = 8$ originating from the unconventional mode \eqref{eq:SchNewModesL2}. }\label{fig:l2n8}
\end{figure*}
\begin{figure*}[htbp]
	\centering
\subfloat[ Incomplete results from Cook\label{fig:l2n9-a} ]{\includegraphics[width=2.6in]{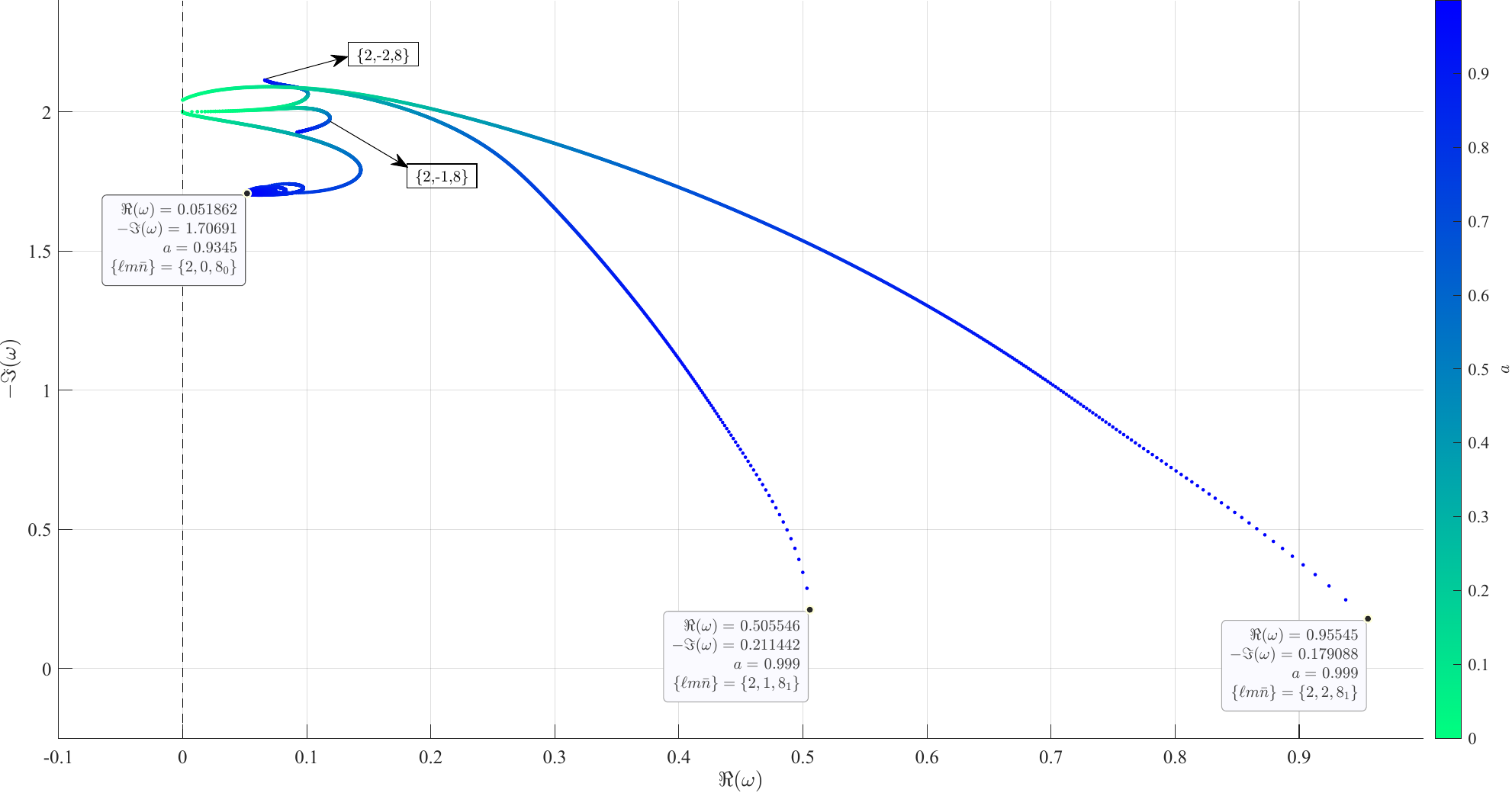}}\,\,
\subfloat[ HeunC-Chen method ]{\includegraphics[width=2.6in]{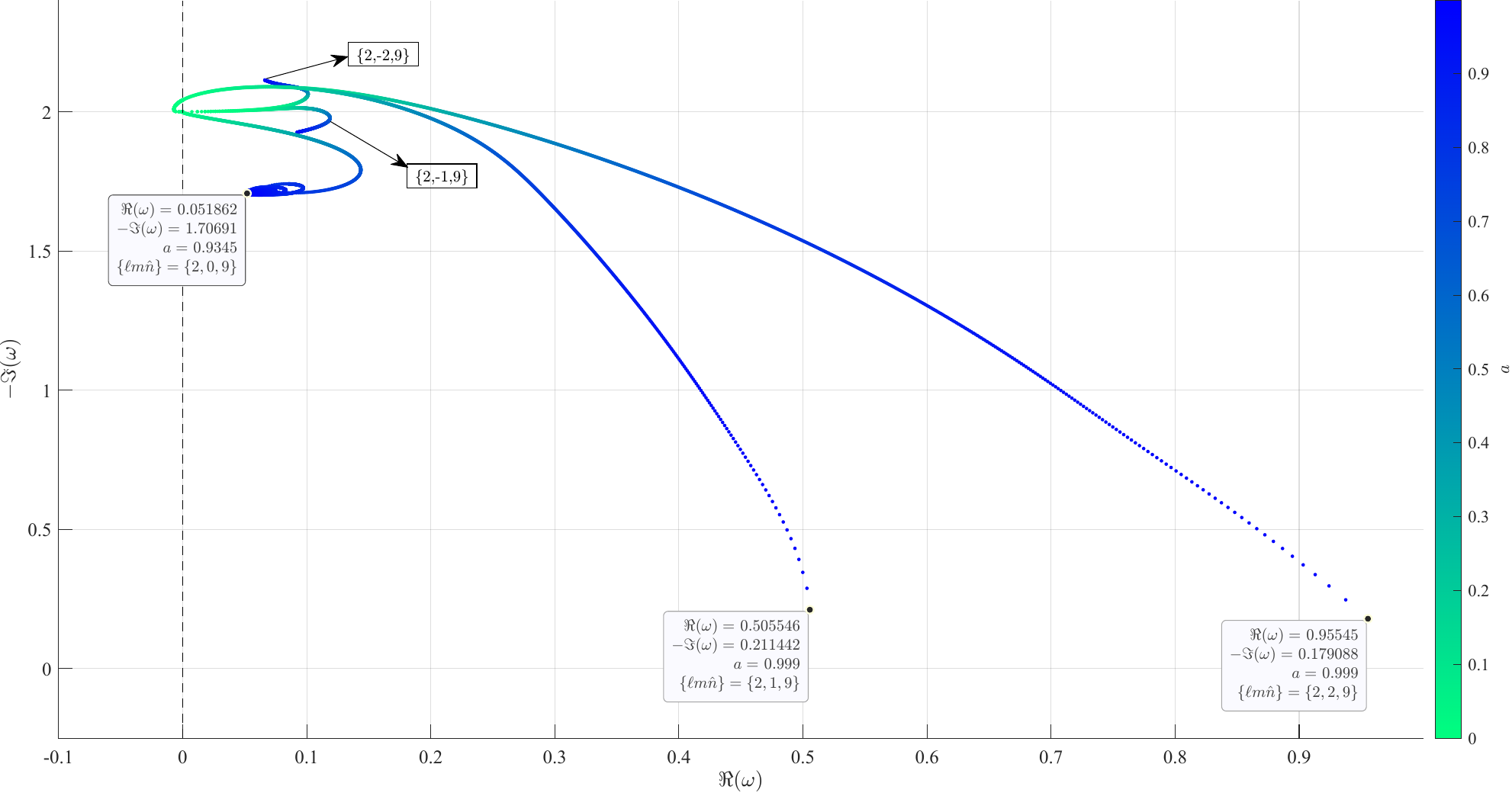}}\,\,
\subfloat[Close-up of Zeeman splitting\label{fig:l2n9-c} ]{\includegraphics[width=1.7in]{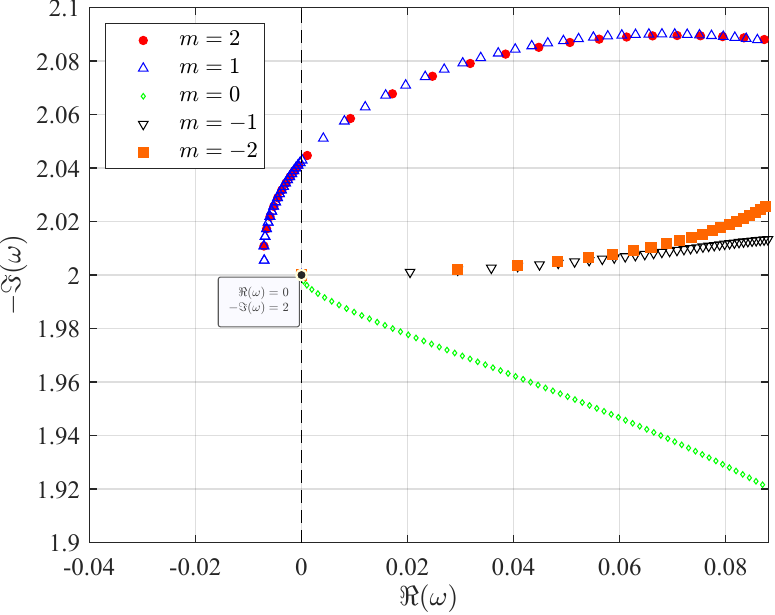}}
\caption{ QNM spectrum for $\ell= 2$ and $ \hat n = 9$ originating from the AS mode \eqref{eq:ASmode}.}\label{fig:l2n9}
\end{figure*}
\begin{figure*}[htbp]
	\centering
\subfloat[Complete spectrum ]{\includegraphics[width=3.4in]{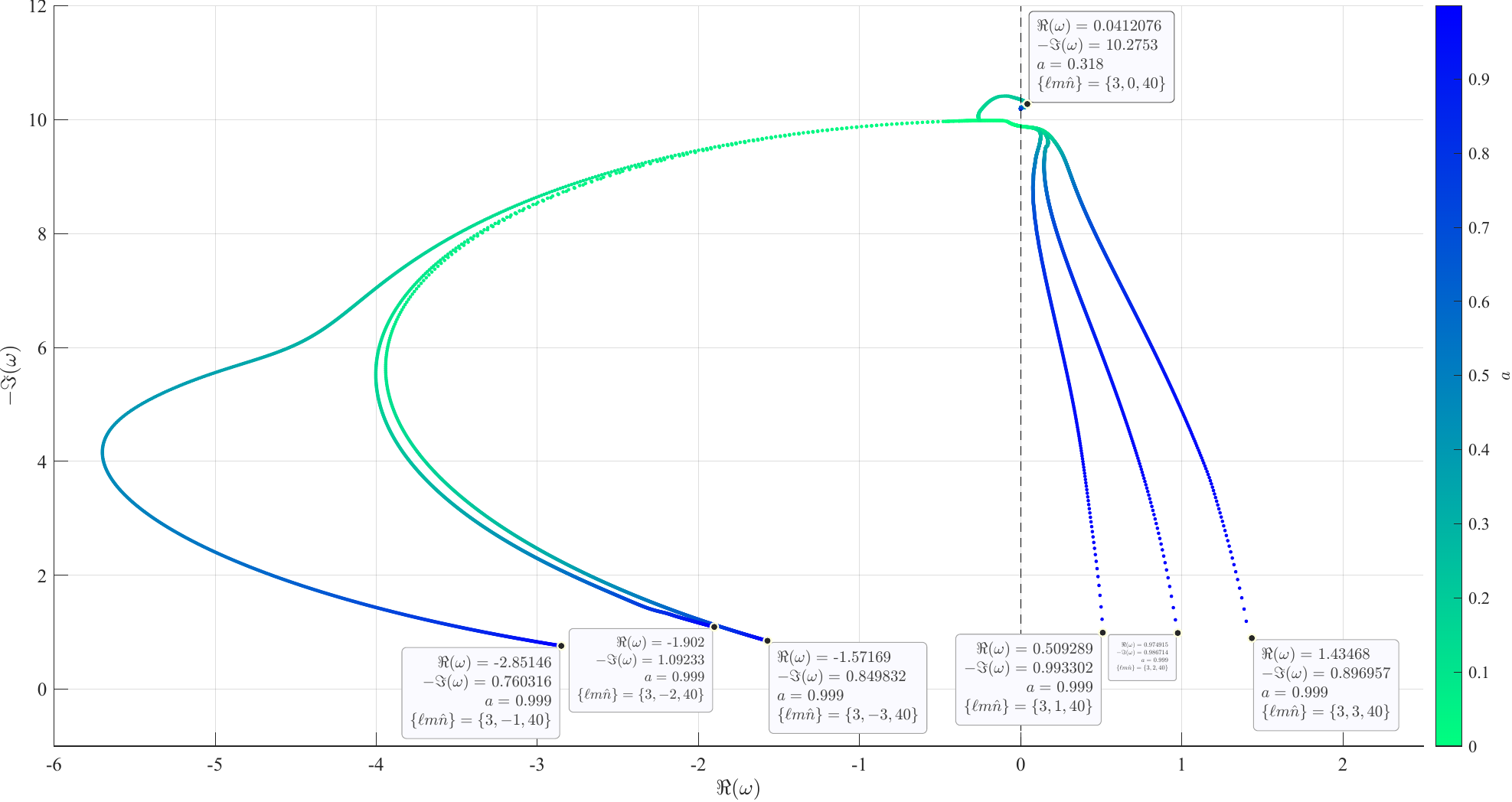}}
\subfloat[Close-up of Zeeman splitting]{\includegraphics[width=3.4in]{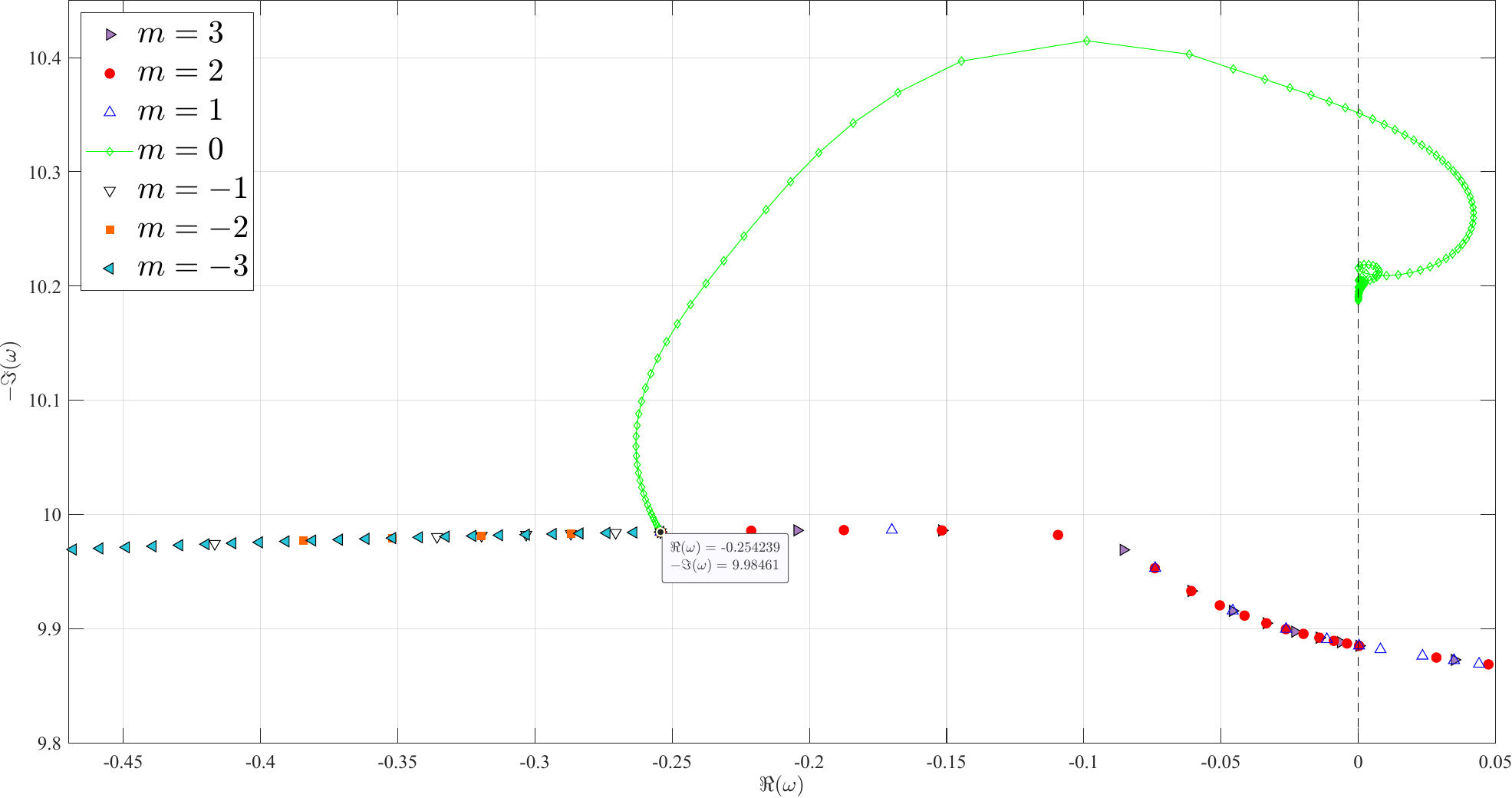}}
\caption{ QNM spectrum for $\ell= 3, \hat n = 40$ originating from the unconventional mode \eqref{eq:SchNewModesL3}. }\label{fig:l3n40}
\end{figure*}

\begin{table*}[htbp]
  \centering
  \caption{Computational times (in seconds) of the first eight QNMs of the Kerr BH with $s=-2$.}
    \begin{tabular}{ccccccccc}
    \toprule
         & \multicolumn{4}{c}{HeunC-Chen} &      & \multicolumn{3}{c}{Cook's CFM} \\
\cmidrule{2-5}\cmidrule{7-9}    ($\ell,m,\hat n$)
         & $L_2$-Error   & $L_\infty$-Error &  1-Core Time & 12-Cores Time &   & $L_2$-Error   & $L_\infty$-Error & 1-Core Time \\
    \midrule
    220  & $6.351 \times 10^{-12}$\,\, & $9.887 \times 10^{-13}$ & 9.587  & 2.083  &      & $9.508 \times 10^{-12}$\,\, & $5.672 \times 10^{-12}$ & 59.480  \\
    221  & $3.702 \times 10^{-12}$ & $7.850 \times 10^{-13}$ & 7.616  & 1.560  &      & $1.594 \times 10^{-12}$ & $3.418 \times 10^{-13}$ & 80.845  \\
    222  & $1.928 \times 10^{-12}$ & $3.771 \times 10^{-13}$ & 8.374  & 1.566  &      & $1.460 \times 10^{-12}$ & $2.246 \times 10^{-13}$ & 90.314  \\
    223  & $3.937 \times 10^{-13}$ & $7.184 \times 10^{-14}$ & 8.441  & 1.537  &      & $3.878 \times 10^{-13}$ & $2.204 \times 10^{-13}$ & 88.494  \\
    224  & $1.425 \times 10^{-13}$ & $3.055 \times 10^{-14}$ & 8.515  & 1.714  &      & $6.962 \times 10^{-12}$ & $7.330 \times 10^{-13}$ & 174.858  \\
    225  & $7.479 \times 10^{-14}$ & $1.659 \times 10^{-14}$ & 8.342  & 1.776  &      & $3.943 \times 10^{-12}$ & $3.325 \times 10^{-13}$ & 189.238  \\
    226  & $5.344 \times 10^{-14}$ & $9.344 \times 10^{-15}$ & 8.756  & 2.017  &      & $2.014 \times 10^{-12}$ & $1.543 \times 10^{-13}$ & 211.323  \\
    227  & $4.518 \times 10^{-14}$ & $7.902 \times 10^{-15}$ & 9.375  & 1.822  &      & $1.120 \times 10^{-12}$ & $7.087 \times 10^{-14}$ & 243.570  \\
    \bottomrule
    \end{tabular}%
  \label{tab:Kerr_L2M2}%
\end{table*}
\begin{table*}[htbp]
  \centering
  \caption{Computational times (single-core) of the TTMs of the Kerr BH with $a\in[0,0.999]$.}
    \begin{threeparttable}
    \begin{tabular}{lccccccccccc}
    \toprule
         & \multicolumn{3}{c}{HeunC-STC} &      & \multicolumn{3}{c}{Dual-HeunC} &      & \multicolumn{3}{c}{Cook’s CFM-STC} \\
\cmidrule{2-4}\cmidrule{6-8}\cmidrule{10-12}    ($\ell,m,\hat n$) & $L_2$-Error   & $L_\infty$-Error & Time(s) &      & $L_2$-Error   & $L_\infty$-Error& Time(s)  &     & $L_2$-Error   & $L_\infty$-Error & Time(s) \\
    \midrule
    (2,2,0)  & $8.589 \times {10^{ - 16}}$ & $3.895 \times {10^{ - 16}}$ & 5.423  &      & $7.071 \times {10^{ - 10}}$ & $5.126 \times {10^{ - 10}}$ & 12.499  &      & $6.331 \times {10^{ - 10}}$ & $1.385 \times {10^{ - 10}}$ & 29.684  \\
    (2,1,0)  & $9.629 \times {10^{ - 16}}$ & $3.052 \times {10^{ - 16}}$ & 5.171  &      & $7.705 \times {10^{ - 12}}$ & $3.340 \times {10^{ - 12}}$ & 11.984  &      & $4.743 \times {10^{ - 11}}$ & $8.582 \times {10^{ - 12}}$ & 44.916  \\
    (2,0,0)$^\bigstar$ & $2.442 \times {10^{ - 11}}$ & $2.440 \times {10^{ - 11}}$ & 3.051  &      & $9.297 \times {10^{ - 12}}$ & $5.078 \times {10^{ - 12}}$ & 9.038  &      & $2.271 \times {10^{ - 11}}$ & $2.268 \times {10^{ - 11}}$ & 44.030  \\
    (2,-1,0) & $6.794 \times {10^{ - 10}}$ & $8.591 \times {10^{ - 11}}$ & 5.240  &      & $3.191 \times {10^{ - 8}}$ & $6.931 \times {10^{ - 9}}$ & 13.044  &      & $1.525 \times {10^{ - 6}}$ & $2.040 \times {10^{ - 7}}$ & 95.807  \\
    (2,-2,0) & $1.502 \times {10^{ - 10}}$ & $1.864 \times {10^{ - 11}}$ & 5.552  &      & $5.651 \times {10^{ - 8}}$ & $1.975 \times {10^{ - 9}}$ & 12.109  &      & $2.747 \times {10^{ - 7}}$ & $5.159 \times {10^{ - 8}}$ & 97.711  \\
    \bottomrule
    \end{tabular}%
    \begin{tablenotes}
    \item[$\bigstar$] For the (2,2,0) sequence, QNMs were computed only for $a \in [0,0.494]$. As shown in \cref{fig:TTM_L2L3}, when $a \in [0,0.494]$, the (2,2,0) sequence lies on the NIA. This configuration facilitates comparison of all three methods at the same step size ($\square a=0.001$).

    \end{tablenotes}

    \end{threeparttable}
  \label{tab:Kerr_TTMs}%
\end{table*}

\begin{figure*}[htbp]
	\centering
\subfloat[$\ell=2$]{\includegraphics[width=3.4in]{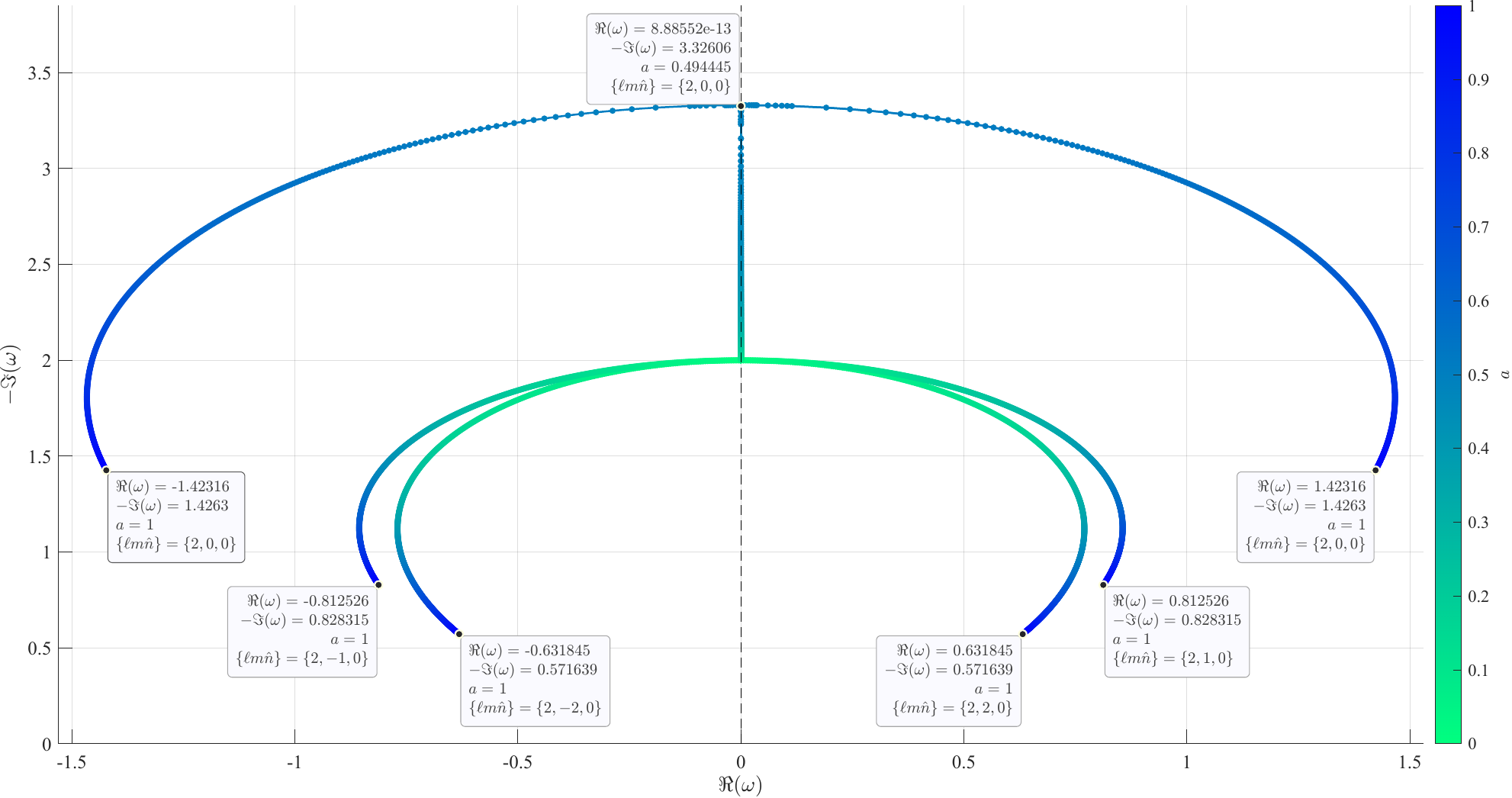}}\,\,
\subfloat[$\ell=3$]{\includegraphics[width=3.4in]{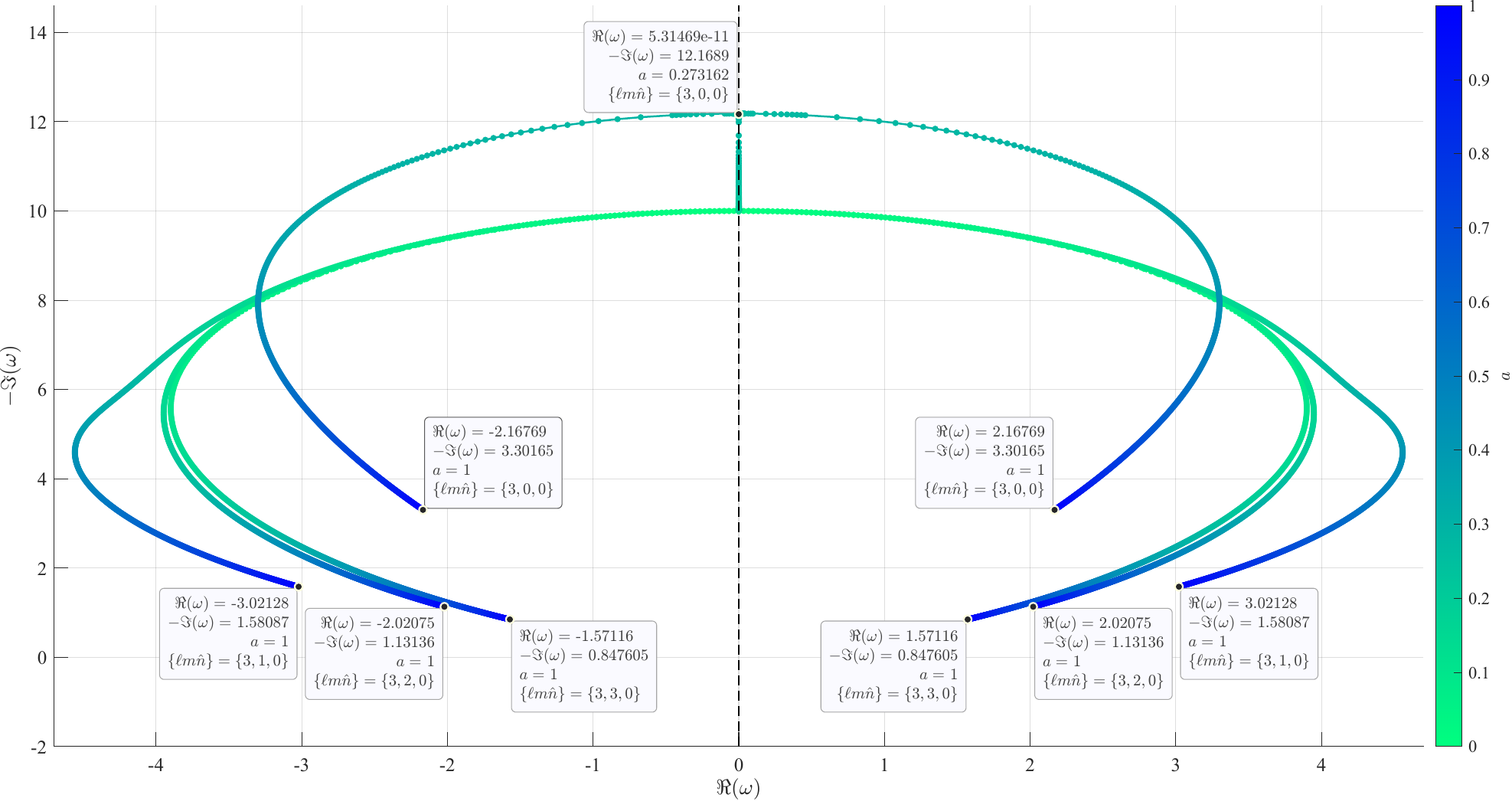}}\,\,
\caption{Kerr algebraically special (TTM) mode sequences for $\ell =2,\,m=\{0,1,2\}$ and $\ell =3,\,m=\{0,1,2,3\}$ with $a\in \mbox{[}0,1\mbox{]}$.}\label{fig:TTM_L2L3}
\end{figure*}

\begin{figure*}[htbp]
	\centering
\subfloat[Behavior of QNMs for $m=\{1,2,3\}$.]{\includegraphics[width=3.4in]{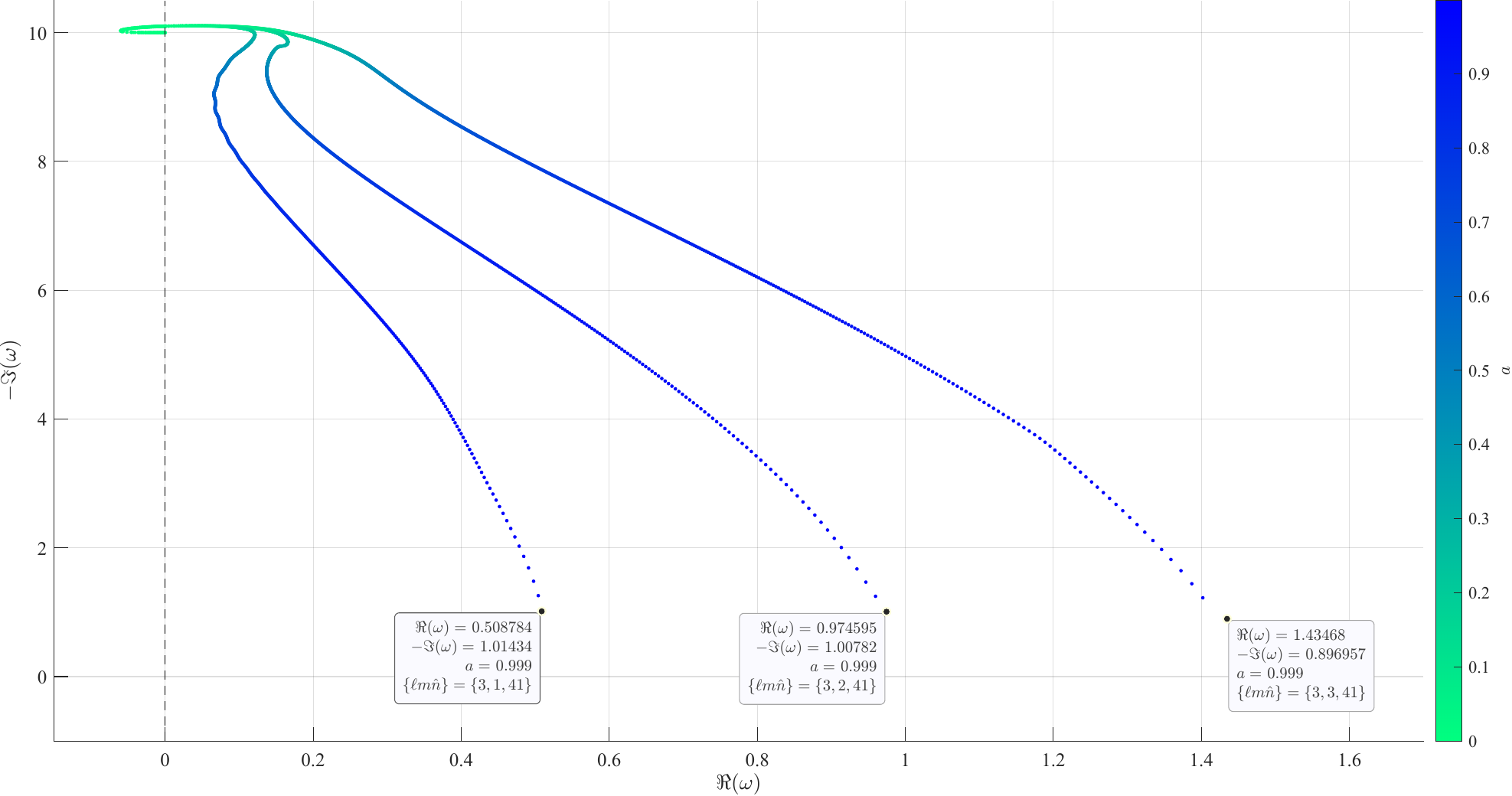}}
\subfloat[Behavior of QNMs for $m=\{0,-1,-3\}$.]{\includegraphics[width=3.4in]{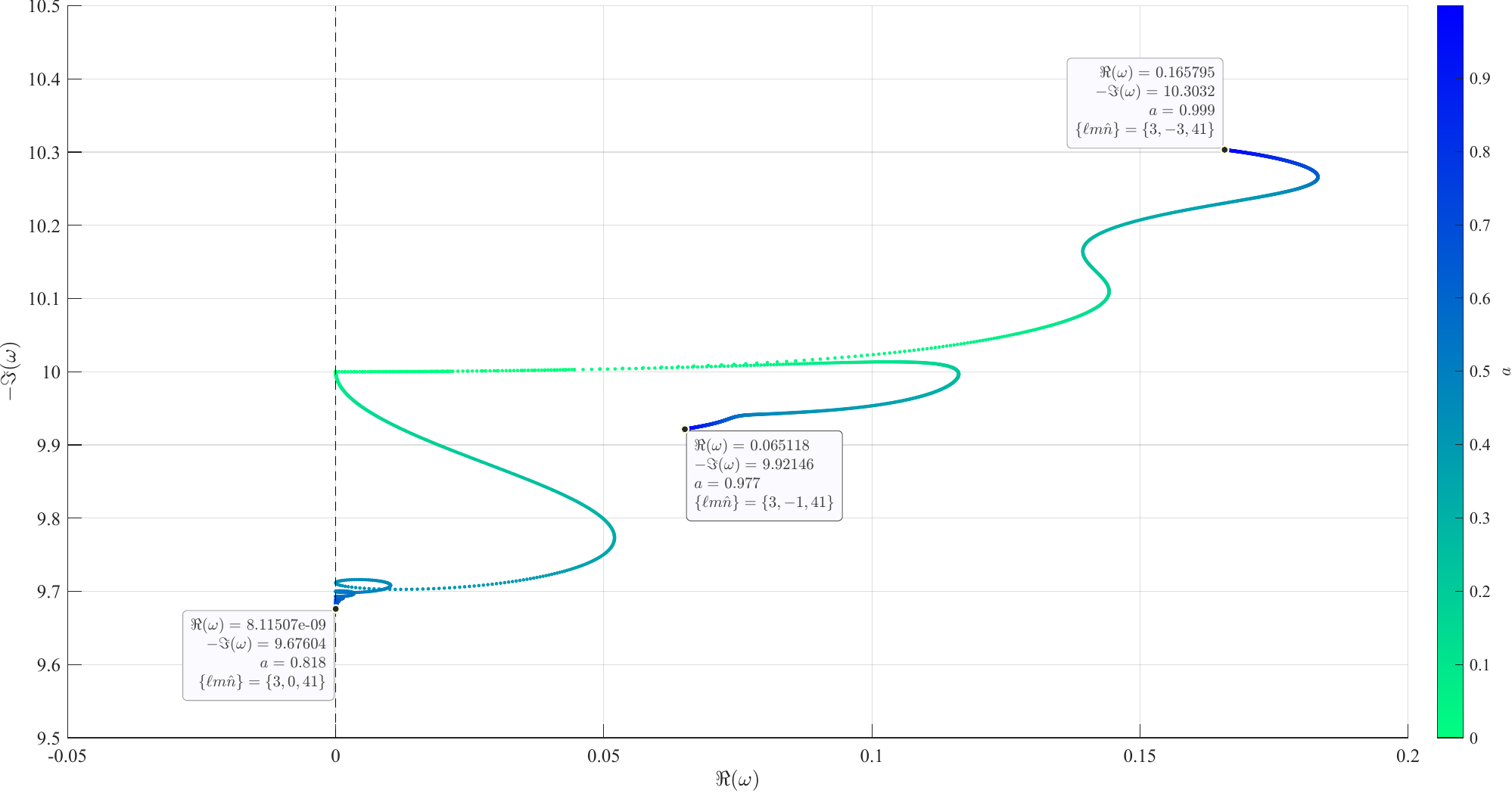}}\\
\subfloat[Behavior of QNMs for $m=-2$.]{\includegraphics[width=3.4in]{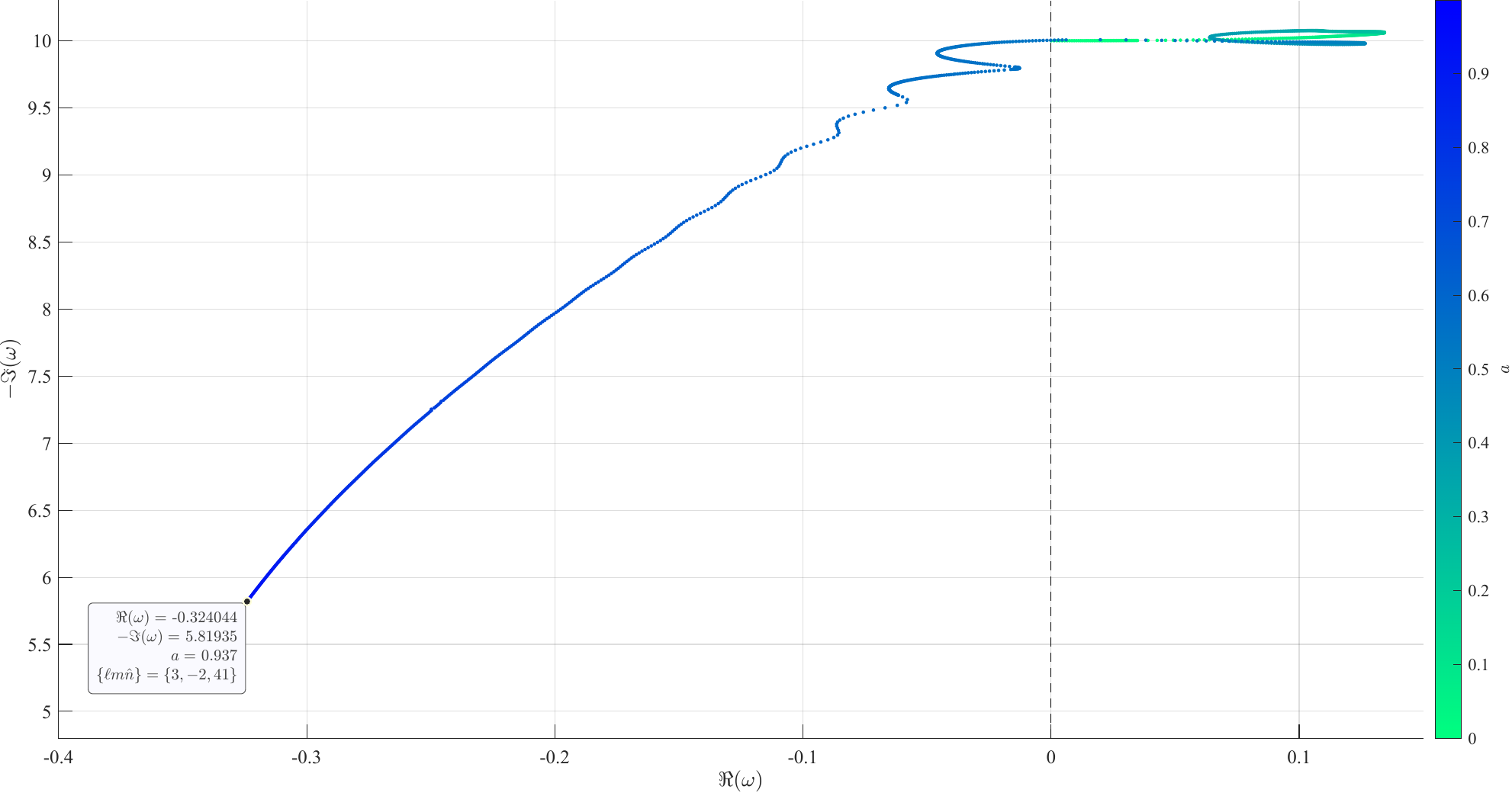}}
\subfloat[Close-up of Zeeman splitting. ]{\includegraphics[width=3.4in]{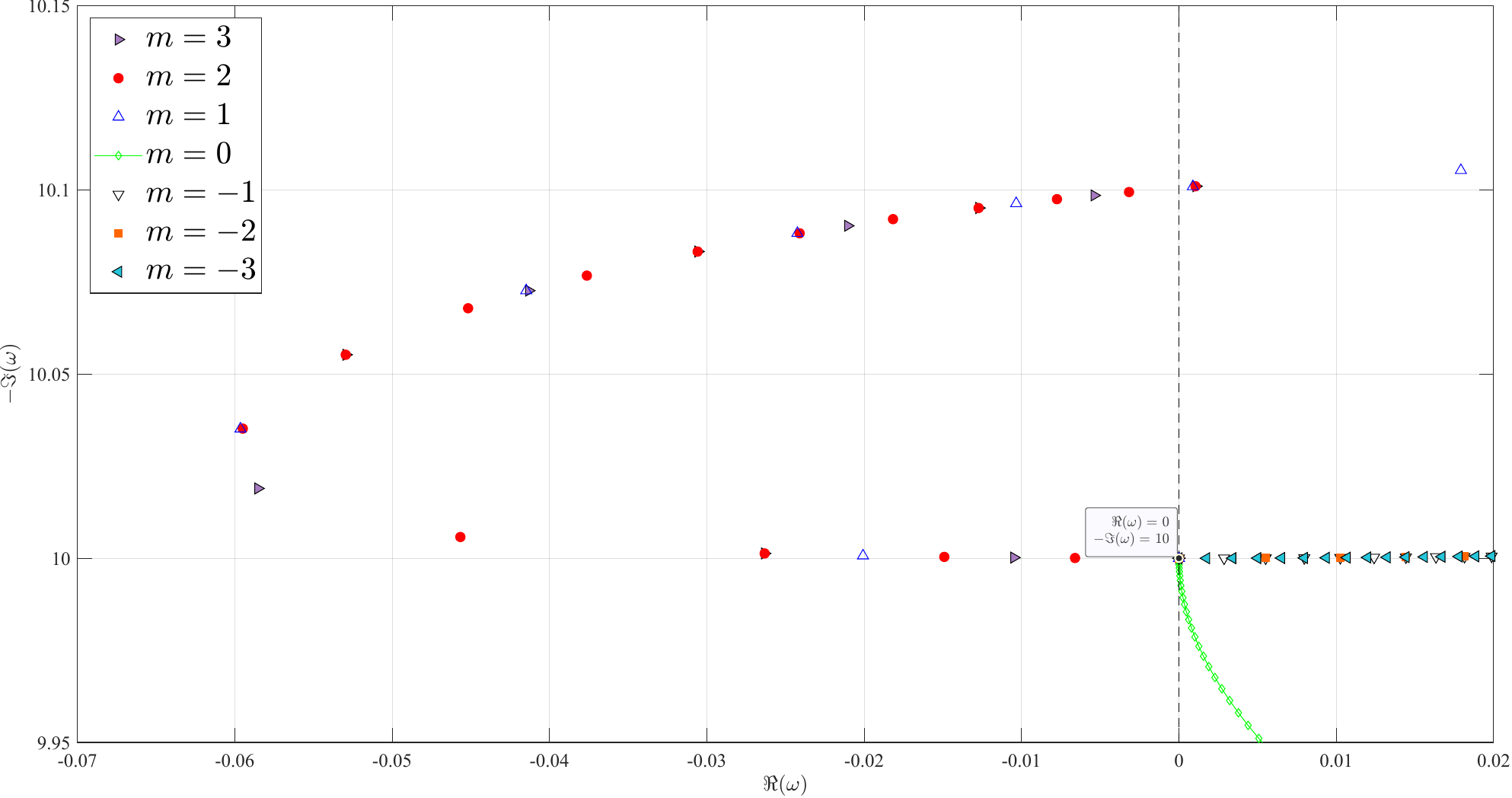}}
\caption{ QNM spectrum for $\ell= 3, \hat n = 41$ originating from the AS mode \eqref{eq:ASmode}.}\label{fig:l3n41}
\end{figure*}

\subsubsection{Weakly Damped and Unconventional Modes}
\cref{tab:Kerr_L2M2} presents the computational time required to calculate the first eight overtones of QNMs of $\ell=m=2$, in which the $L_2$-error and $L_\infty$-error of $\xi(\omega)$ are employed.
The results demonstrate significantly superior computational efficiency compared to Cook's method at equivalent errors.
Our algorithm achieves further acceleration through parallel computation (12 cores), whereas Cook's method cannot be parallelized due to its inherent sequential dependency\textemdash each iteration of QNM frequency calculations requires the results from the preceding iteration.

The overtone number is an artificially introduced integer index labeling QNMs by increasing $|\mathrm{Im}(\omega)|$.
Schwarzschild QNMs have one mode per overtone, making labeling straightforward.
Even when missing unconventional modes, only one overtone per $\ell$ was absent.
However, Kerr QNM overtones exhibit extreme complexity.
Although Fiziev \cite{Fiziev:2010yy,Fiziev:2011mm} provided the high-precision unconventional mode \eqref{eq:SchNewModesL2} for $\ell=2$, subsequent studies \cite{Berti:2003jh,Cook:2014cta,Cook:2016fge,Cook:2016ngj} failed to compute the complete QNM spectrum using this mode as initial conditions.
Theoretically, unconventional modes for $\ell=2$ should split into 5 branches with increasing spin.
Because the QNMs crossing the NIA were not handled \cite{Berti:2003jh,Cook:2014cta,Cook:2016fge,Cook:2016ngj}, only incomplete spectra were obtained: (i) three branches ($m=0,1,2$) split from the unconventional modes, and (ii) five branches ($m=0,\pm1,\pm2$) split from the AS mode \eqref{eq:ASmode}.

Subsequently, Cook \cite{Cook:2014cta,Cook:2016fge,Cook:2016ngj,Berti:2025hly} classified these eight branches into the QNM sequences of overtone $\bar{n} = 8$.
For the $\bar{n} = 8$ QNM sequences with the same $m$ value, new indices were introduced according to their magnitudes and used as the subscripts of the overtone. This is the so-called ``overtone multiplets".
 These eight branches respectively correspond to \Cref{fig:l2n8-a,fig:l2n9-a}.
\cref{fig:l2n8} displays the complete QNM spectrum for $\ell=2$ and $\hat{n}=8$.
The detailed Zeeman-like splitting structure of the QNM spectrum can be observed in \cref{fig:l2n8-c}.
This reveals that each overtone should correspond to exactly $2\ell+1$ spectral branches in the complete QNM spectrum.
Moreover, it is evident that Cook's results indeed lack two complete QNM sequences for $m = -1, -2$ and three partial QNM sequences with $m = 0, 1, 2$.
Similarly, the Kerr QNM sequences for unconventional modes with $\ell=3$  and $\hat{n}=40$ are presented in \cref{fig:l3n40}.
It can also be seen that the $\hat{n}=40$ QNM spectrum exhibits Zeeman splitting with exactly $2\ell+1=7$ branches, showing no evidence of ``overtone multiplets" phenomena.

\subsubsection{Intersections of TTMs and QNMs}\label{sec:TTM_QNM}
This paper provides two methods for solving TTMs that share the same eigenvalue equation \eqref{eq:ATE_Wronskian} for angular separation constants.
The difference lies in their radial equations:
\begin{itemize}
      \item The first method solves the frequency of radial equations using the Starobinsky-Teukolsky constant \eqref{eq:Starobinsky_const}  (named as HeunC-STC).
    \item The second method solves the frequency of radial equations via the HeunC method described in \cref{sec:HeunC-TTM}  (named as Dual-HeunC).
\end{itemize}
Cook also proposed a method to solve the TTM \cite{Cook:2014cta,Cook:2018ses,Cook:2022kbb}. The angular separation constant uses the modified CFM, while the frequency of the radial equation uses the Starobinsky-Teukolsky constant (named as CFM-STC).

Using the HeunC-STC method, we show the Kerr TTMs for $\ell=2$ and $\ell=3$ in \cref{fig:TTM_L2L3}.
\cref{tab:Kerr_TTMs} presents the computational time of the Kerr TTMs for $\ell=2$ and $\ell=3$, in which the $L_2$-error and $L_\infty$-error of $\xi(\omega)$ are employed.
The results indicate that the HeunC-STC method demonstrates better computational efficiency and accuracy than the Dual-HeunC method and Cook's CFM-STC method \cite{Cook:2014cta,Cook:2018ses,Cook:2022kbb}.
This is expected because the HeunC-STC method employs the exact analytical expression \eqref{eq:Starobinsky_const} for the TTM equations, while the Dual-HeunC method uses approximate analytical expressions.
Even though the Dual-HeunC method does not adopt the Starobinsky-Teukolsky constant, its computational efficiency still outperforms Cook’s CFM-STC method. Both the HeunC-STC method and the Dual-HeunC method owe their high efficiency to the high-performance computation \cite{Motygin2018} of the confluent Heun function, which surpasses the CFM method.
According to Cook's results \cite{Cook:2018ses,Cook:2022kbb}, the TTMs for $m=0$ exhibit exceptionally complicated behavior.
While this paper focuses on QNM spectra, the HeunC-Chen method for solving TTMs and detailed related discussions can be found in Ref. \cite{ChenTTMs}.
\Cref{fig:l2n9,fig:l3n41} show the complete QNM spectrum for $\ell=2,\hat{n}=9$ and $\ell=3,\hat{n}=41$.
It can also be seen that Cook's results indeed lack certain QNM sequences corresponding to $\ell=2,\hat{n}=9$.

\begin{figure*}[htbp]
	\centering
\subfloat[ Incomplete results from Cook ]{\includegraphics[width=3.4in]{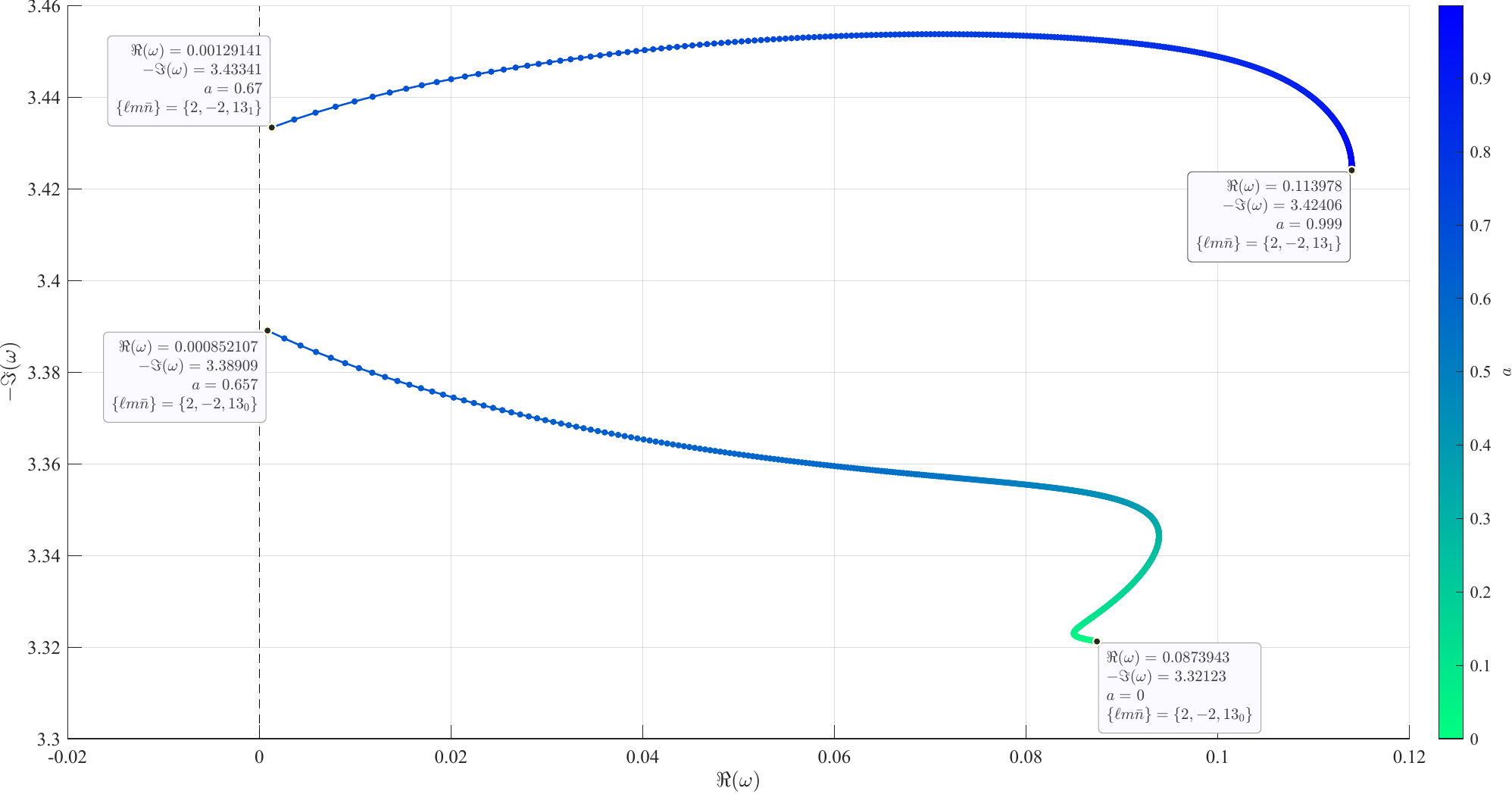}}\,\,
\subfloat[ HeunC-Chen method ]{\includegraphics[width=3.4in]{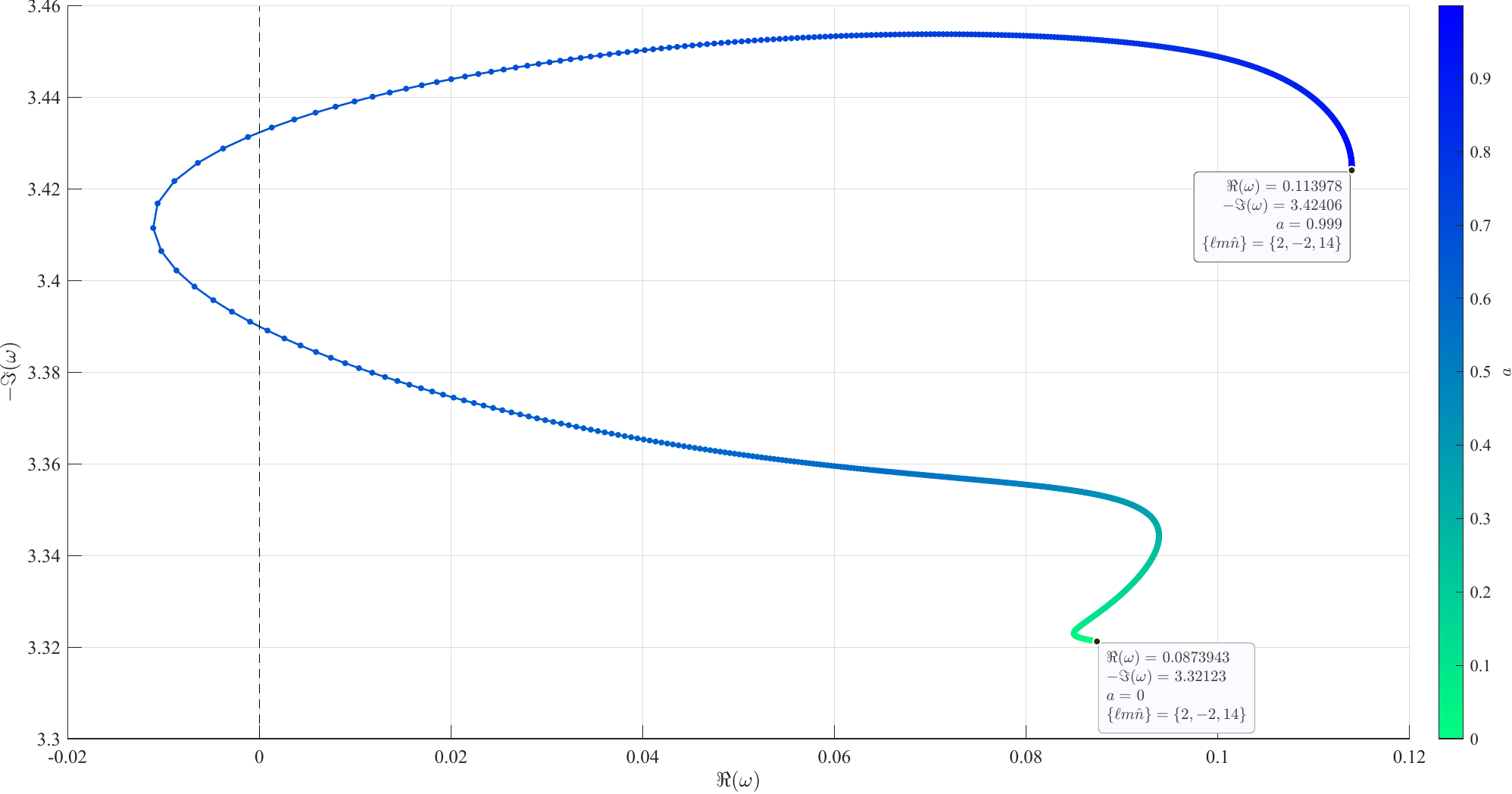}}\,\,
\subfloat[ Incomplete results from Cook \label{figc:l2mf2n14-17}]{\includegraphics[width=3.4in]{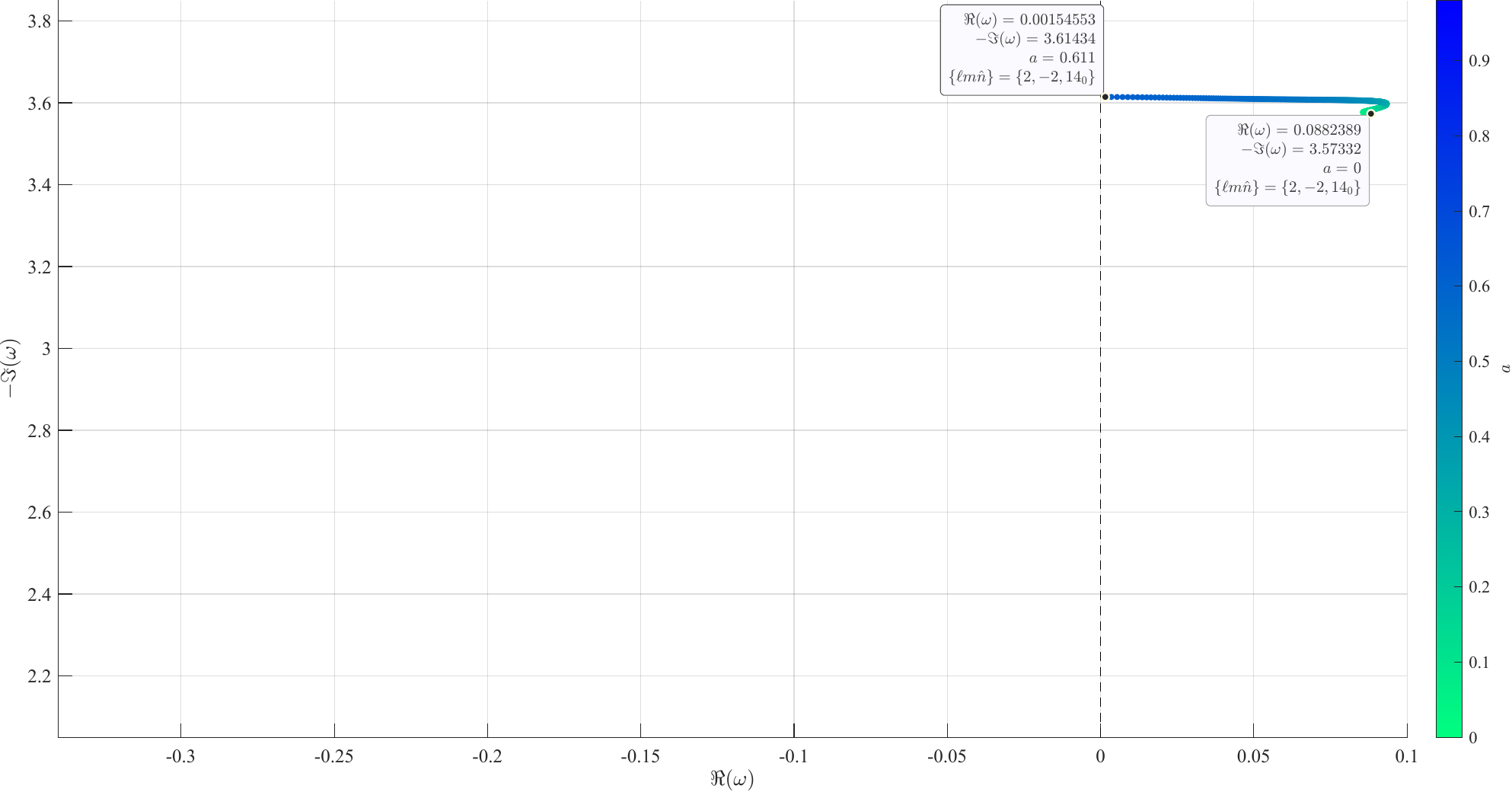}}\,\,
\subfloat[ HeunC-Chen method ]{\includegraphics[width=3.4in]{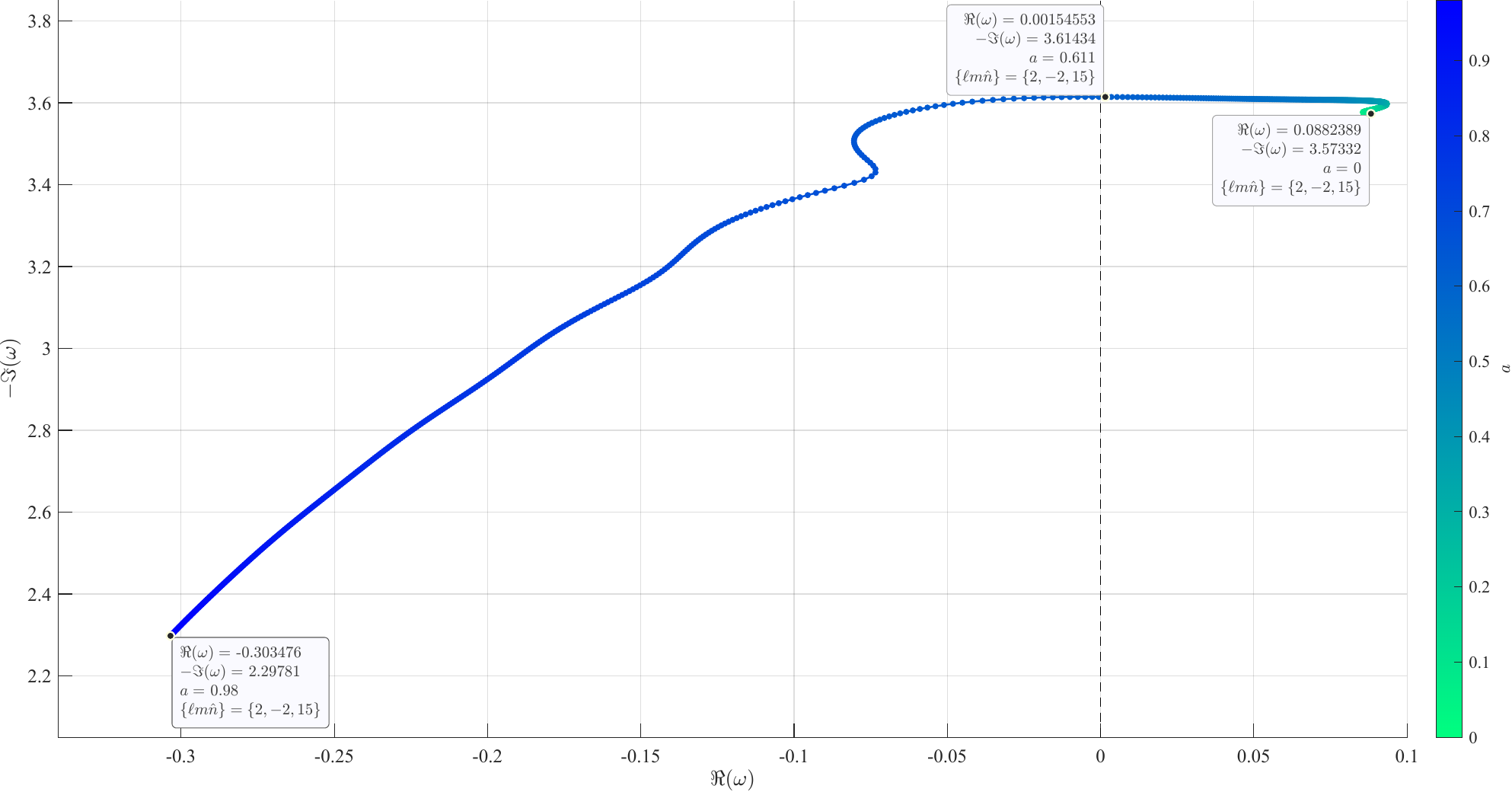}}\,\,
\subfloat[ Incomplete results from Cook \label{fige:l2mf2n14-17}]{\includegraphics[width=3.4in]{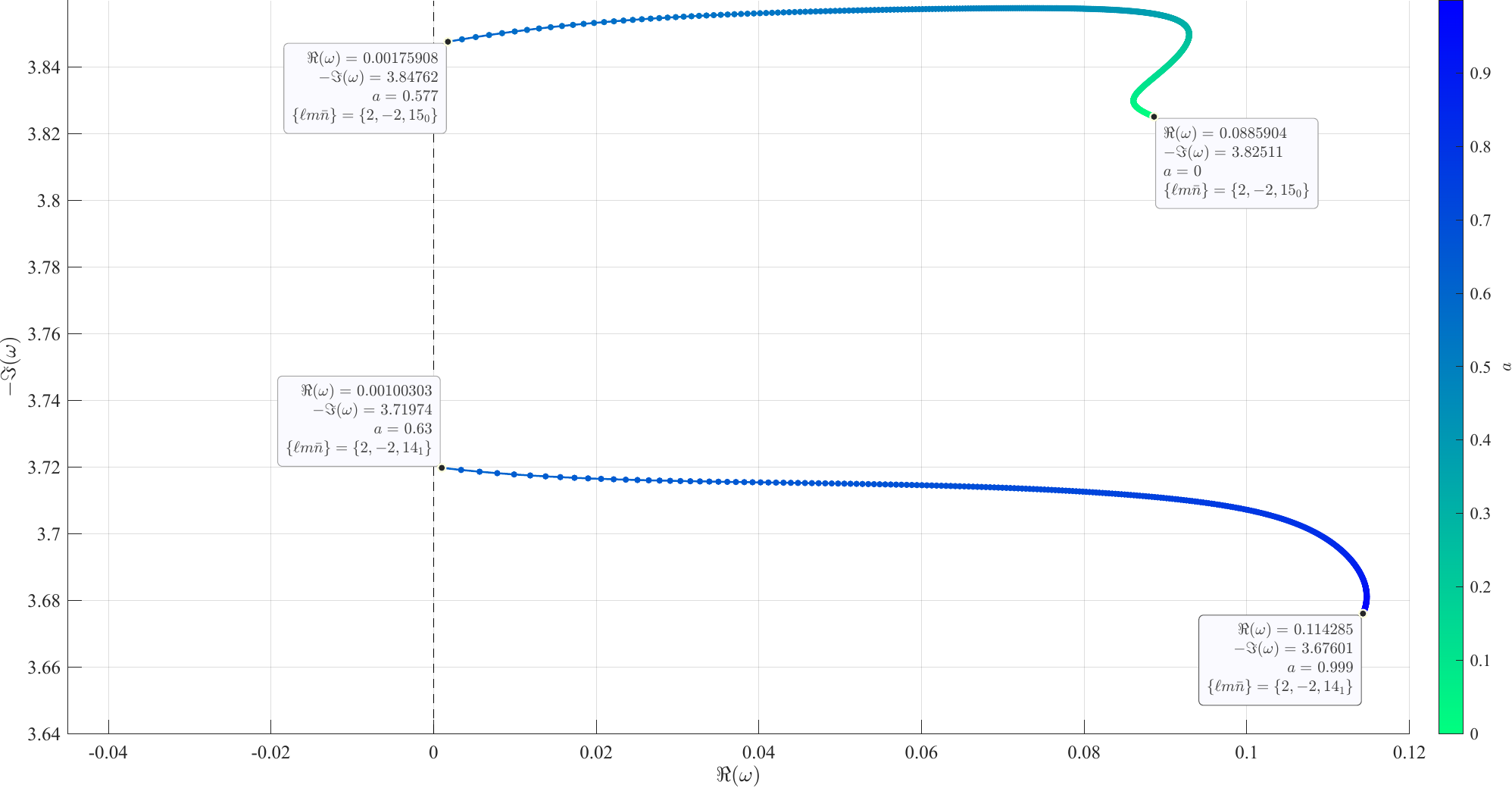}}\,\,
\subfloat[ HeunC-Chen method ]{\includegraphics[width=3.4in]{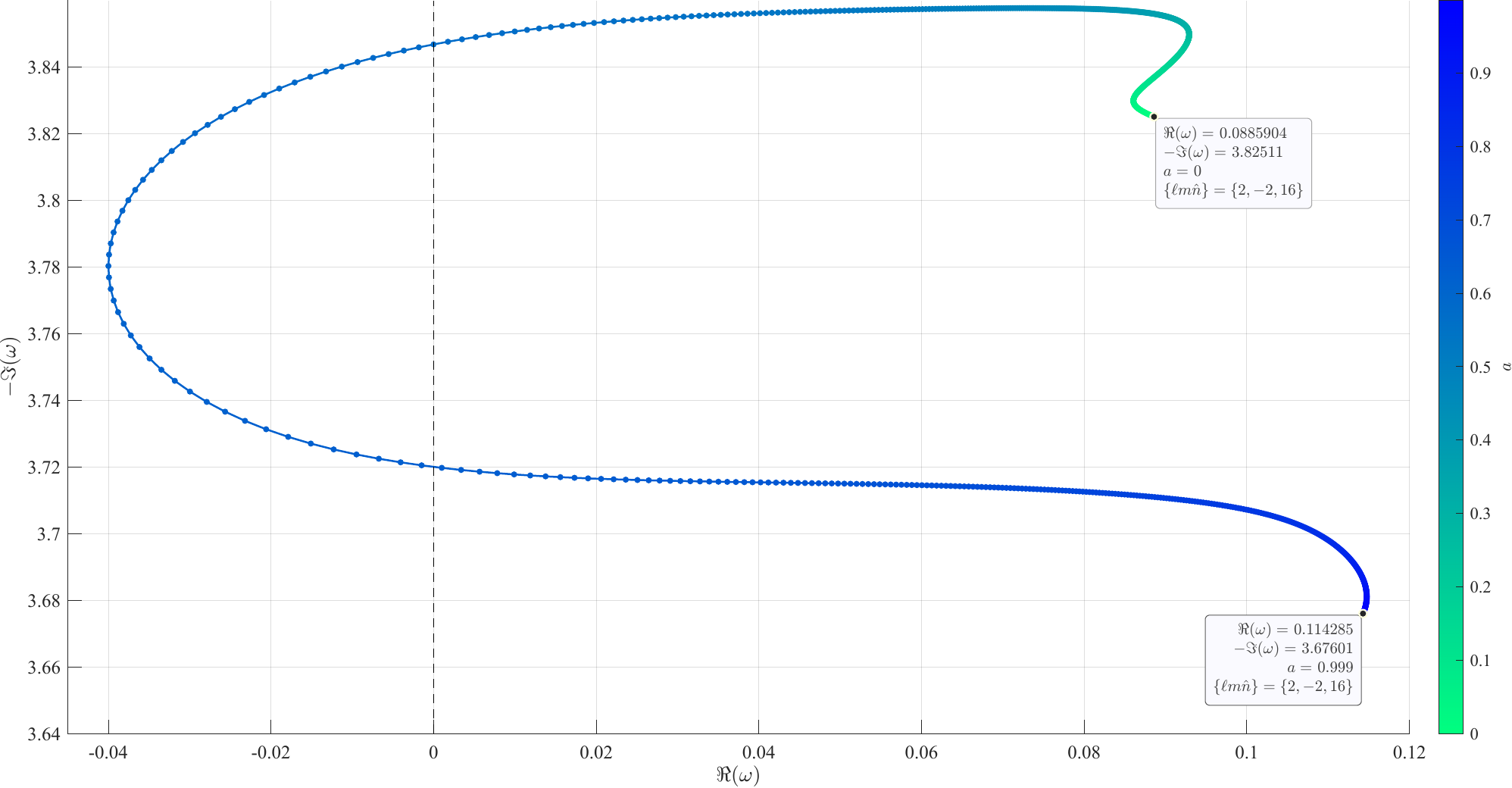}}\,\,
\subfloat[ Incomplete results from Cook ]{\includegraphics[width=3.4in]{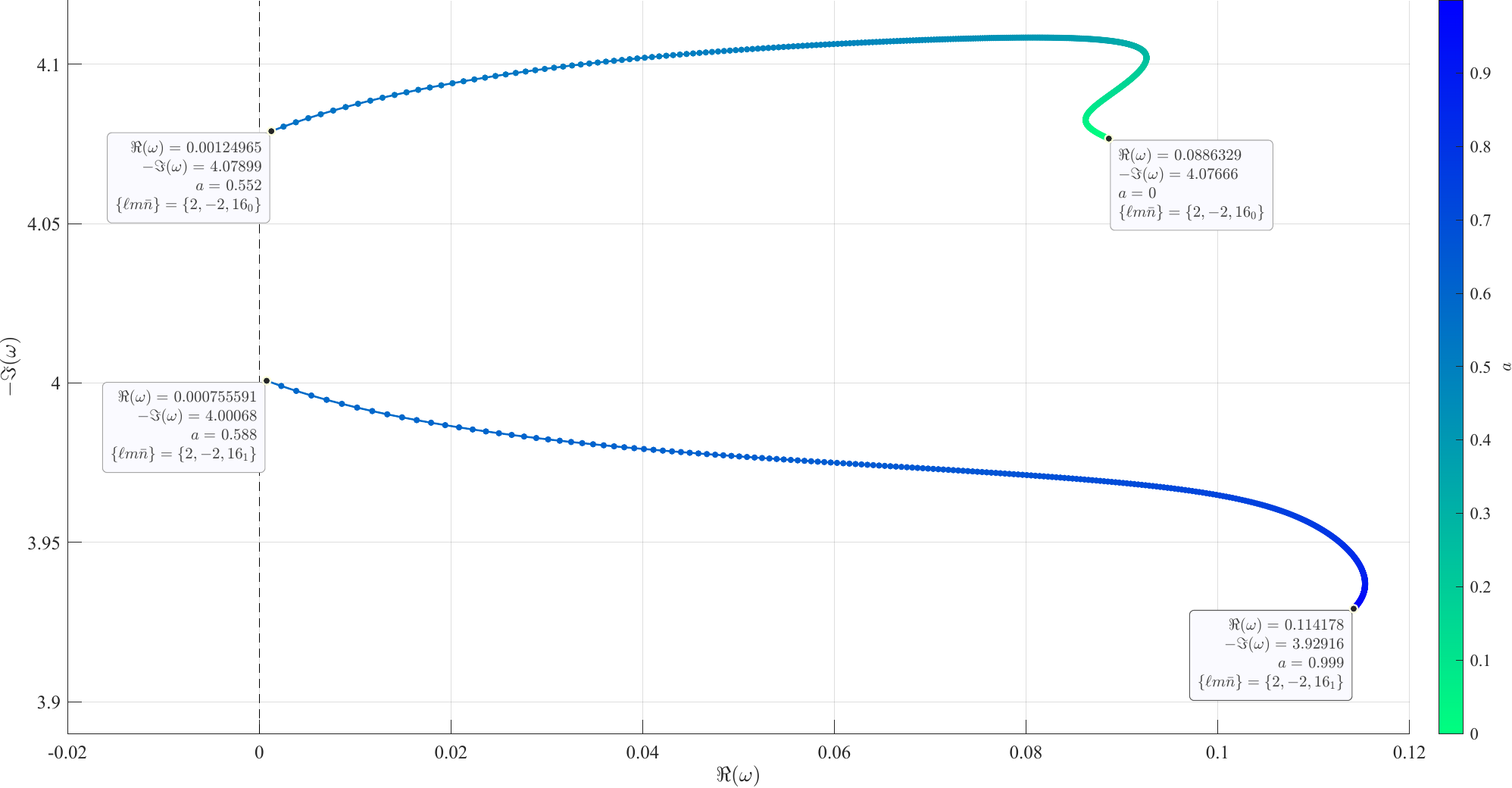}}\,\,
\subfloat[ HeunC-Chen method ]{\includegraphics[width=3.4in]{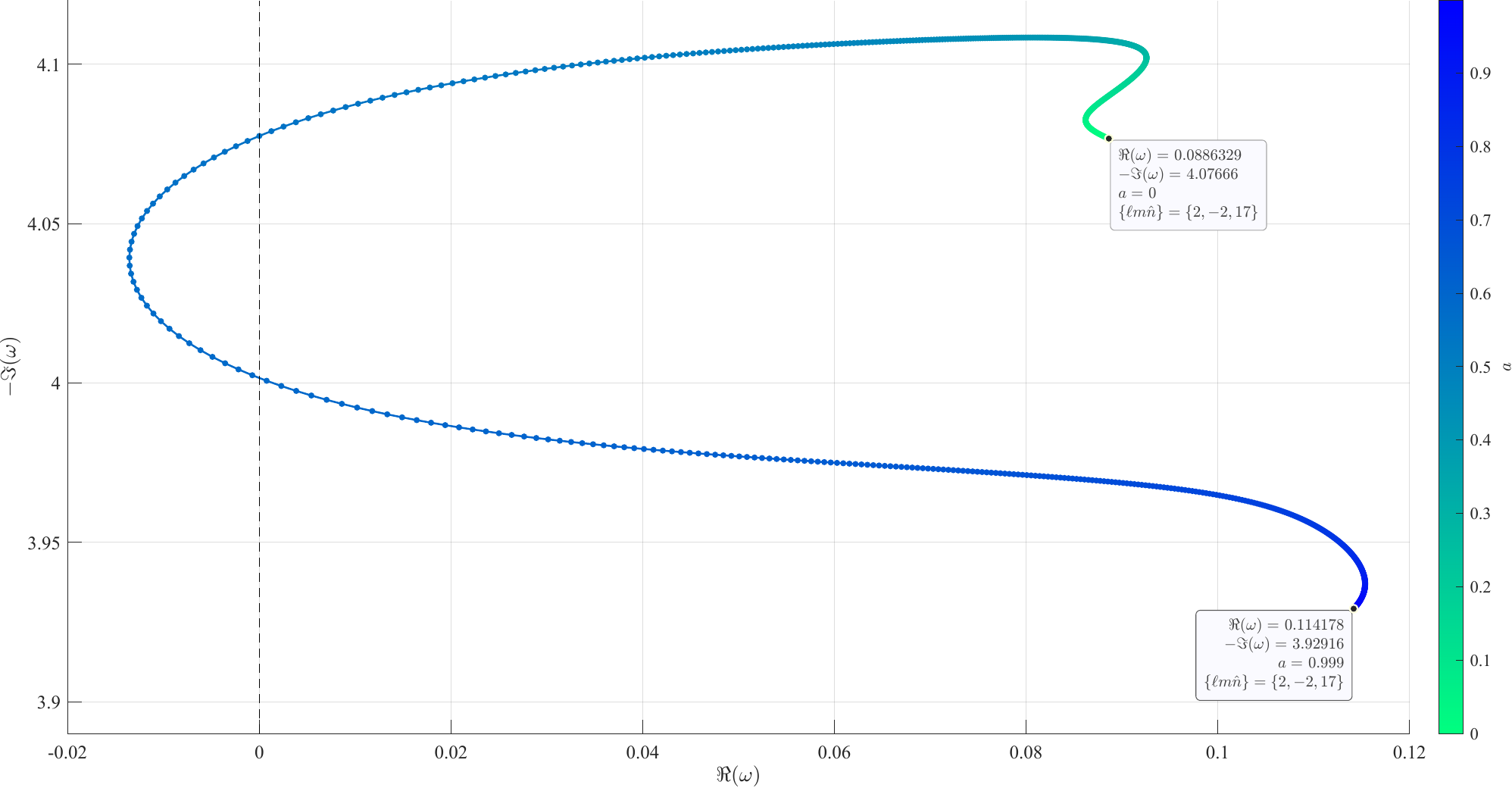}}\,\,
\caption{Detail view near the NIA of Kerr QNM sequences for $\ell= 2,m=-2, 14\leq \hat n \leq 17$. }\label{fig:l2mf2n14-17}
\end{figure*}
\begin{figure*}[htbp]
	\centering
\subfloat[ Incomplete results from Cook \label{figa:l3mf2n26-29}]{\includegraphics[width=3.4in]{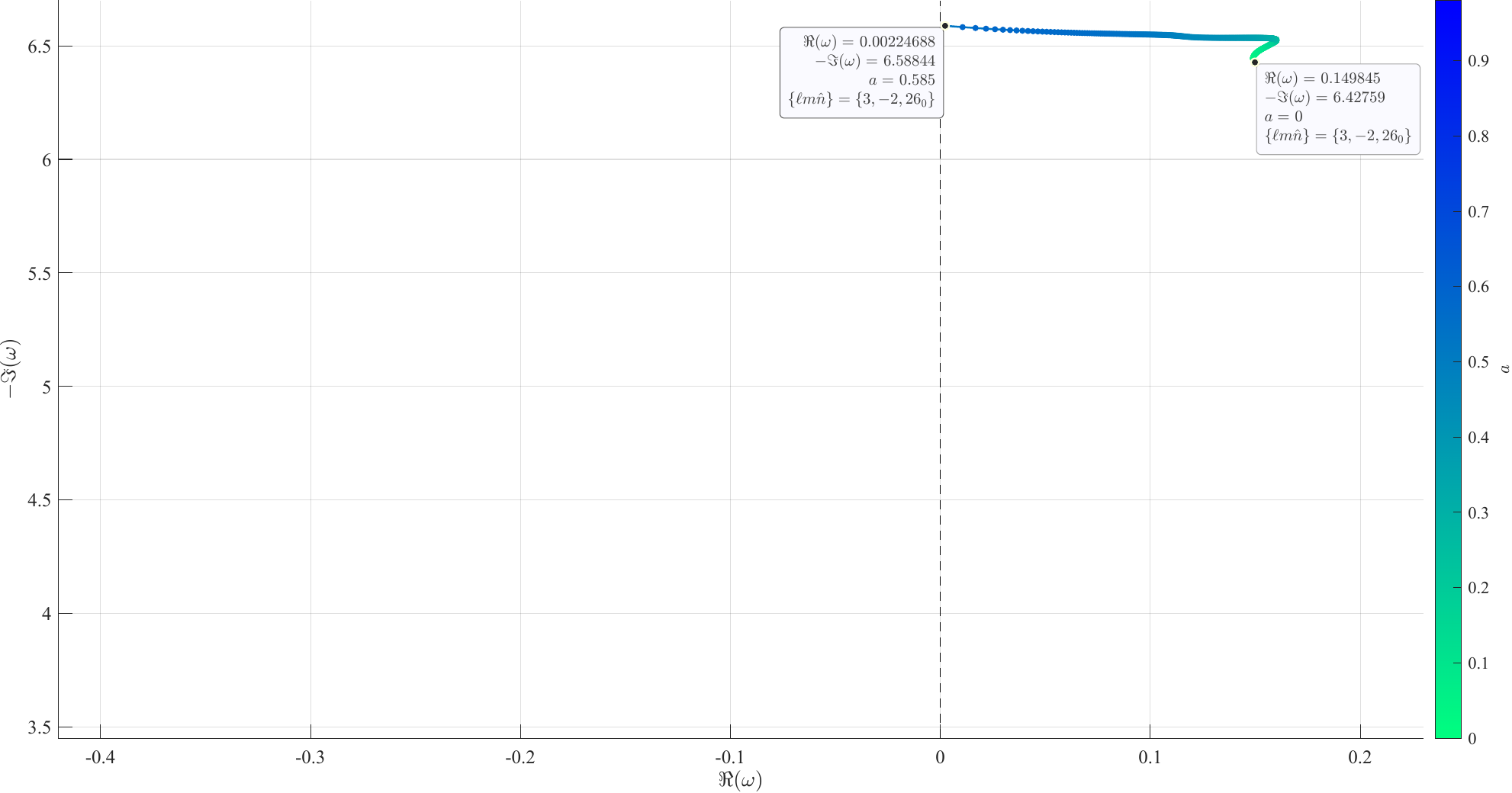}}\,\,
\subfloat[ HeunC-Chen method ]{\includegraphics[width=3.4in]{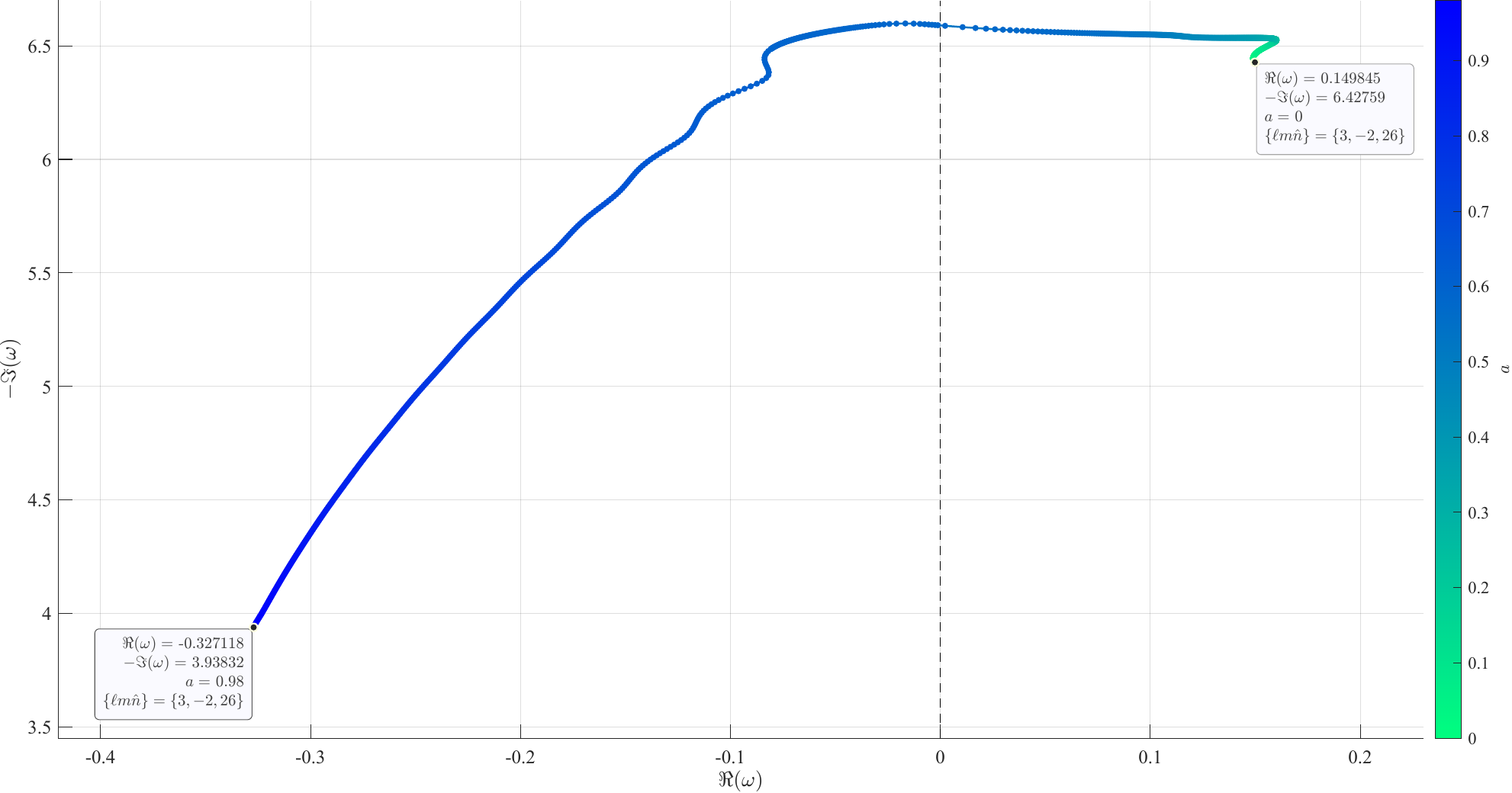}}\,\,
\subfloat[ Incomplete results from Cook \label{figc:l3mf2n26-29}]{\includegraphics[width=3.4in]{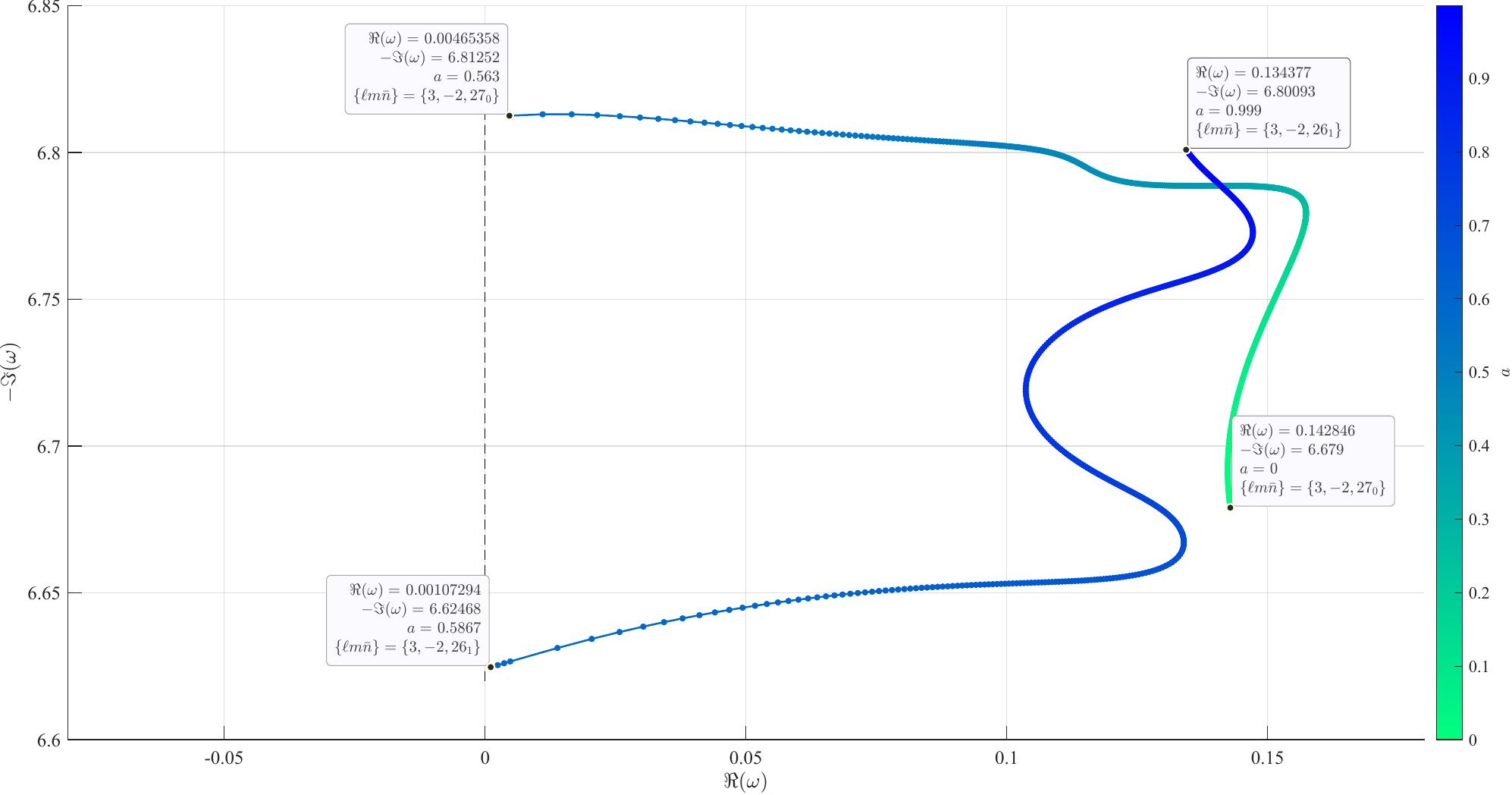}}\,\,
\subfloat[ HeunC-Chen method ]{\includegraphics[width=3.4in]{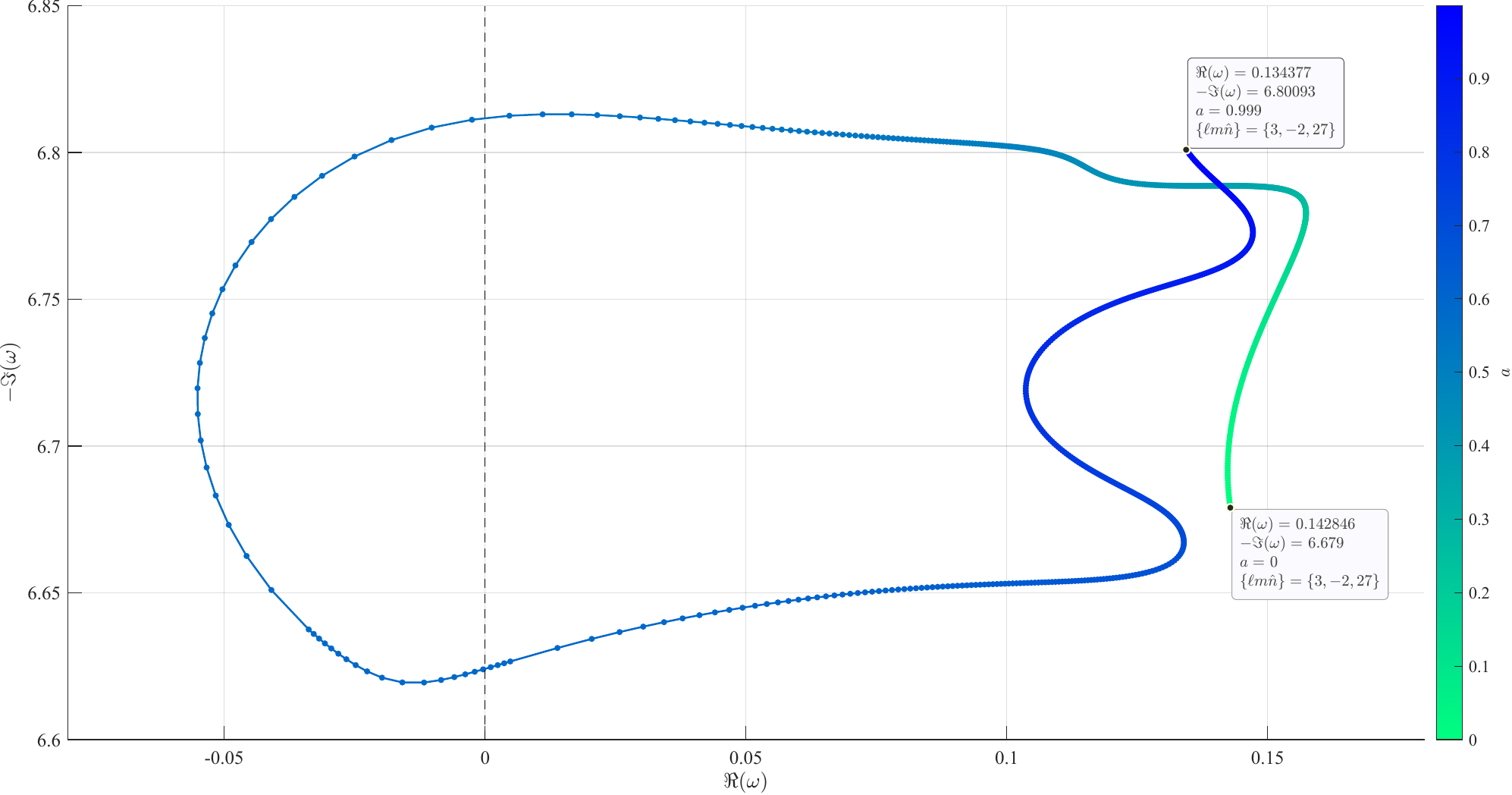}}\,\,
\subfloat[ Incomplete results from Cook \label{fige:l3mf2n26-29}]{\includegraphics[width=3.4in]{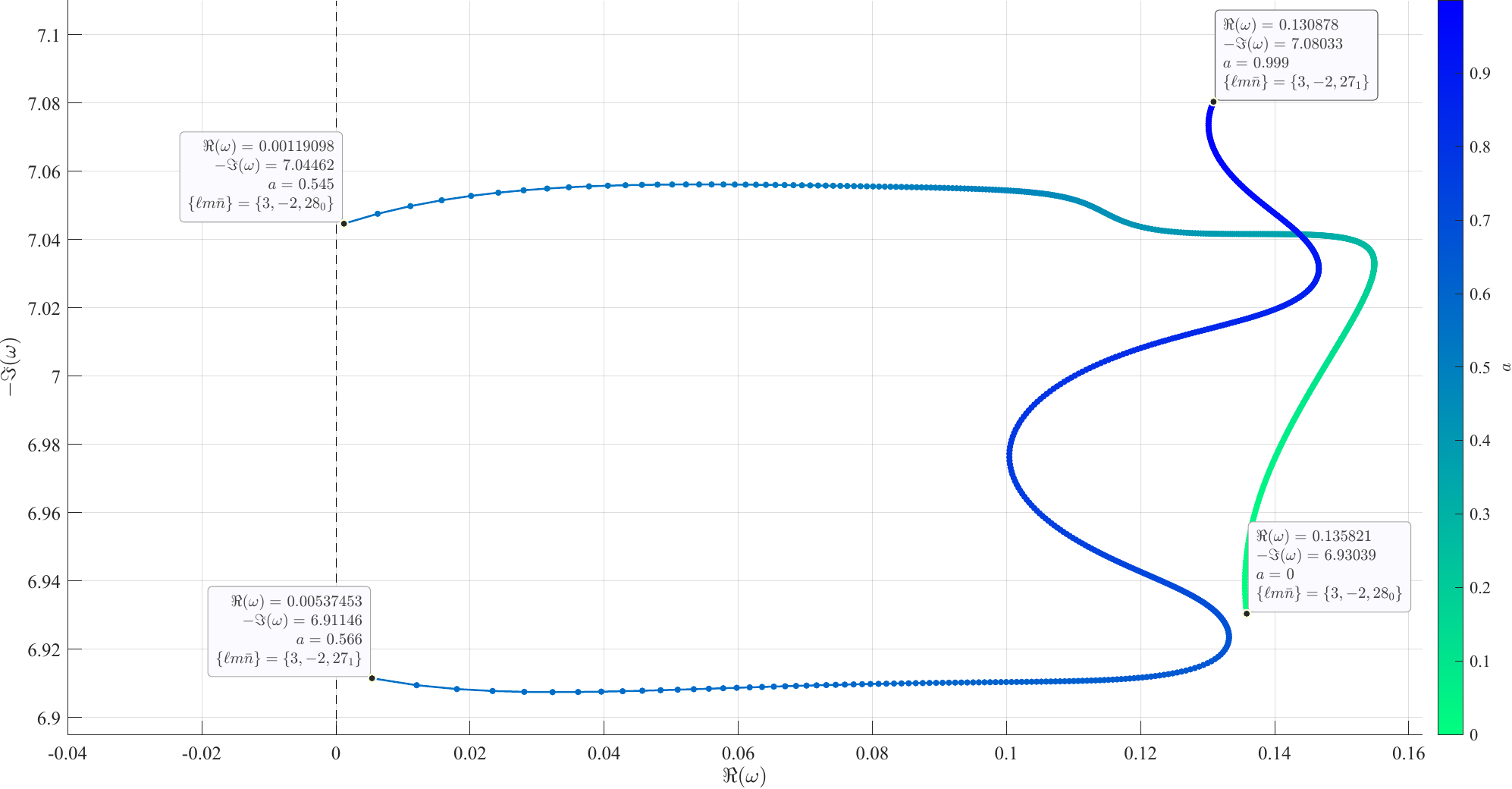}}\,\,
\subfloat[ HeunC-Chen method ]{\includegraphics[width=3.4in]{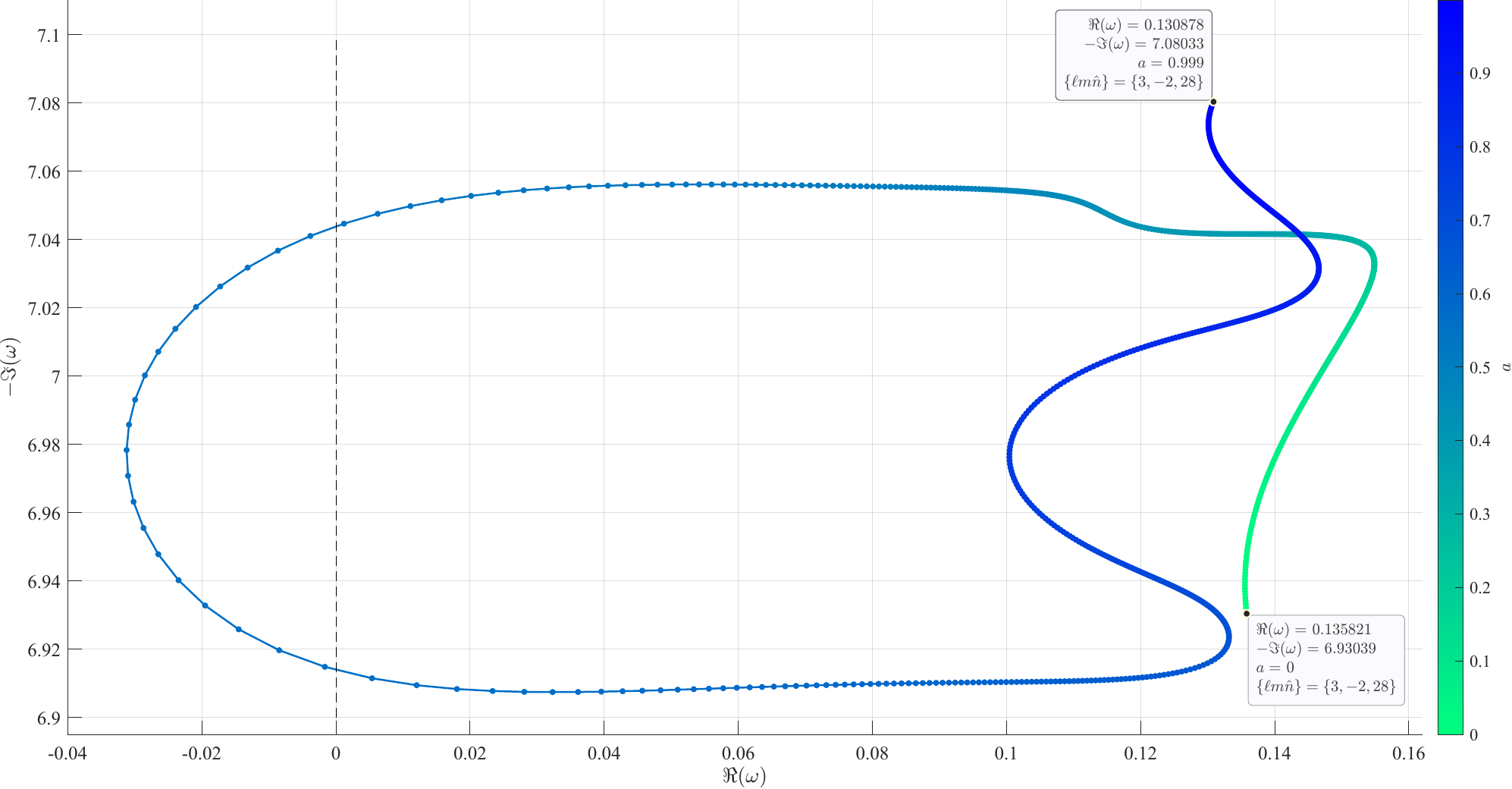}}\,\,
\subfloat[ Incomplete results from Cook ]{\includegraphics[width=3.4in]{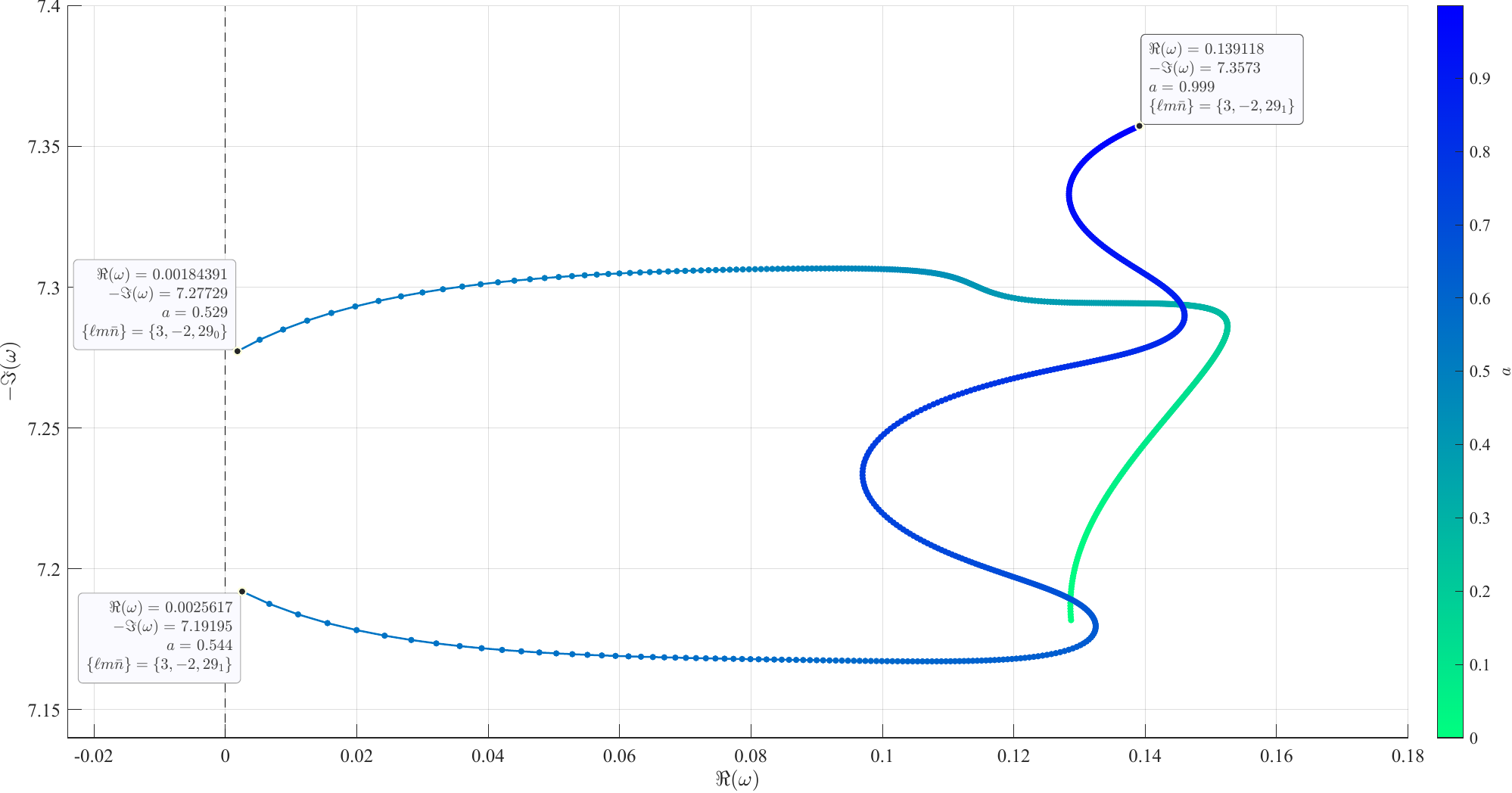}}\,\,
\subfloat[ HeunC-Chen method ]{\includegraphics[width=3.4in]{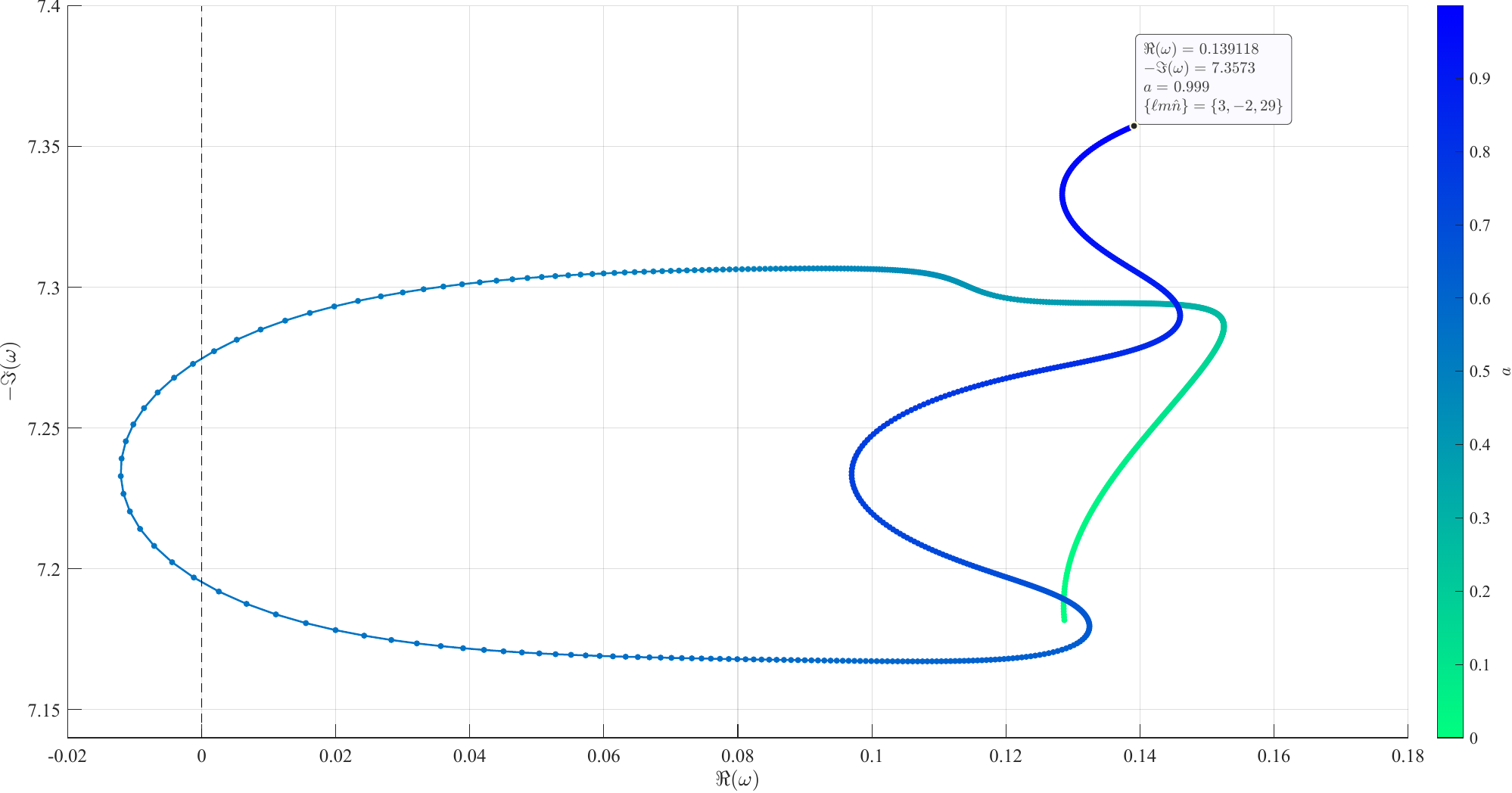}}\,\,
\caption{Detail view near the NIA of Kerr QNM sequences for $\ell= 3,m=-2, 26\leq \hat n \leq 29$. }\label{fig:l3mf2n26-29}
\end{figure*}
\begin{figure*}[htbp]
	\centering
\subfloat[ Incomplete results from Cook ]{\includegraphics[width=3.4in]{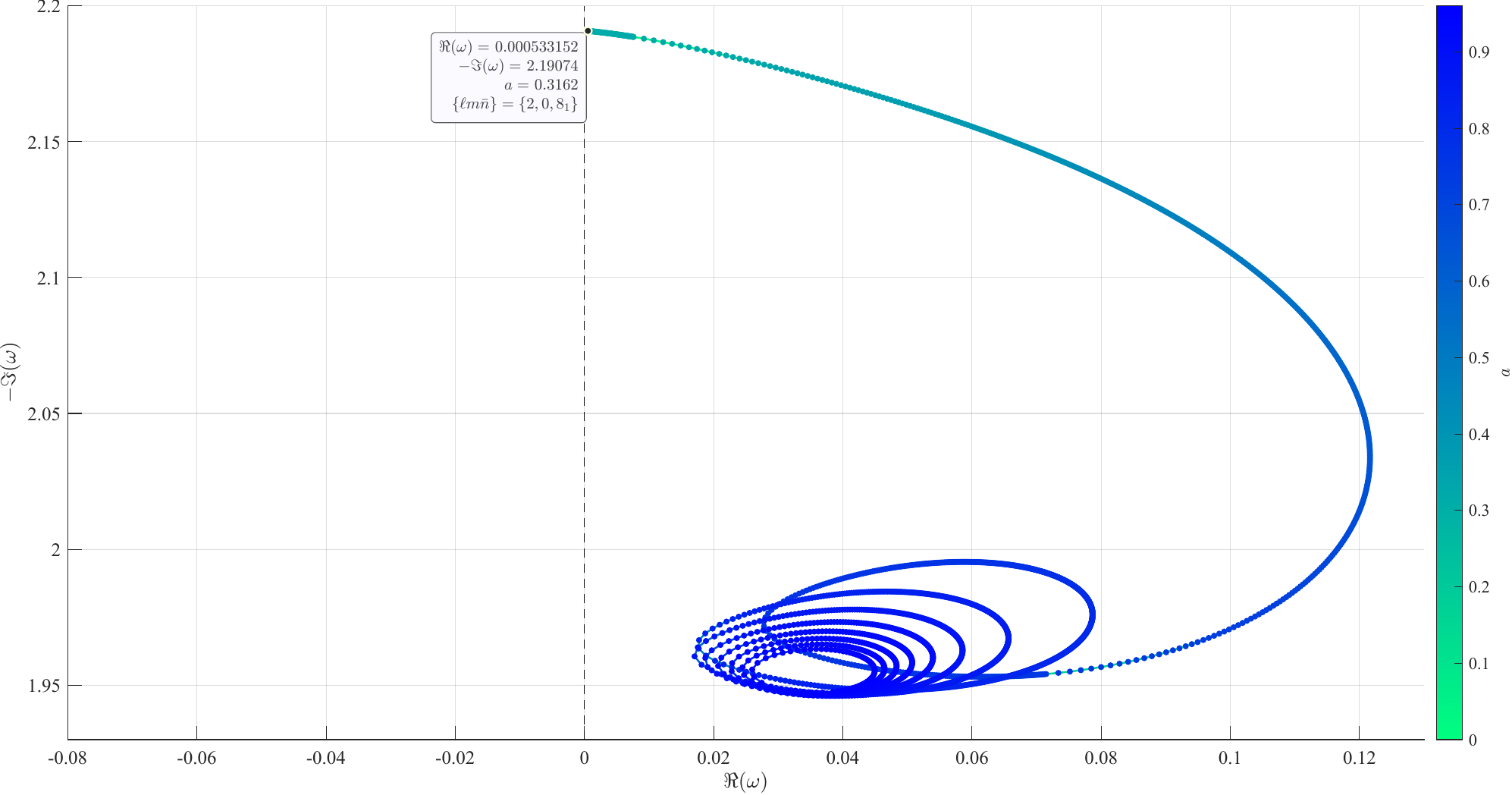}}\,\,
\subfloat[ HeunC-Chen method ]{\includegraphics[width=3.4in]{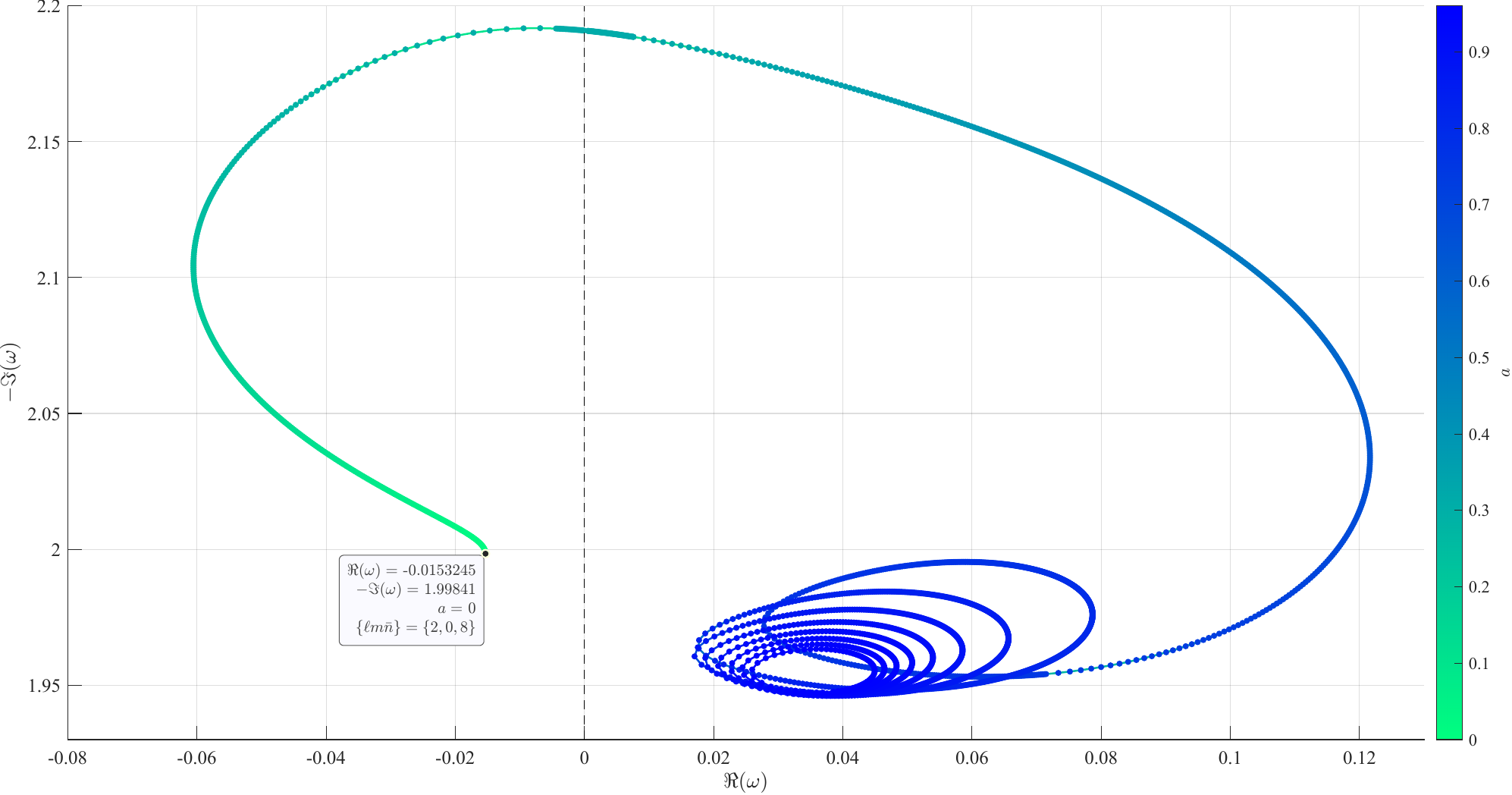}}\,\,
\subfloat[ Incomplete results from Cook (missing $\omega_{\text{I}}=-2i$ at $a=0$) ]{\includegraphics[width=3.4in]{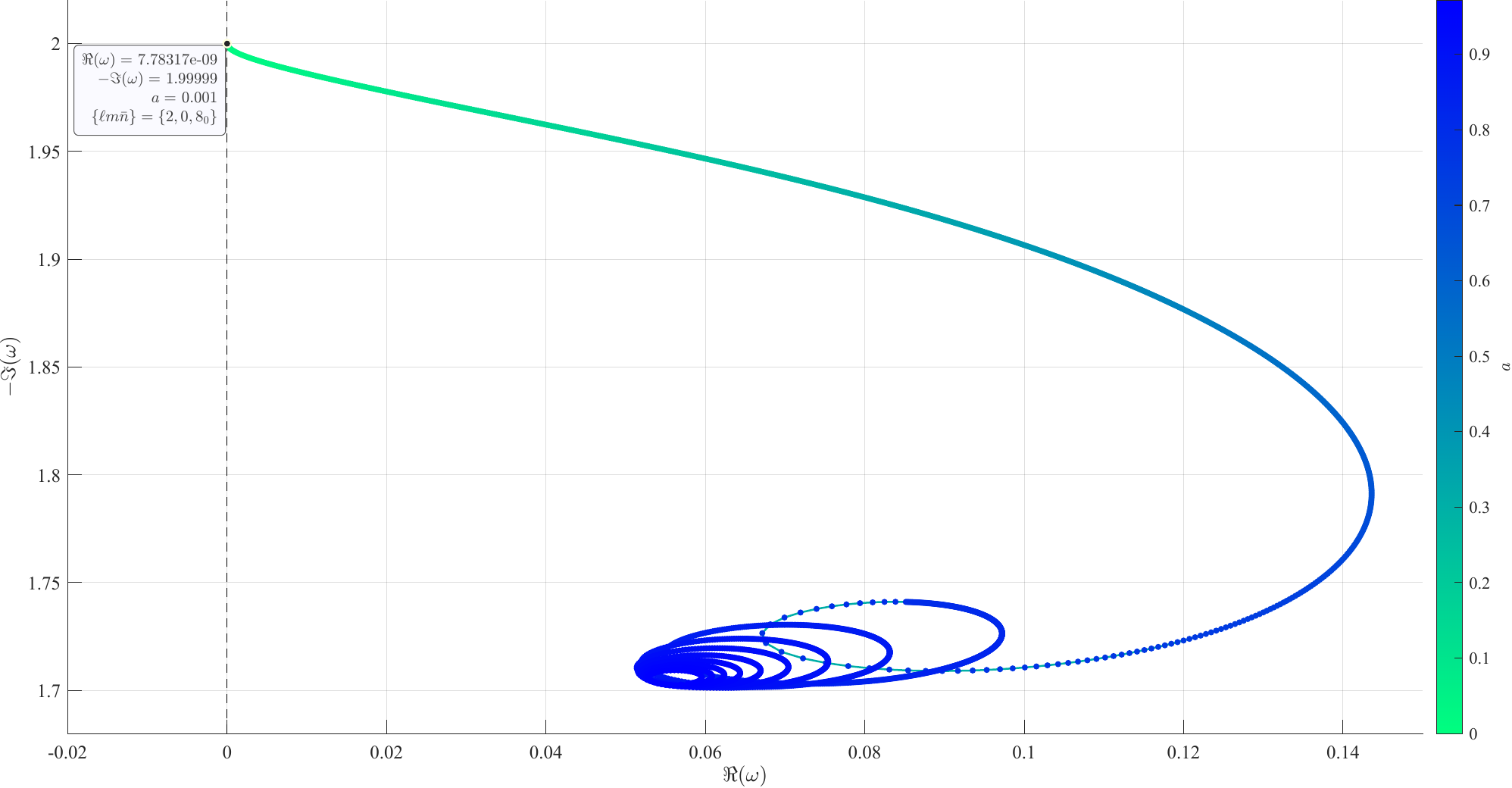}}\,\,
\subfloat[ HeunC-Chen method ]{\includegraphics[width=3.4in]{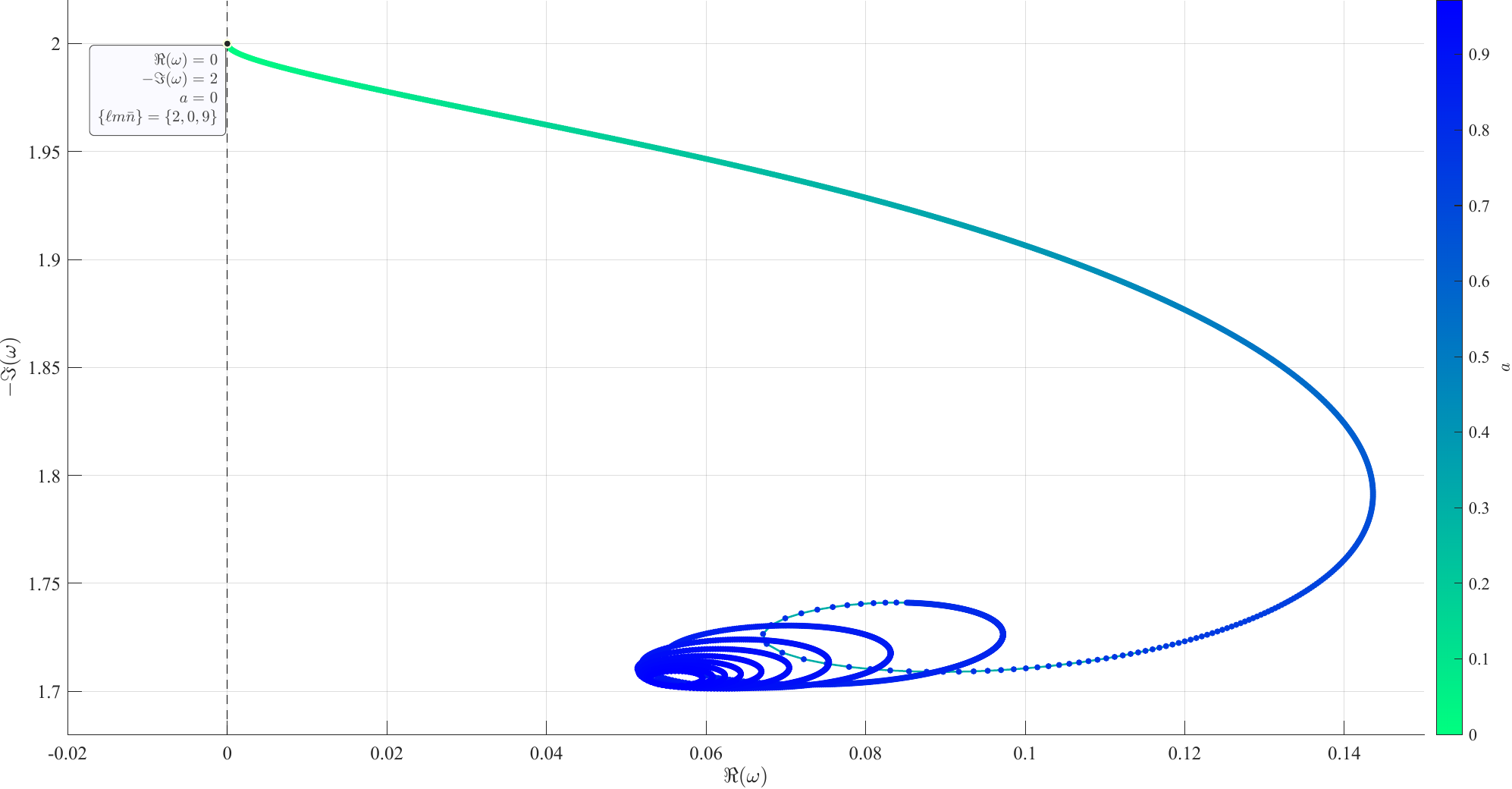}}\,\,
\subfloat[ Incomplete results from Cook ]{\includegraphics[width=3.4in]{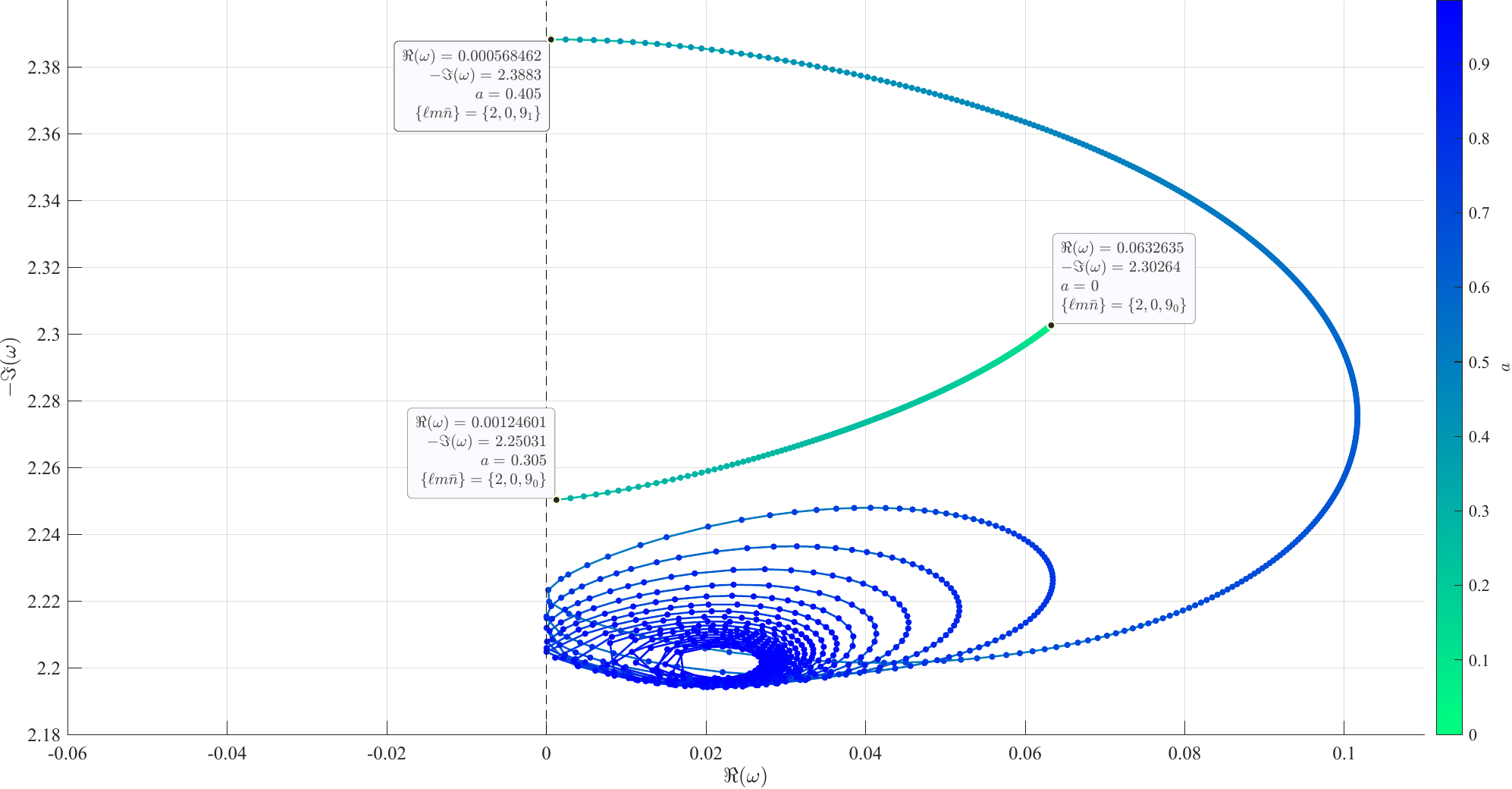}}\,\,
\subfloat[ HeunC-Chen method ]{\includegraphics[width=3.4in]{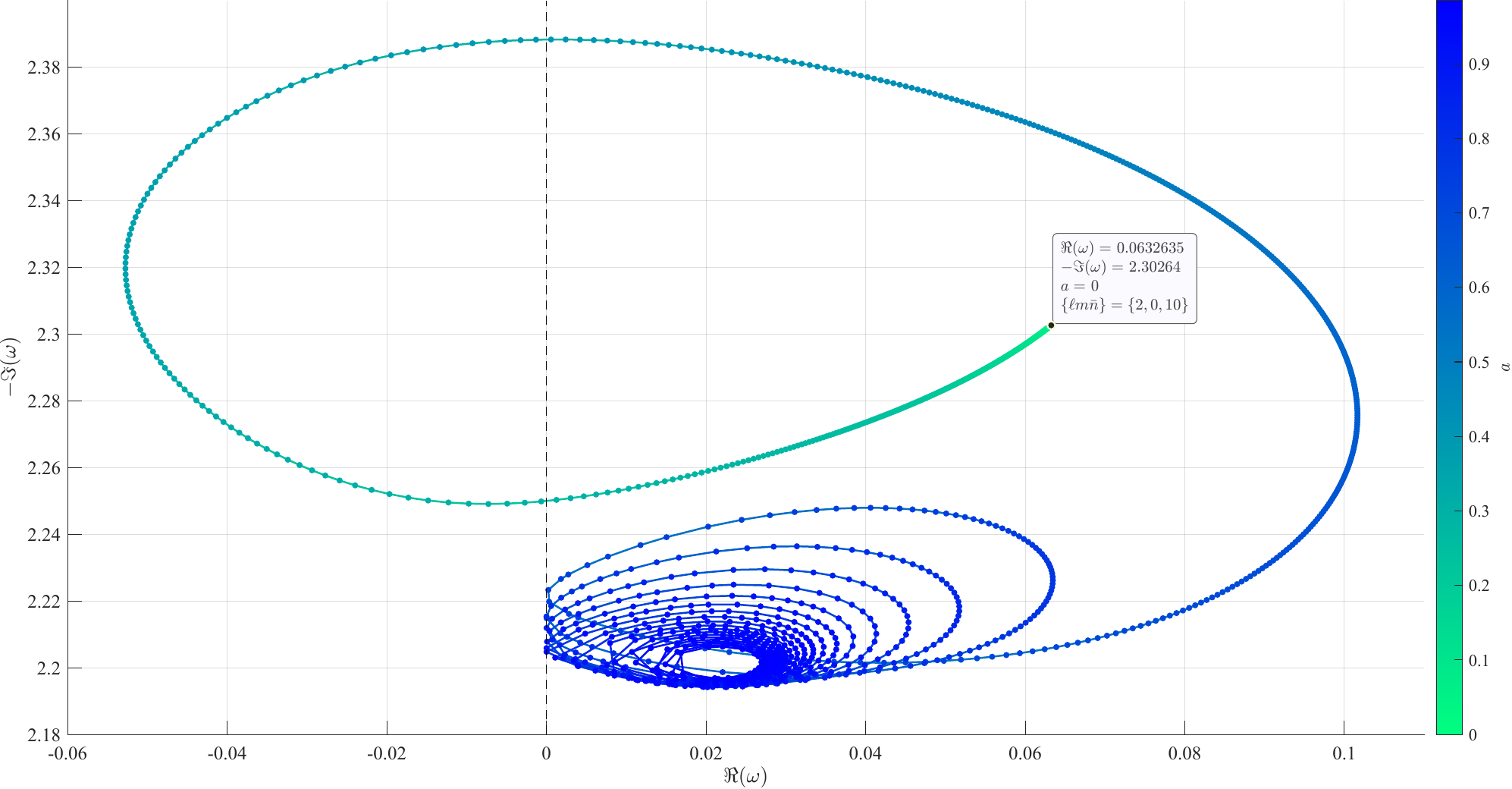}}\,\,
\subfloat[ Incomplete results from Cook ]{\includegraphics[width=3.4in]{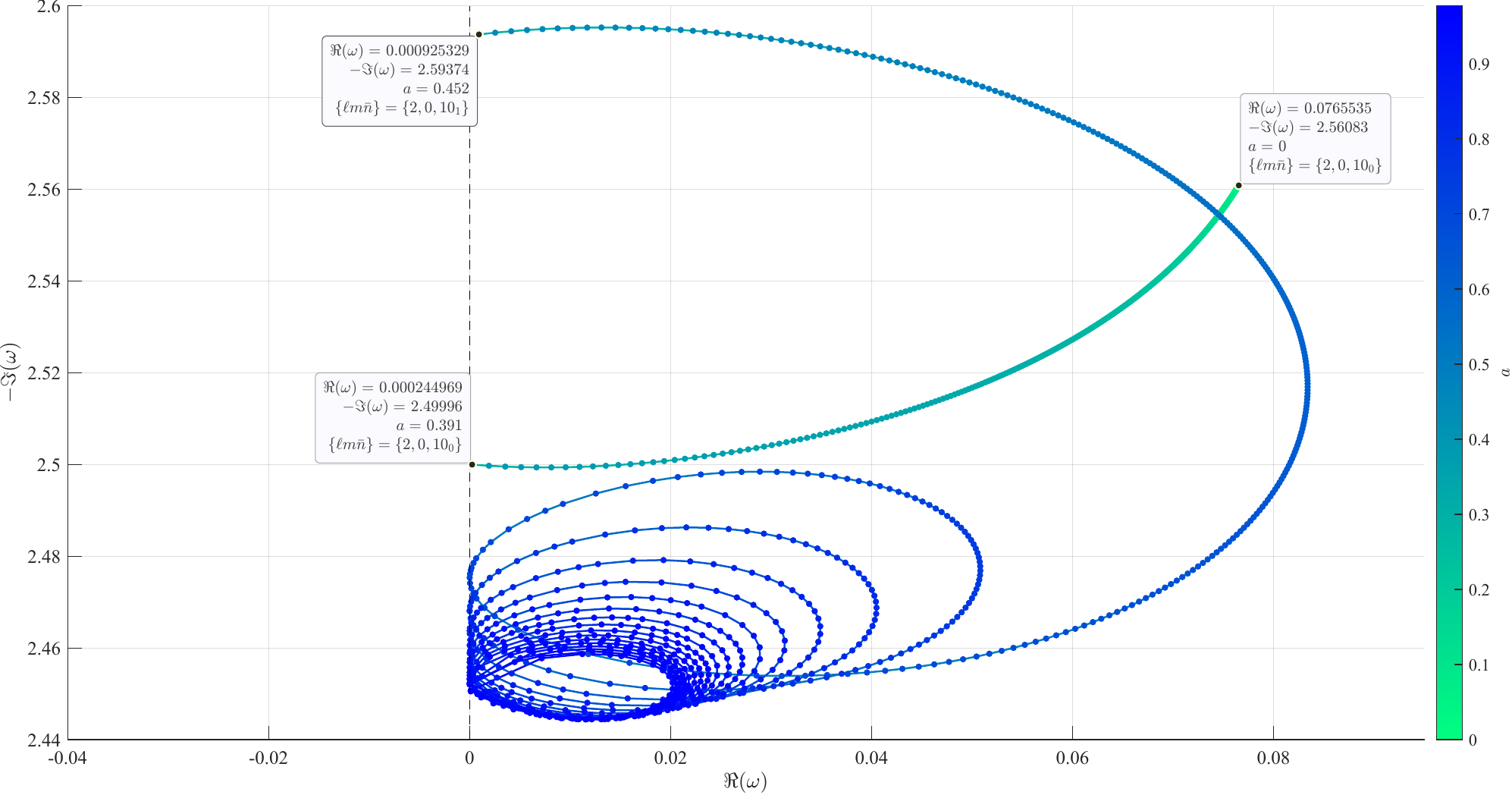}}\,\,
\subfloat[ HeunC-Chen method ]{\includegraphics[width=3.4in]{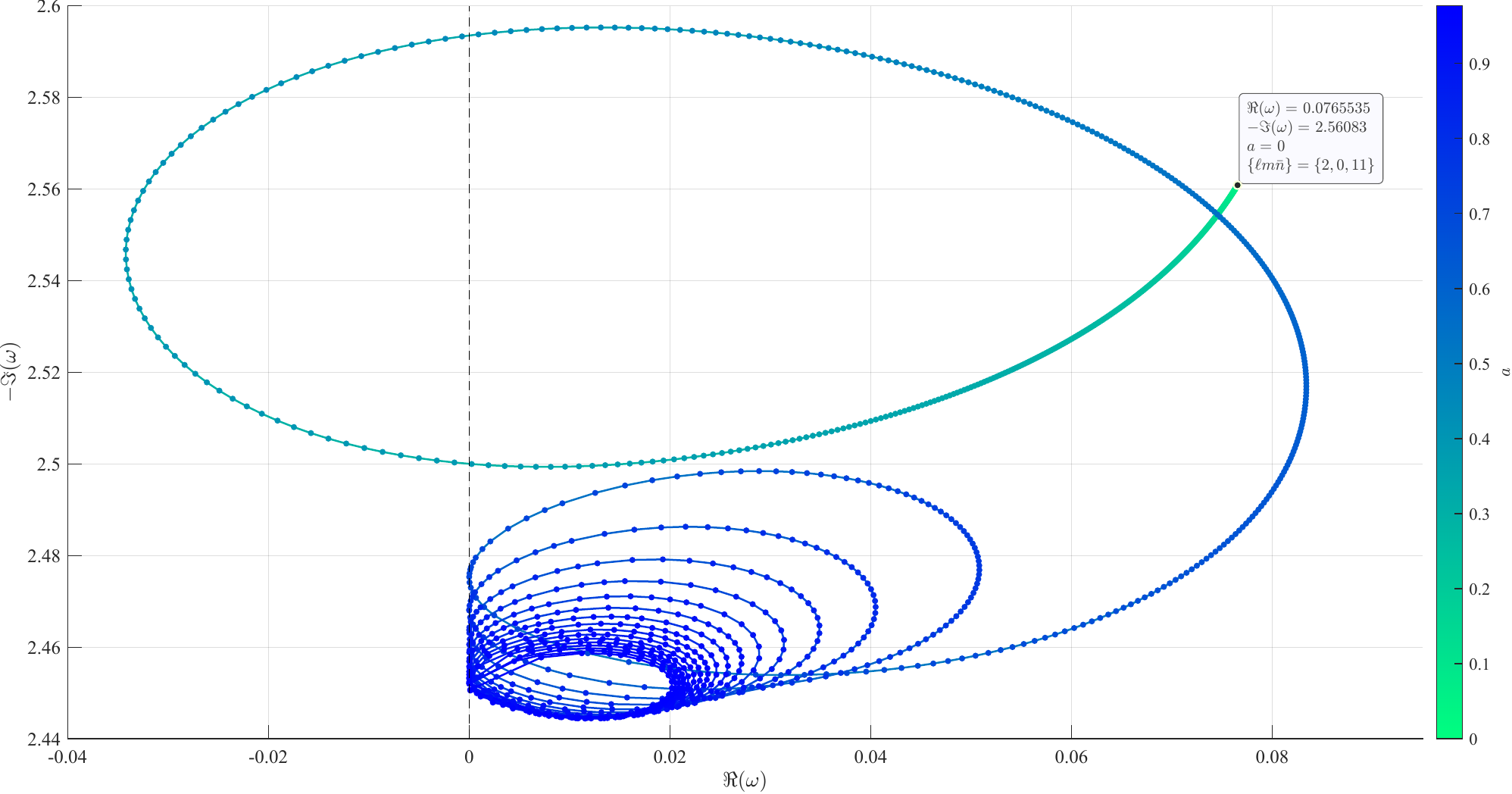}}\,\,
\caption{Detail view near the NIA of Kerr QNM sequences for $\ell= 2,m=0, 8\leq \hat n \leq 11$ (variable step sizes in $a$ resolve the spectrum's spiral features). }\label{fig:l2m0n8-11}
\end{figure*}
\subsubsection{Highly Damped Modes}

For highly damped modes with $\hat{n} < 40$, it is primarily the QNM sequences with $m \leq 0$ that cross the NIA. The closer the Schwarzschild QNMs are to the NIA, the more likely the corresponding Kerr QNM sequences are to cross the NIA.
As shown in \Cref{fig:l2mf2n14-17,fig:l3mf2n26-29}, some of the complete QNM sequences with $m = -2$ cross the NIA.
If the QNMs crossing the NIA are not solved, this leads to an incorrect classification of Cook's overtones \cite{Cook:2016fge,Cook:2016ngj}, where several incomplete QNM sequences are artificially assigned to a particular overtone (with artificial subscripts 0 and 1 added for distinction).
For example, in \cref{fig:l2mf2n14-17}, the sequences  $\{2, -2, 14_0\}$, $\{2, -2, 14_1\}$, and $\{2, -2, 15_0\}$ of Cook's results are three misclassified incomplete QNM sequences.
Similarly, in \cref{fig:l3mf2n26-29}, the sequences $\{3, -2, 26_0\}$, $\{3, -2, 26_1\}$, $\{2, -2, 27_0\}$, $\{3, -2, 27_1\}$, and $\{2, -2, 28_0\}$ of Cook's results are also misclassified sequences.
The spectral structure becomes significantly more complicated for $\ell=2$ and $m=0$ QNM sequences.
As the spin parameter increases, these sequences first cross the NIA (a feature missing in Cook's incomplete structural evolution) and subsequently develop spiral trajectories.
Notably, higher overtone sequences correlate with spiral structures that progressively approach the NIA, eventually becoming tangent to it.
Cook \cite{Cook:2016ngj,Berti:2025hly} argued that the tangential points in the $m=0$ QNM sequence are neither QNMs nor TTMs. However, our calculations of the errors \eqref{eq:err1_QNM} at these tangential points confirm that they are indeed QNMs, not TTMs. Therefore, Cook's assertion on this issue is incorrect.
From the evolution of the QNM sequences with $\ell=2$ and $m=0$, we can infer that the $m=0$ sequences of highly damped modes corresponding to higher $\ell$ values should exhibit spectra similar to those of $\ell=2$.
When $\hat{n}$ exceeds the overtone of $\omega_{\ell}^{\rm{AS}}$, the QNM sequences first cross the NIA and then spirally approach tangency with it.
Although this paper does not present the highly damped modes for higher $\ell$ values, our method is capable of computing them.
These results will be incrementally incorporated into our dataset \cite{ChenQNM}.

\subsubsection{Extreme Kerr QNMs}
The QNM spectrum of Kerr BHs exhibits accumulation points (degeneracies) in the extremal spin limit, which coincide with the critical frequency for superradiance. Specifically,
\begin{subequations}\label{eq:ExtremeKerr_QNM}
\begin{align}
\lim_{a \to 1} \omega_{\mathrm{I}} &= \lim_{a \to 1} m\Omega = \frac{m}{2}, \quad m > 0,   \\
\lim_{a \to 1} \omega_{\mathrm{II}} &= \lim_{a \to 1} m\Omega = \frac{m}{2}, \quad m < 0,
\end{align}
\end{subequations}
where $\Omega = \tfrac{a}{2Mr_+}$ denotes the horizon's angular velocity.
This limiting behavior was first predicted by Detweiler \cite{Detweiler:1980gk}.
Berti and Cardoso \cite{Berti:2003jh,Cardoso:2004hh} subsequently extended and refined his analysis, ultimately deriving \cref{eq:ExtremeKerr_QNM}.
As shown in \cref{eq:ExtremeKerr_QNM}, the imaginary part of the QNM frequencies vanishes in the limit $a \to 1$, leaving only the real component. This implies that QNMs at extremal spin lose their damping behavior, retaining only oscillations. Yang et al. \cite{Yang:2012pj,Yang:2013uba} termed these modes {zero damped modes} (ZDMs), in contrast to {damped modes} (DMs) where the real part remains finite. Using matched asymptotic expansions, Yang et al. derived an approximate formula for frequencies near the accumulation points. Subsequently, Cook provided more accurate fitting expressions based on high-precision QNM spectra in near-extremal Kerr \cite{Cook:2014cta}. Using the MST method, Casals et al. \cite{Casals:2019vdb} computed both the QNMs and superradiant amplification factors for near-extremal and extremal spins, while demonstrating the formation of additional superradiant branch cuts.

\begin{figure*}[htbp]
	\centering
\subfloat[$(\ell,m)=(2,2)$ ]{\includegraphics[width=3.4in]{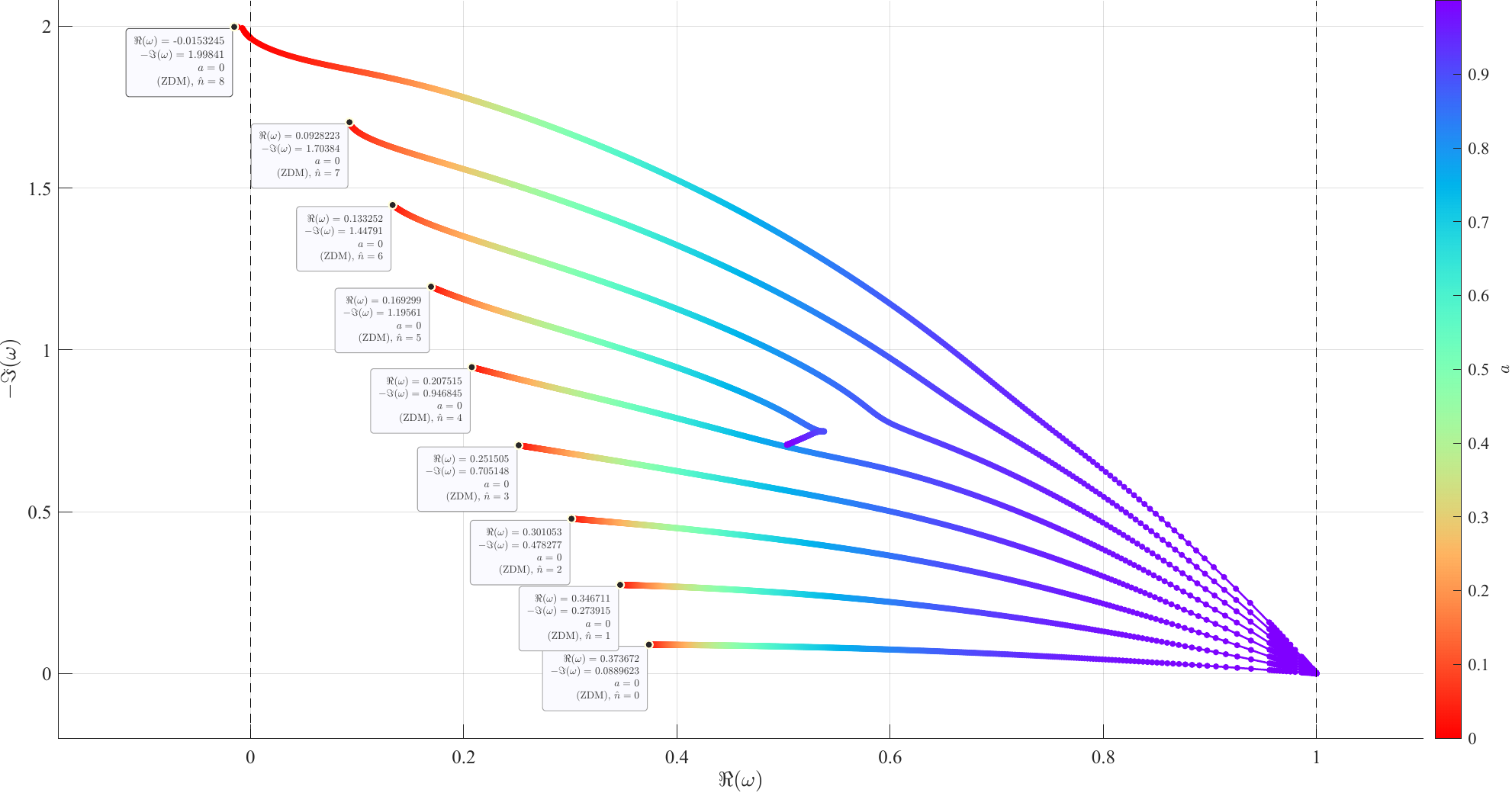}}\,\,
\subfloat[Close-up of $a \in \mathopen{[}1-10^{-3},1-10^{-8}\mathclose{]}$ for $(\ell,m)=(2,2)$]{\includegraphics[width=3.4in]{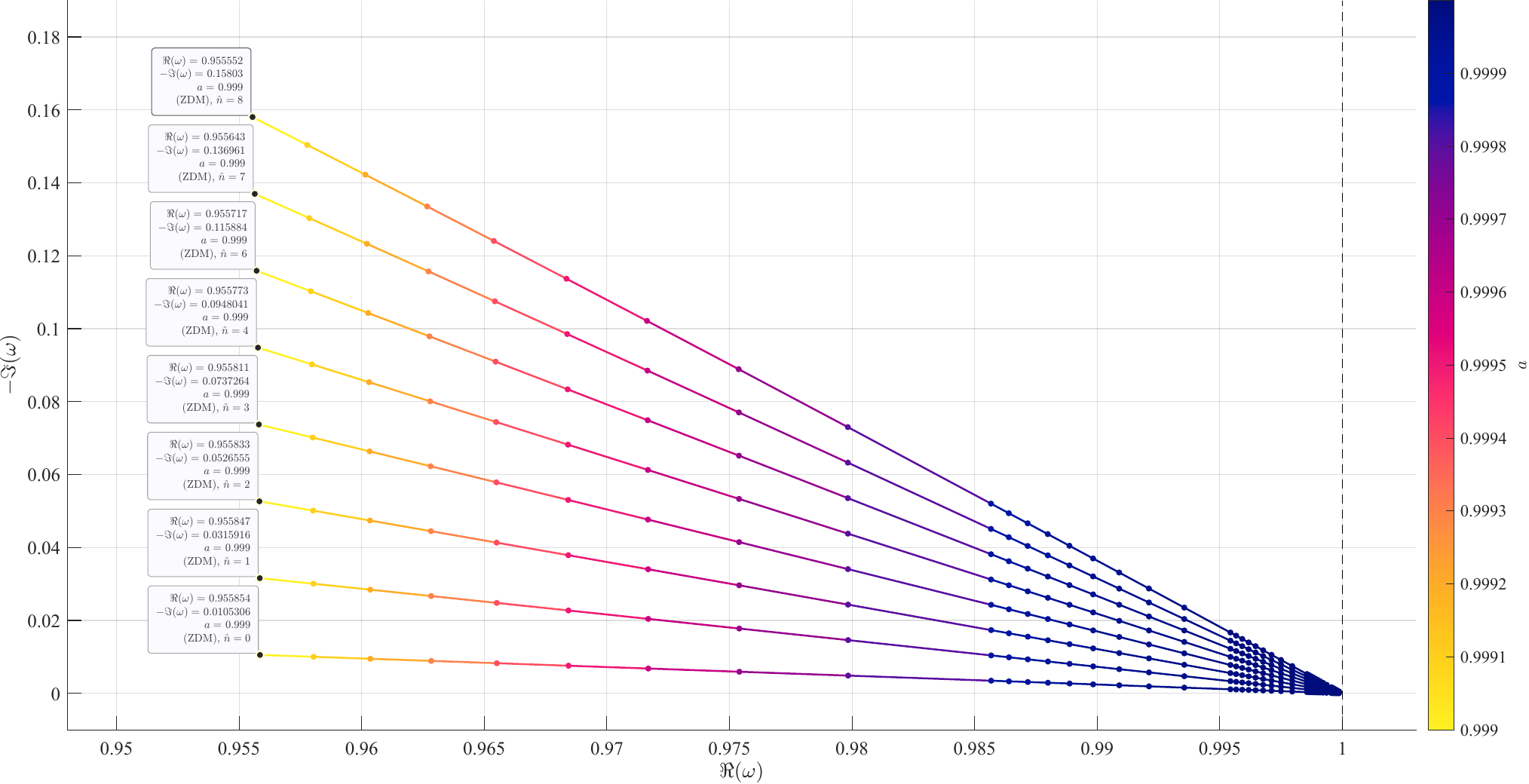}}\,\,
\caption{Kerr QNM sequences for $\ell=m=2$ and $\hat{n} = [0,8]$ with $a\in[0,1-10^{-8}]$. In the data tip, the ZDM denotes a zero-damping mode while the DM represents a damped mode.}\label{fig:QNM_l2m2}
\end{figure*}

\cref{fig:QNM_l2m2} displays the QNM spectrum for all weakly damped modes with $\ell=m=2$, evolving from the Schwarzschild limit to near-extremal spin ($a=1-10^{-8}$). The results show that all modes are ZDMs except for the $(2,2,5)$ mode, which remains a DM.
Although our numerical simulations adopt $a=1-10^{-8}$ as the near-extremal spin, the extremal limit ($a\to1$) can be approached closely by increasing floating-point precision.
For spins beyond $1-10^{-8}$, the resulting spectra are visually indistinguishable from those in \cref{fig:QNM_l2m2}. Thus, we present only the $a=1-10^{-8}$ case here, with QNM spectra for more extreme spins to be incrementally added to our dataset \cite{ChenQNM}.

Within the confluent Heun equation framework, no numerical method (including our method and CFM) can directly compute extremal-spin ($a \to 1$) physics.
As evident from the radial HeunC parameters \eqref{eq:RTE_HCparameters_Kerr}, the coalescence of horizons ($r_- \to r_+$) leads to divergent denominators in key parameters. However, the radial eigenvalue equation in \cref{sec:TTM_QNM} successfully calculates extremal-spin TTMs as demonstrated in \cref{fig:TTM_L2L3} precisely because it employs the Starobinsky-Teukolsky constant.

\begin{table*}[htbp]
  \centering
  \caption{Computational times (in seconds) of the scalar QNM with $a\in[0.99,0.99999]$ and $\square a=10^{-5}$.}
    \begin{tabular}{lccccccccc}
    \toprule
         &      & \multicolumn{4}{c}{HeunC-Chen} &      & \multicolumn{3}{c}{Isomonodromic Method \cite{Cavalcante:2024swt,Cavalcante:2024kmy}} \\
\cmidrule{3-6}\cmidrule{8-10}    ($\ell,m,\hat n$)\quad
&$\mu$& $L_2$-Error   & $L_\infty$-Error &  1-Core Time & 12-Cores Time &   & $L_2$-Error   & $L_\infty$-Error & 1-Core Time \\
    \midrule
    \multirow{4}[2]{*}{(1,1,0)} & 0    & $1.700 \times 10^{-10}$ & $1.612 \times 10^{-10}$ & 12.946  & 3.419  &      & $5.833 \times 10^{-10}$ & $3.844 \times 10^{-10}$ & 123966.569  \\
         & 0.1  & $1.187 \times 10^{-10}$ & $8.691 \times 10^{-11}$ & 14.311  & 3.043  &      & $3.838 \times 10^{-9}$ & $2.098 \times 10^{-9}$ & 119612.359  \\
         & 0.2  & $2.754 \times 10^{-10}$ & $2.046 \times 10^{-10}$ & 15.679  & 3.302  &      & $6.238 \times 10^{-9}$ & $5.459 \times 10^{-9}$ & 119612.359  \\
         & 0.3  & $2.038 \times 10^{-10}$ & $1.750 \times 10^{-10}$ & 14.193  & 3.488  &      & $2.648 \times 10^{-6}$ & $1.407 \times 10^{-6}$ & 120960.031  \\
    \midrule
    \multirow{4}[2]{*}{(1,1,1)} & 0    & $1.618 \times 10^{-10}$ & $1.606 \times 10^{-10}$ & 13.934  & 3.177  &      & $5.270 \times 10^{-9}$ & $2.717 \times 10^{-9}$ & 120710.853  \\
         & 0.1  & $7.172 \times 10^{-11}$ & $6.830 \times 10^{-11}$ & 14.769  & 3.233  &      & $4.617 \times 10^{-9}$ & $2.647 \times 10^{-9}$ & 114725.083  \\
         & 0.2  & $1.407 \times 10^{-10}$ & $1.354 \times 10^{-10}$ & 15.667  & 3.051  &      & $1.243 \times 10^{-8}$ & $6.570 \times 10^{-9}$ & 119215.635  \\
         & 0.3  & $5.549 \times 10^{-9}$ & $5.516 \times 10^{-9}$ & 14.637  & 3.371  &      & $4.867 \times 10^{-9}$ & $8.189 \times 10^{-10}$ & 113720.470  \\
    \bottomrule
    \end{tabular}%
\label{tab:KerrS0_L2M2}%
\end{table*}%

\begin{figure*}[htbp]
	\centering
\subfloat[$\hat{n} = 0$ ]{\includegraphics[width=3.4in]{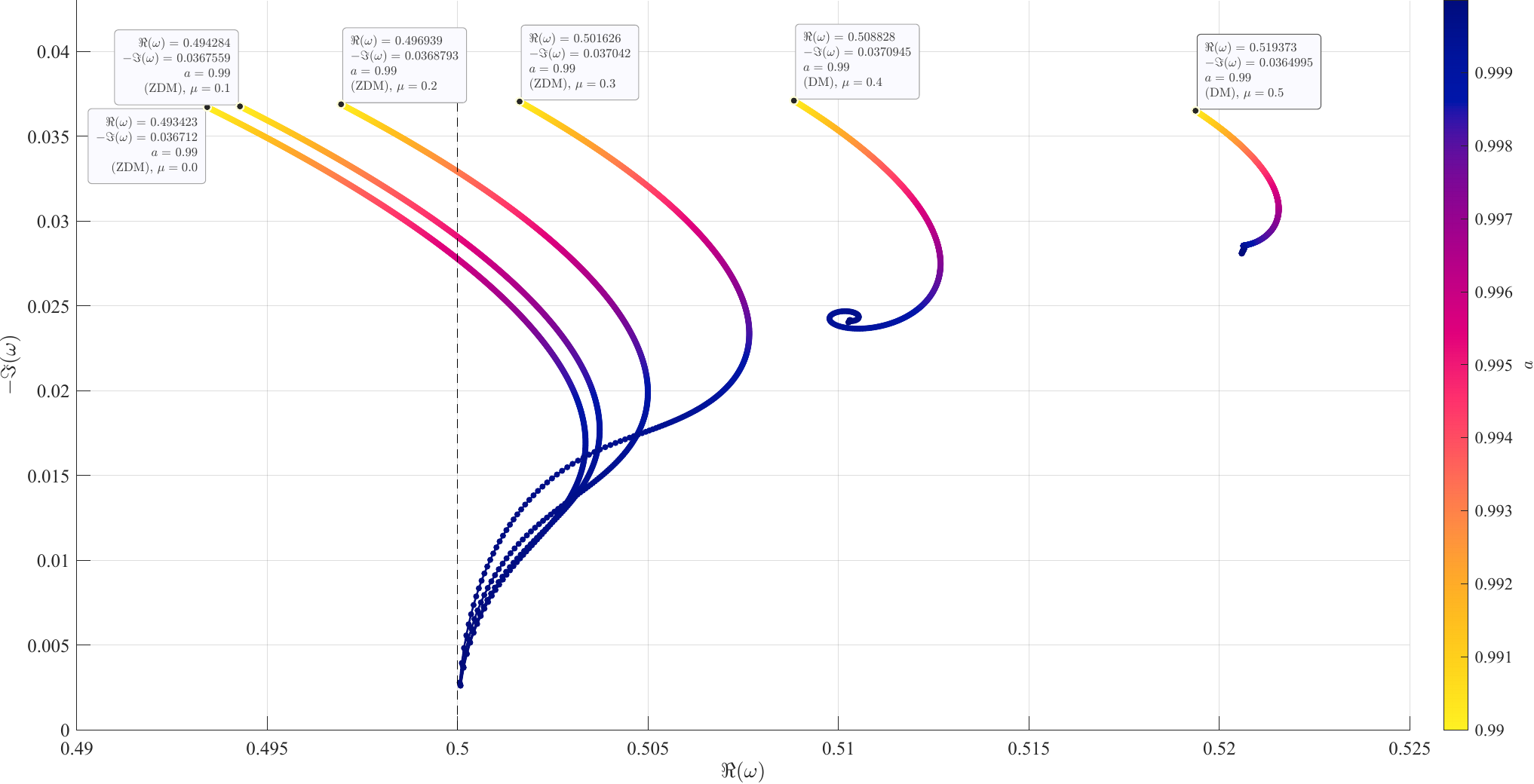}}\,\,
\subfloat[$\hat{n} = 1$\label{figb:QNMs0_l1m1}]{\includegraphics[width=3.4in]{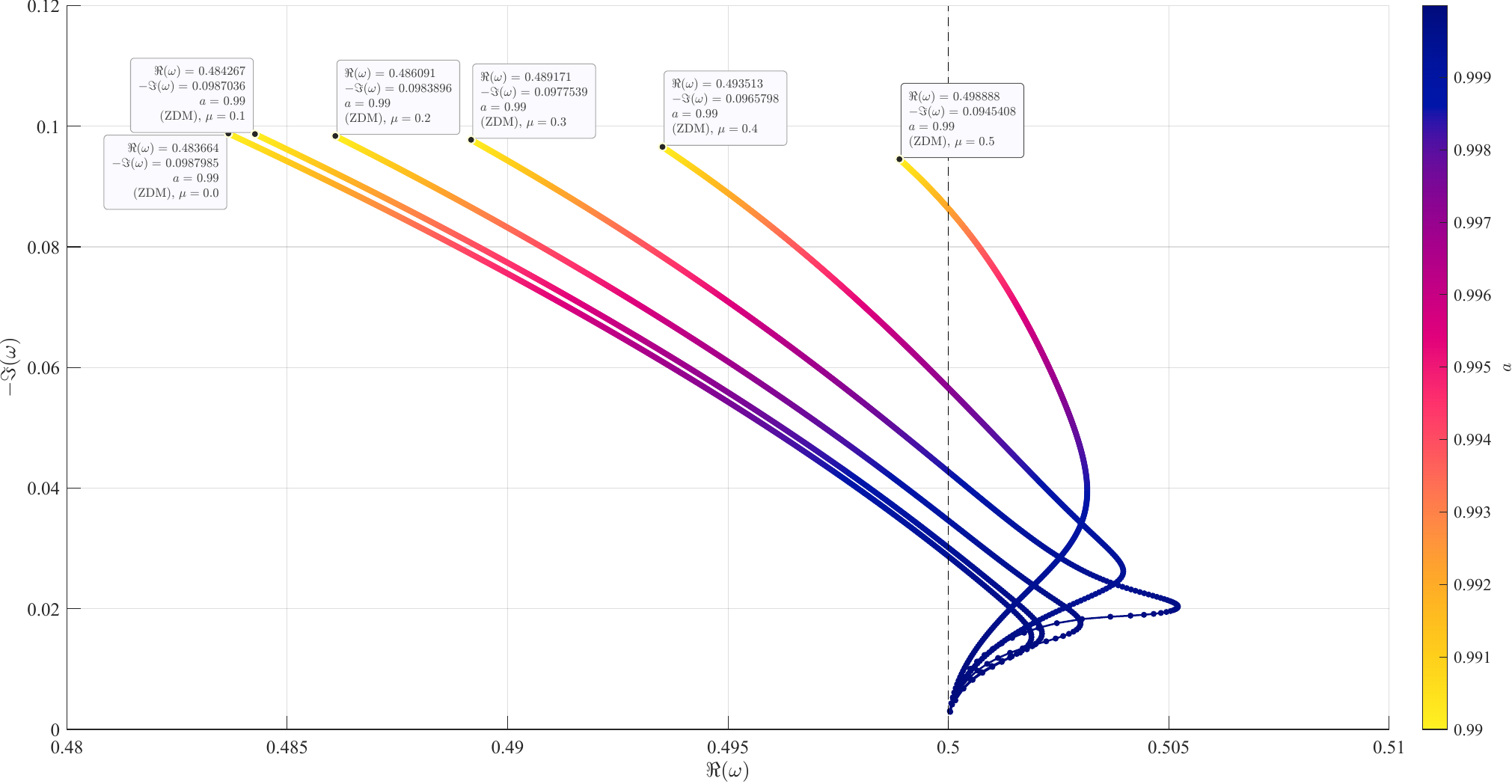}}\,\,
\caption{Kerr scalar QNMs for $\ell=m=1$ and $\hat{n} = \{0,1\}$ with $a\in[0.99,0.99999]$.}\label{fig:QNMs0_l1m1}
\end{figure*}

\subsection{Massive Scalar Kerr QNMs}

\begin{figure*}[htbp]
	\centering
\subfloat[Kerr Scalar QNMs with $a\in \mbox{[}0,0.999999\mbox{]}$\label{figa:s0l1m1n0}]{\includegraphics[width=3.4in]{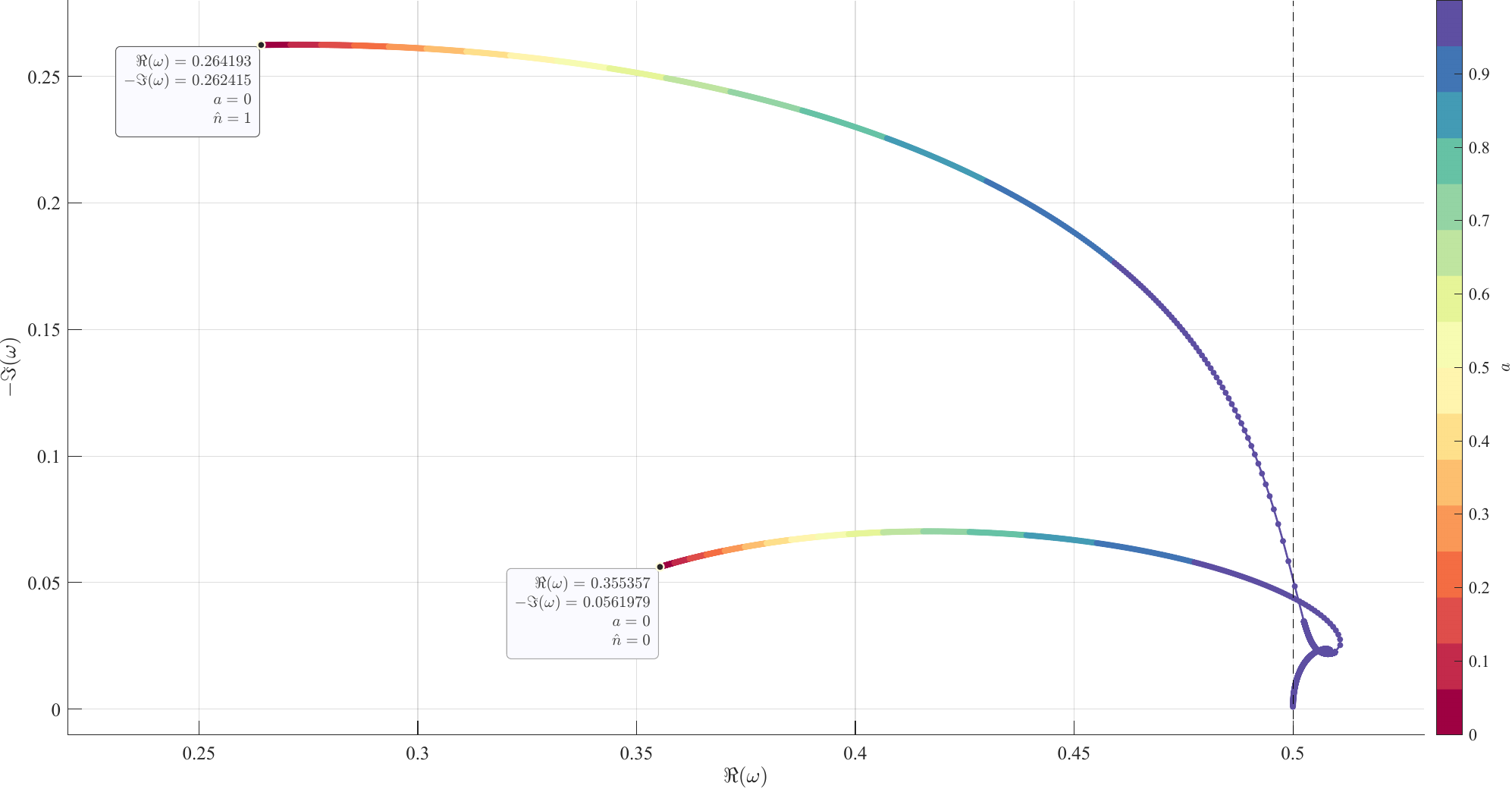}}\,\,
\subfloat[ Close-up for Kerr Scalar QNMs\label{figb:s0l1m1n0}]{\includegraphics[width=3.4in]{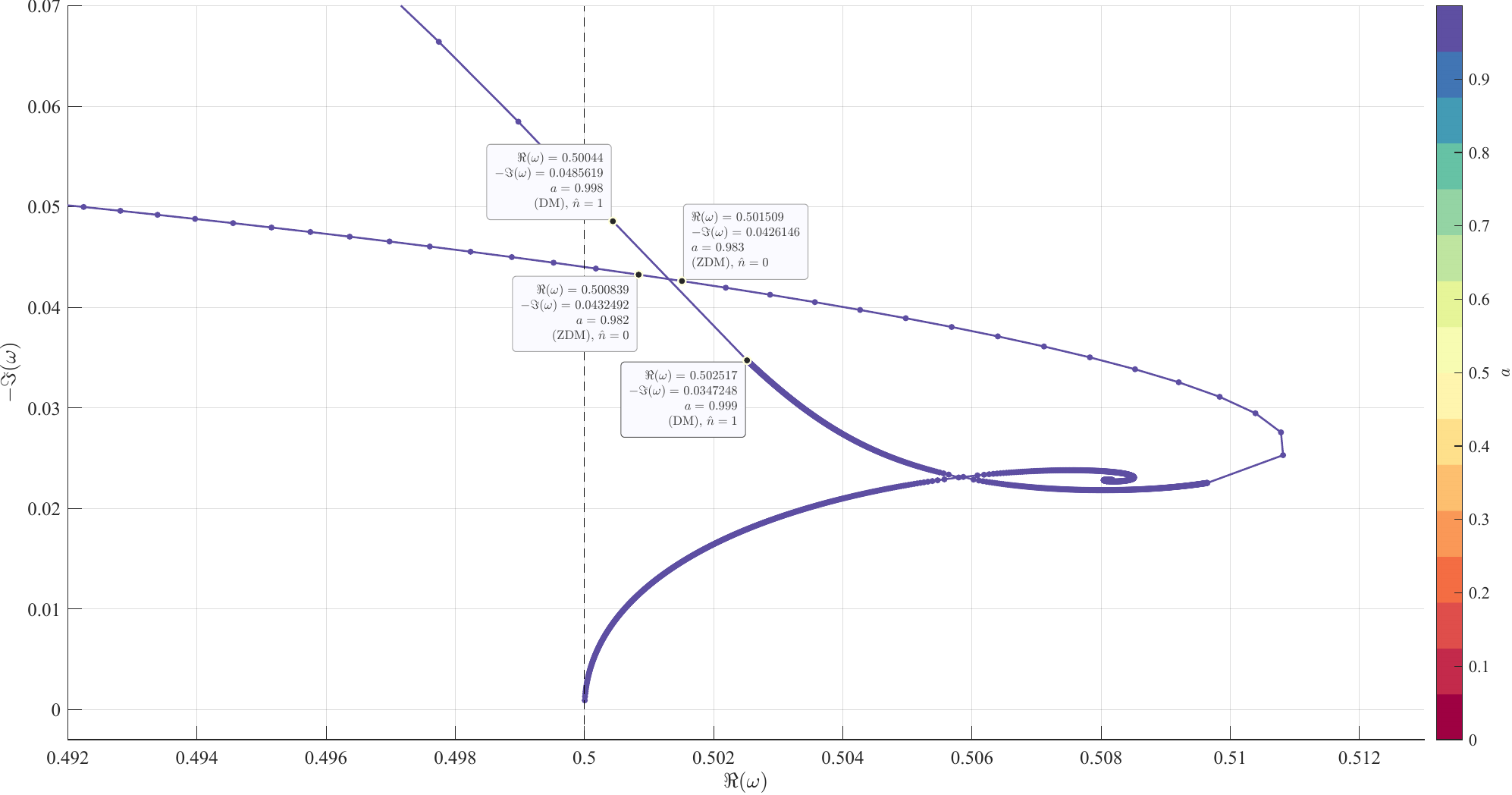}}\,\,
\subfloat[Kerr Scalar QNMs with $\mu = 0.37049813$ and $a\in \mbox{[}0.999,0.999999\mbox{]}$ \label{figc:s0l1m1n0}]{\includegraphics[width=3.4in]{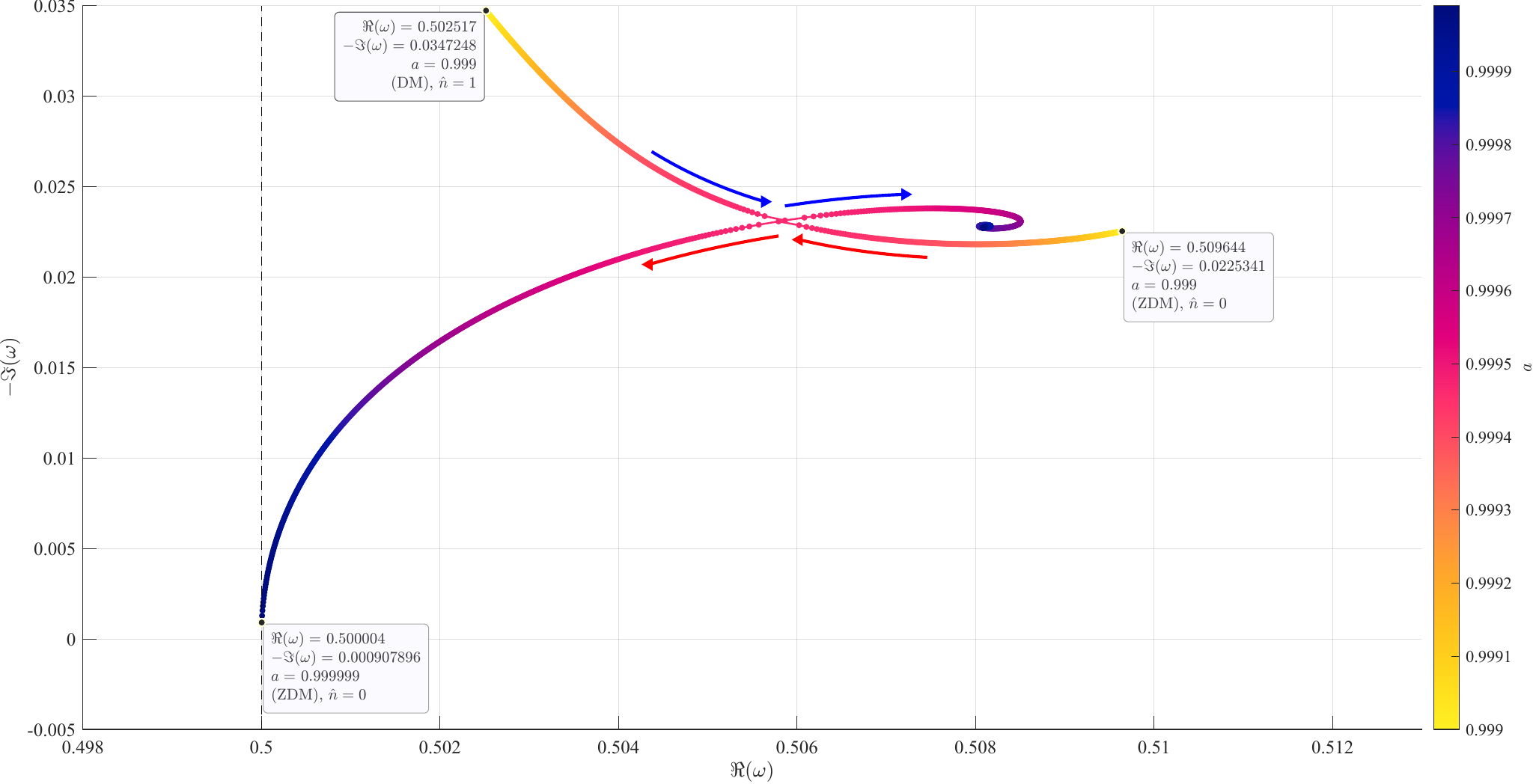}}\,\,
\subfloat[Kerr Scalar QNMs with $\mu = 0.37049814$ and $a\in \mbox{[}0.999,0.999999\mbox{]}$\label{figd:s0l1m1n0}]{\includegraphics[width=3.4in]{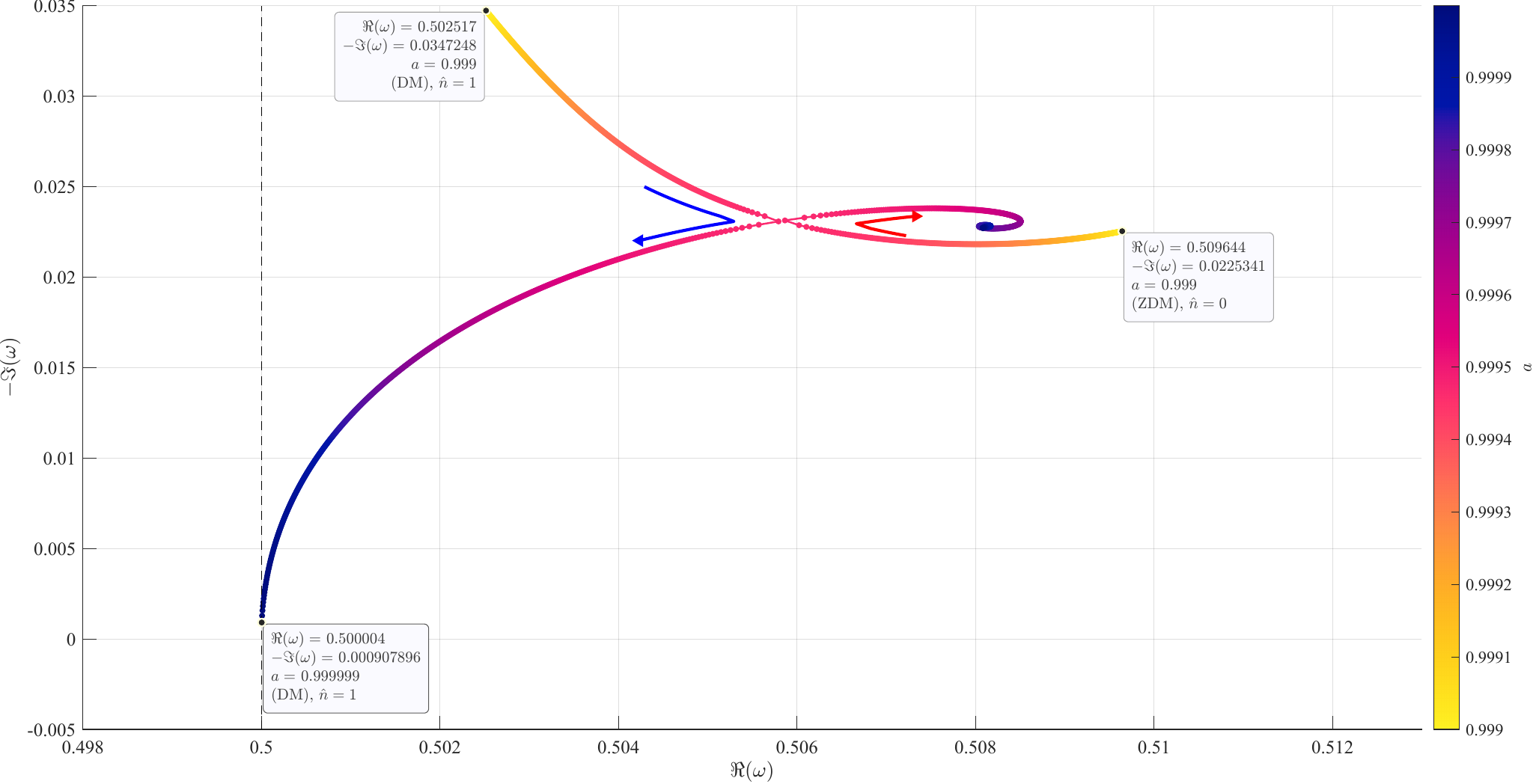}}\,\,

\caption{Massive Scalar Perturbations $s=0$ of Kerr QNMs for $\ell=m=1, \hat n =\{0,1\}$ with $\mu = \{0.37049813,0.37049814\}$. The QNMs for $\mu = \{0.37049813, 0.37049814\}$ are both depicted in panel (a), as the minimal mass difference leads to small numerical deviations.
Their distinction lies in whether the mass parameter yields a ZDM or a DM.
 Panel (b) shows a zoomed-in view of panel (a) near $\mathrm{Re}(\omega)=0.5$. In panels (c)-(d), red arrows indicate the evolutionary direction of the $\hat{n}=0$ QNM sequence, while blue arrows denote the $\hat{n}=1$ sequence's evolution. The computation employs adaptive step sizes for the spin parameter $a$, with $\square a = 10^{-3}$ in the range $a \in [0, 0.999]$ and $\square a = 10^{-6}$ for $a \in [0.999, 0.999999]$.}\label{fig:s0l1m1n0}
\end{figure*}

\begin{figure*}[htbp]
	\centering
\subfloat[$\{2,2,5\}$ and $\{2,2,6\}$\label{figa:QNM_Resonant} ]{\includegraphics[width=2.2in]{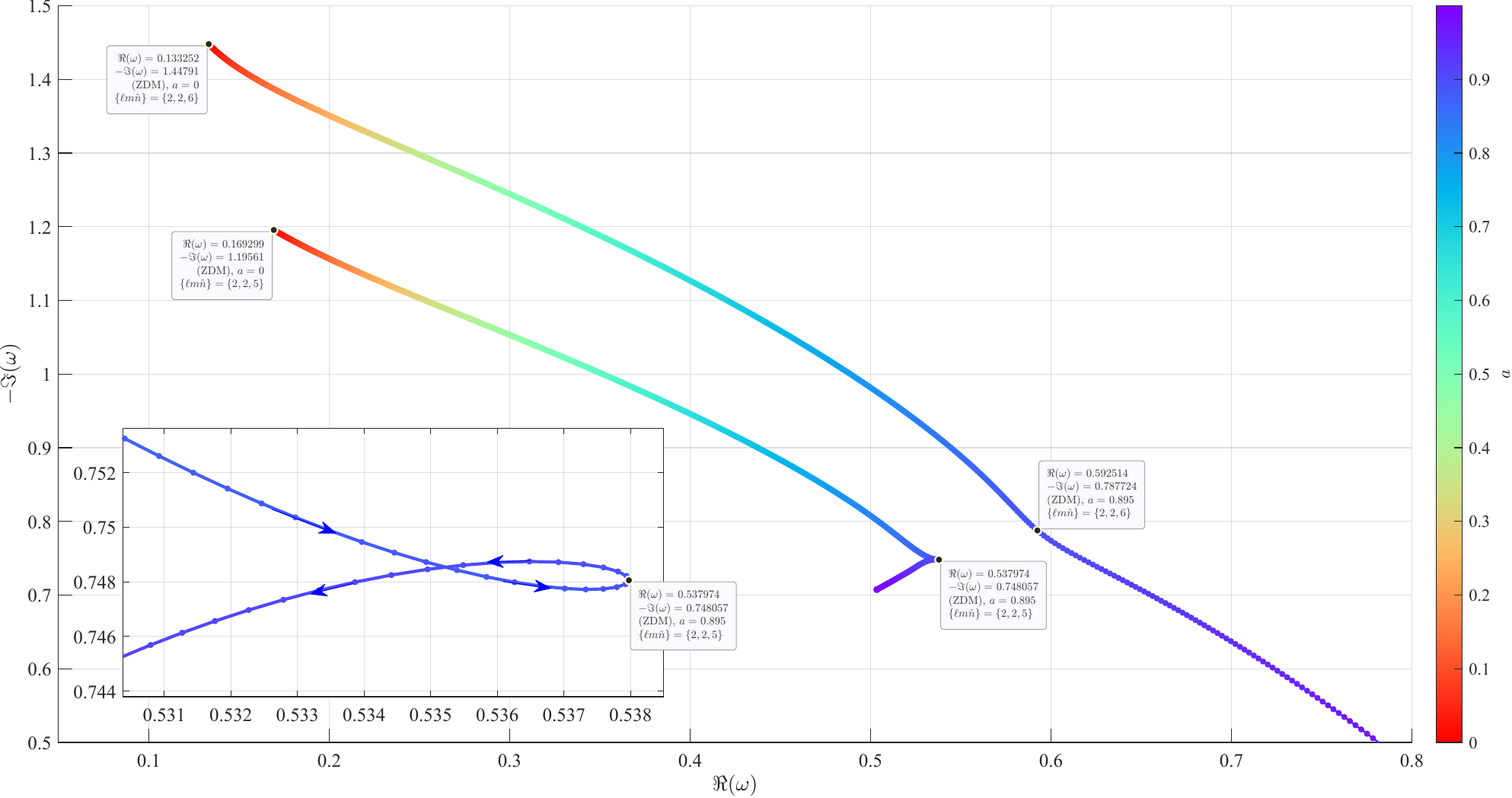}}\,
\subfloat[$\{3,1,4\}$,$\{3,1,5\}$,$\{3,1,6\}$ and $\{3,1,7\}$\label{figb:QNM_Resonant} ]{\includegraphics[width=2.2in]{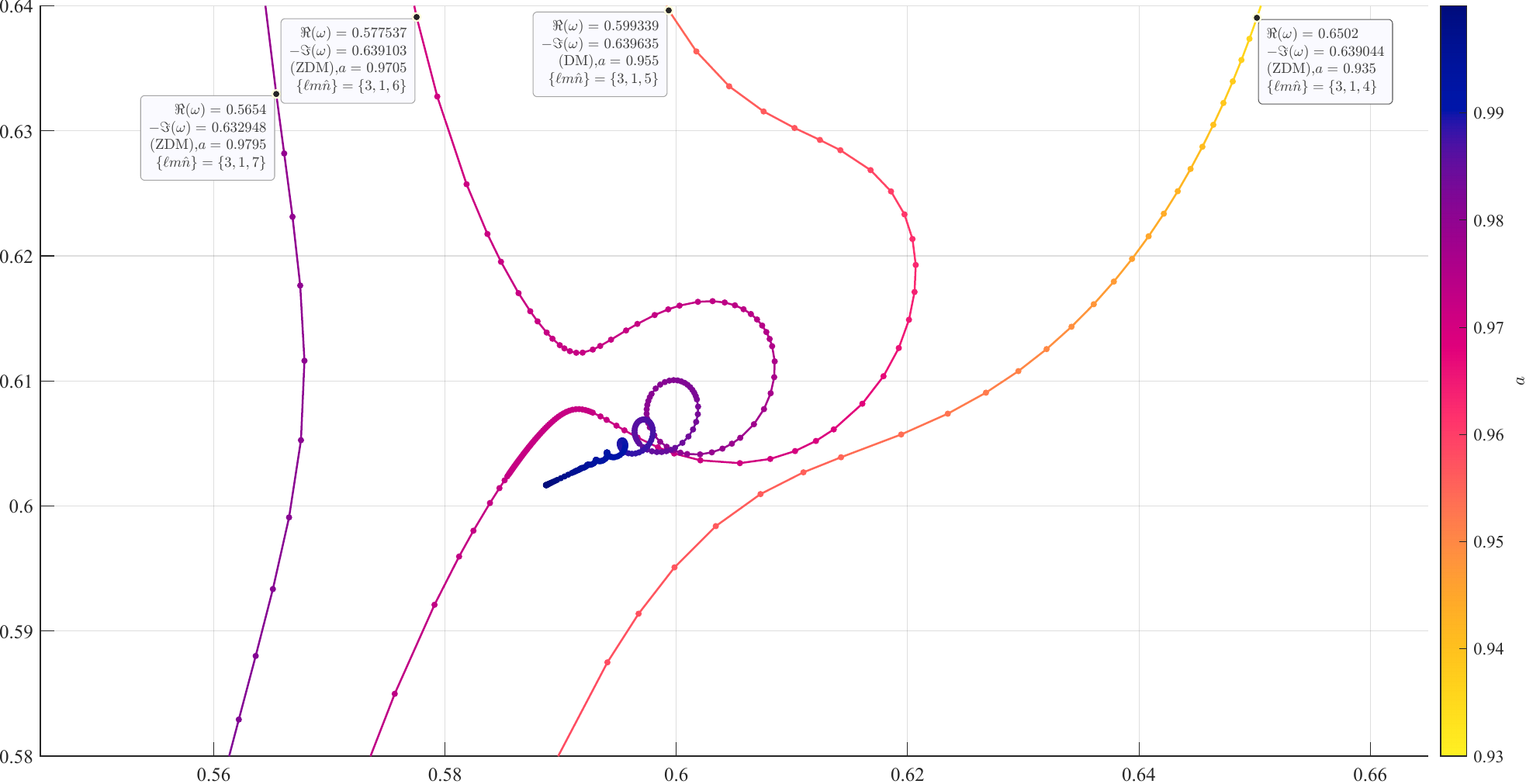}}\,
\subfloat[$\{3,-2,26\}$ and $\{3,-2,27\}$\label{figc:QNM_Resonant}]{\includegraphics[width=2.2in]{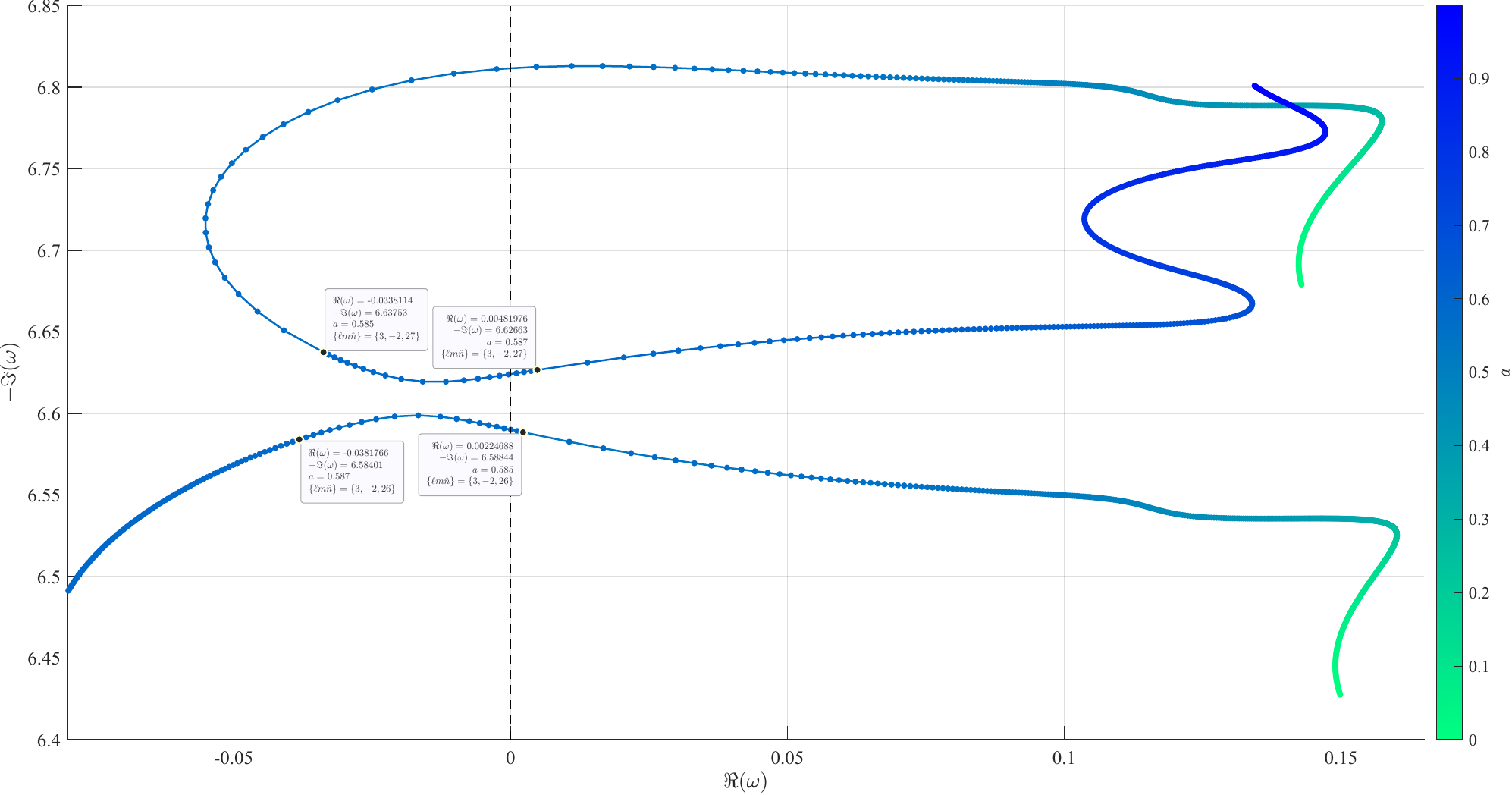}}\,
\caption{Resonance of massless Kerr QNMs.}\label{fig:QNM_Resonant}
\end{figure*}

Initially, Simone and Will \cite{Simone:1991wn} investigated the QNMs of massive scalar fields in Schwarzschild and Kerr BHs using higher-order WKB approximations. Subsequently, Konoplya and Zhidenko \cite{Konoplya:2004wg,Konoplya:2006br} provided accurate QNM spectra for Schwarzschild and Kerr BHs using the CFM. Moreover, they discussed the stability of QNM spectra and the asymptotic behavior of highly damped modes \cite{Konoplya:2004wg,Konoplya:2006br}. Later, they also studied the QNM spectra and stability of massless/massive scalar fields and Dirac fields in the Kerr-Newman family \cite{Konoplya:2007zx,Konoplya:2013rxa}. Dolan also utilized the CFM to study the instability in massive scalar Kerr QNM spectra \cite{Dolan:2007mj}, which originates from the amplification effect caused by classical superradiance.

For massive scalar perturbations in the Kerr BHs, the potential of the ATE \eqref{eq:GFoATE} is
\begin{equation}\label{eq:U_ATE_scalarKerr}
U(\Theta ) = {c^2}{\Theta ^2} - \frac{{{m^2}}}{{1 - {\Theta ^2}}} + {\kern 1pt} {{\kern 1pt} _0}{A_{\ell m}},\end{equation}
and the potential of the RTE \eqref{eq:GFoRTE} is
\begin{equation}\label{eq:V_RTE_scalarKerr}
V(r) = {K^2} - \lambda \Delta  - {\mu ^2}{r^2}\Delta ,
\end{equation}
where $c = a\sqrt {{\omega ^2} - {\mu ^2}} ,$ $\lambda  = {\kern 1pt} {{\kern 1pt} _0}{A_{\ell m}}(a\omega ) + {a^2}{\omega ^2} - 2am\omega $, and $K = ({r^2} + {a^2})\omega - am$.

The HeunC parameters of the ATE \eqref{eq:GFoATE} can be derived in the following form,
\begin{subequations}\label{eq:ATE_HCprameters_scalarKerr}
 \begin{align}
&{\alpha _{ - 1}} = 4c,\quad {\alpha _{ + 1}} =  - 4c ,\quad \\
&{\beta _{ - 1}} = m,\quad {\beta _{ + 1}} = m ,\quad \\
&{\gamma _{ - 1}} = m,\quad {\gamma _{ + 1}} = m,\quad \\
&{\delta _{ - 1}} = 0,\quad\quad {\delta _{ + 1}} =  0 ,\\
&{\eta _{ - 1}} =  - {c^2} + {\textstyle{1 \over 2}}{m^2} - {\kern 1pt} {{\kern 1pt} _0}{A_{\ell m}},\\
&{\eta _{ + 1}} =  - {c^2} + {\textstyle{1 \over 2}}{m^2} - {\kern 1pt} {{\kern 1pt} _0}{A_{\ell m}}.
\end{align}
\end{subequations}

The HeunC parameters of the RTE \eqref{eq:GFoRTE} can be derived in the following form,
\begin{subequations}\label{eq:RTE_HCprameters_scalarKerr}
 \begin{align}
&\alpha _r^ +  = 2({r_+ } - {r_ - })\sqrt {{\mu ^2} - {\omega ^2}} ,\\
&\beta _r^ +  =  - \frac{{2i\omega (r_ + ^2 + {a^2}) - 2iam}}{{{r_ + } - {r_ - }}},\\
&\gamma _r^ +  =  - \frac{{2i\omega (r_ - ^2 + {a^2}) - 2iam}}{{{r_ + } - {r_ - }}},\\
&{\delta _r} =  ({\mu ^2} - 2{\omega ^2})\Delta ,\\
&{\eta _r} =  - \frac{{2\left[ {({a^2} + r_ + ^2)\omega  - am} \right]\left[ {({a^2} + 2{r_ - }{r_ + } - r_ + ^2)\omega  - am} \right]}}{{{{({r_ - } - {r_ + })}^2}}} \nonumber\\
&\quad \quad - \lambda  - {\mu ^2}r_ + ^2,\\
&{r_ \pm } = M \pm \sqrt {{M^2} - {a^2}}.
\end{align}
\end{subequations}

\subsubsection{Comparison of Connection Formula}

Cavalcante et al. \cite{Cavalcante:2024swt,Cavalcante:2024kmy} employed the isomonodromic method \cite{ablowitz1981solitons,its2006isomonodromic} to study QNMs of massive scalar perturbations in near-extremal Kerr BHs.
As discussed in \cref{sec:HeunC_Sol_RTE}, constructing the eigenvalue equation of QNM frequencies via analytic connection formulas inevitably requires additional constraint equations.
These additional equations are typically transcendental and necessitate computation of two-sided infinite series using the CFM.
The isomonodromic method likewise faces this inherent limitation, requiring substantial computational resources to evaluate auxiliary parameters analogous to the renormalized angular momentum $\nu$ in the MST method.
As shown in \cref{tab:KerrS0_L2M2}, our method demonstrates superior computational efficiency compared to the isomonodromic method.
While both approaches fundamentally compute the connection coefficients, our method employs analytic continuation to bypass redundant calculations of auxiliary parameters like $\nu$. The QNM frequencies can be computed directly via \cref{eq:HeunC_QNM} using only the HeunC parameters\textemdash which are  well-established parameters of the Teukolsky equation.
However, our connection formula is restricted to calculating zero transmission amplitude at infinity ($B_{\ell m}^{\rm{ref}} = 0$), and cannot handle non-zero transmission amplitudes.
To obtain the complete amplitudes $(B_{\ell m }^{{\rm{trans}}}, B_{\ell m }^{{\rm{inc}}},B_{\ell m }^{{\rm{ref}}})$ at both the horizon and infinity, one still needs to perform complicated calculations via power series methods such as the MST method, including the calculation of $\nu$.

\subsubsection{Resonant Excitation of Kerr QNMs}

For the Kerr QNM spectrum of massive perturbations, Cavalcante et al. \cite{Cavalcante:2024swt,Cavalcante:2024kmy} found that the mass parameter $\mu$ differentially affects ZDMs and DMs in the extremal limit.
\cref{fig:QNMs0_l1m1} displays QNMs of scalar fields in near-extremal Kerr spacetime for various $\mu$ values, demonstrating that $\mu$ determines whether the $\{1,1,0\}$ sequence becomes DMs or ZDMs.
Although no DMs are observed in \cref{figb:QNMs0_l1m1}, the $\{1,1,1\}$ sequence contains DMs for $\mu\in[0.3,0.4]$.
As discovered by Cavalcante et al. \cite{Cavalcante:2024kmy,Cavalcante:2024swt}, a critical mass $\mu_c \in[0.37049813,0.37049814]$ governs the DM-ZDM transition for both $\{1,1,0\}$ and $\{1,1,1\}$ sequences.
\cref{fig:s0l1m1n0} demonstrates the resulting hysteresis when $\mu_c$ takes values in $\{0.37049813,0.37049814\}$ for these mode sequences.
\Cref{figa:s0l1m1n0,figb:s0l1m1n0} show that at $\mu = \mu_c$, the trajectories of both $\{1,1,0\}$ and $\{1,1,1\}$ sequences exhibit two coincidence points.
The first coincidence point occurs at $a \in [0.982,0.983]$ for the $\{1,1,0\}$ sequence and $a \in [0.998,0.999]$ for the $\{1,1,1\}$ sequence, which represents a simple intersection of trajectories. In contrast, the second coincidence point appears at $a \in [0.9994655,0.999466]$ for both sequences.
At this identical spin parameter, their trajectories become extremely close, generating resonant effects in the QNM spectrum.
\Cref{figc:s0l1m1n0,figd:s0l1m1n0} display the mutual transformation between DMs and ZDMs under two distinct mass parameters.
The DM-ZDM transition exhibits $\mu$-dependent behavior: at $\mu = 0.37049813$, the $\{1,1,0\}$ sequence becomes the ZDM while the $\{1,1,1\}$ sequence becomes the DM; conversely at $\mu = 0.37049814$, the $\{1,1,0\}$ sequence transforms into the DM while the $\{1,1,1\}$ sequence becomes the ZDM.
Classification of these sequences as DMs or ZDMs was confirmed using high-resolution sampling ($\square a = 10^{-11}$).

Cavalcante et al. refer to the second coincidence point as \textit{exceptional points}. These points emerge naturally in the general perturbation theory of non-Hermitian operators~\cite{kato2013perturbation}, where one examines eigenvalues of operators parameterized by a complex variable ${}_s A_{\ell m}$. Due to non-Hermiticity, these eigenvalues need not be real-valued. While locally analytic in ${}_s A_{\ell m}$, special degeneracies occur at certain parameter values where two or more eigenvalues coalesce. Known as exceptional points, such degeneracies find broad applications across condensed matter physics~\cite{Berry:2004ypy,Bergholtz:2019deh,Ashida:2020dkc}.
Compared with the isomonodromic method proposed by Cavalcante et al., our method can quickly obtain exceptional points, as shown in \cref{tab:KerrS0_L2M2}.

Such spectral resonances are not restricted to massive scalar fields, but also exist for massless gravitational perturbations.
As shown in \Cref{figa:QNM_Resonant,figa:QNM_Resonant}, the $\{2,2,5\}$ sequence influences the $\{2,2,6\}$ sequence. Although less prominent, the $\{2,2,6\}$ sequence shows slight distortion at $a \approx 0.895$. The $\{3,1,6\}$ sequence produces the strongest effect on adjacent QNM sequences in the near-extremal regime.
These spectral features were initially identified in earlier studies \cite{Onozawa:1996ux,Cook:2014cta}.
Motohashi \cite{Motohashi:2024fwt} numerically analyzed these effects through QNM excitation factors (calculated via the MST method), employing the non-Hermitian physics framework of QNMs to explain the resonance phenomenon.
These spectra indicate higher resonance probability when $a \to 1$ with $m > 0$.
Our complete QNM spectrum analysis reveals resonance phenomena also exist for $m < 0$ at intermediate spins.
\Cref{figc:QNM_Resonant,fig:l3mf2n26-29} reveal that the high-spin QNMs of the $\{3,-2,26\}$ sequence exhibit distinct behavior compared to other sequences, appearing on the left side of the NIA.
A distinct deformation appears in the $\{3,-2,27\}$ sequence trajectories when approaching $\{3,-2,26\}$ at spins $a \in [0.585,0.587]$.
Furthermore, resonant phenomena between different sequences introduce numerical challenges. When sequences approach too closely, spectral segments from distinct overtones may be erroneously connected, forming artificial mode sequences.
Our method successfully addresses this issue by providing the required high-resolution capabilities.

\section{CONCLUSION}\label{sec:Conclusion}
{ QNM spectroscopy of BHs provides a direct probe of strong-field gravity accessible through GW observations with LIGO, Virgo, and the future LISA mission. Complete QNM spectra are essential for testing GR and the no-hair theorem through ringdown analysis. However, theoretical predictions have been limited by incomplete spectral knowledge: two open problems identified by Berti et al.~\cite{Berti:2004md,Berti:2009kk} have persisted for two decades, preventing full utilization of QNM data for precision tests of GR. This work resolves both problems by computing the first complete QNM spectra—including all modes crossing or lying on the negative imaginary axis—for Schwarzschild and Kerr BHs across a wide range of overtones ($0 \leq \hat{n} \leq 41$) and multipoles ($2 \leq \ell \leq 16$).
These results were enabled by our analytic continuation method for the RTE \eqref{eq:GFoRTE}, which eliminates the dependence on auxiliary parameters (such as the renormalized angular momentum $\nu$) and systematically handles the branch cut in the QNM spectrum. This approach surpasses both Cook's continued fraction method (which fails at the NIA) and isomonodromic techniques of the same type \cite{Cavalcante:2024swt,Cavalcante:2024kmy} in computational efficiency while providing complete spectral coverage.
}

{The results from our complete QNM spectrum provide solutions to two long-standing \cref{open1,open2}.
The following presents the solutions to these two problems, together with some additional discussions.}

{For \cref{open1}, the complete QNM spectrum confirms the continuity between Schwarzschild and Kerr solutions as $a\to0$. As the overtone number increases, the asymptotic spectrum preserves this continuity.
Specifically, conventional methods failed to compute QNM families that cross the NIA as the spin parameter varies. Our results show that each Kerr overtone at $a = 0$ corresponds to exactly one Schwarzschild mode, preserving the fundamental principle that the spectrum of the limit equals the limit of the spectrum.}

{For \cref{open2}, the AS mode $\omega_\ell^{{\rm{AS}}}$ is no longer enigmatic.
At the AS frequency $\omega^{\rm AS}_{\ell}$ of Schwarzschild BHs, our high-precision calculations ($< 10^{-10}$ error) confirm that: (i) A genuine QNM exists at exactly $\omega^{\rm AS}_{\ell}$, coinciding with Chandrasekhar's analytic formula~\cite{Chandrasekhar:1984mgh}. (ii) This QNM coexists with a TTM$_{\rm L}$, but not with TTM$_{\rm R}$—resolving the long-standing ambiguity about which mode types exist at this frequency. (iii) An additional unconventional mode appears near $\omega^{\rm AS}_{\ell}$ (e.g., Eq.~\eqref{eq:SchNewModesL2} for $\ell=2$, and our first-ever computation Eq.~\eqref{eq:SchNewModesL3} for $\ell=3$), confirming Leung et al.'s 2003 conjecture~\cite{Leung:2003eq}.
(iv) For Kerr BHs with $\ell=2$, we demonstrate that the overtone sequences originating from both the unconventional mode ($\hat{n}=8$) and the AS mode ($\hat{n}=9$) each exhibit precisely $2\ell+1=5$ branches under Zeeman-like splitting. No anomalous multiplets or supersymmetry breaking occurs.
 The additional sequences reported by Cook et al.~\cite{Cook:2014cta} were incomplete branches missing their $a \in [0, a_{\rm critical}]$ portions due to NIA crossings.
(v) Our analytic expressions for scattering amplitudes provide deeper understanding of QNMs and TTMs at $\omega_\ell^{{\rm{AS}}}$.
The scattering amplitudes reveal that only ${\rm QNM}_{-2}$ and ${\rm TTM}_{\rm L}$ can exist at $\omega_\ell^{{\rm{AS}}}$. Furthermore, we deduce a key inference: \textit{no complex frequency can simultaneously be both a ${\rm{TTM}_{\rm{R}}}$ mode and a QNM}.}

{\cref{fig:l2n8,fig:l3n40,fig:l2n9,fig:l3n41} provide strong numerical evidence supporting our conclusion.
These two open problems have similar descriptions for RN BHs as for Kerr BHs. Our analytic continuation theory can also solve the RN QNMs.
Although the wave equation of RN BHs cannot be transformed into a confluent Heun equation, we only need to construct the power series solution of pure ingoing wave $R_{\ell m }^{{\rm{in}}}$ and the asymptotic solution at infinity, then apply the analytic continuation method to solve the RN QNMs.
The final results show the same conclusions as the Kerr QNMs. Detailed discussions can be found in Ref. \cite{Chen_RN_QNM}.}

\vspace{8pt}

{In summary, this work provides the first complete QNM spectra for Type-D BHs, resolving two decades-old open problems and removing a theoretical bottleneck in GW physics. Our results reveal new phenomena relevant for GW astronomy: ZDMs and exceptional points in near-extremal BHs (important for superradiance studies), and resonance effects between overtone families that may affect high-overtone ringdown modeling for no-hair theorem tests.
Our analytic continuation framework extends naturally to Type-D BHs in modified gravity theories, where complete spectral characterization is essential for distinguishing GR from beyond-GR physics through QNM measurements. The computational efficiency gains make systematic parameter space surveys feasible, establishing a new paradigm for precision tests of strong-field gravity in the GW era.}

\section*{Acknowledgements}
This work makes use of the Black Hole Perturbation Toolkit \cite{BHPToolkit}, Cook's \texttt{KerrModes} Package \cite{cookgb_kerrmodes_2023}, and Motygin's HeunC Code \cite{motygin_confluent_Heun_functions}.

J. Jing is supported by the Grant of NSFC No. 12035005, and National Key Research and Development Program of China No. 2020YFC2201400.
Z. Cao is supported in part by the National Key Research and Development Program of China Grant No. 2021YFC2203001, in part by ``the Fundamental Research Funds for the Central Universities".
M. Wang is supported by the National Natural Science Foundation of China under Grant No. 12475050 and by the Scientific Research Fund of Hunan Provincial Education Department Grant No. 22A0039.

\appendix
{
\section{List of Uncommon Abbreviations}\label{app:Abbr}
To facilitate readers' understanding, the following list provides the full names of uncommon abbreviations.}
\begin{table}[htp!]
    \centering
    {
    \begin{tabular}{|l|l|}
        \hline
        \textbf{Abbreviations} & \textbf{Full Name} \\
        \hline
        NIA & Negative Imaginary Axis \\
        \hline
        MST & Mano–Suzuki–Takasugi \\
        \hline
        TTM & Total Transmission Mode \\
        \hline
        TTM$_{\rm L}$ & Total Transmission Mode-Left \\
        \hline
        TTM$_{\rm R}$ & Total Transmission Mode-Right \\
        \hline
        CFM & Continued Fraction Method \\
        \hline
        HeunC & Confluent Heun Function \\
        \hline
        AS & Algebraically Special \\
        \hline
        RWE & Regge–Wheeler Equation \\
        \hline
        CHE & Confluent Heun Equation \\
        \hline
        ATE & Angular Teukolsky Equation \\
        \hline
        RTE & Radial Teukolsky Equation \\
        \hline
        STC & Starobinsky-Teukolsky Constant \\
        \hline
        ZDM & Zero Damped Mode \\
        \hline
        DM & Damped Mode \\
        \hline
        Dual-HeunC & Dual Confluent Heun Method \\
        \hline
        PM & Post-Minkowskian \\
        \hline
        BHPToolkit & Black Hole Perturbation Toolkit \\
        \hline
    \end{tabular}
    }
\end{table}

{
\section{Computational Implementation Details}\label{app:CompDetail}
This section provides some computational details to facilitate readers in reproducing the results of this paper.
\subsection{Complex Conjugation of QNMs}\label{app:CompDetail_A}
The $\pm$ sign of the complex frequency's imaginary part governs the boundary conditions, requiring systematic characterization of this correspondence with the HeunC parameters.
To ensure the stability of complex frequencies, the imaginary part of the frequency  must satisfy
\begin{subequations}
\begin{align}
{\rm Im} (\omega ) < 0\quad {\rm{for}}\quad & \psi \sim{e^{ - i\omega t}},\\
{\rm Im} (\omega^* ) > 0\quad {\rm{for}}\quad & \psi \sim{e^{  i\omega^* t}}.
\end{align}
\end{subequations}
Asymptotic solutions at the horizon and infinity can be obtained by performing asymptotic analysis of the RTE \eqref{eq:GFoRTE} at these boundaries.
\begin{subequations}
  \begin{align}
&{f_1}(r - {r_ + }){e^{ - ik{r_*}}}\, {\rm and}&  \,{f_2}(r - {r_ + }){e^{ik{r_*}}},\,\, &r \to {r_ + },\\
&{g_1}({r^{ - 1}}){{\rm{e}}^{-i\omega {r_*}}}\quad \,\, {\rm and}&  {g_2}({r^{ - 1}}){{\rm{e}}^{ i\omega {r_*}}},\,\, &r \to \infty.
  \end{align}
\end{subequations}
where both $f$ and $g$ are power functions. For their derivation in Type-D spacetimes, one may refer to Teukolsky's original derivation for Kerr spacetime \cite{Teukolsky:1973ha} and Table 1 in Ref. \cite{Teukolsky:1974yv}.
$r_*$ is a function of $r$ known as the tortoise coordinate in Kerr spacetime, while $k$ denotes the wavenumber.
When $\psi \sim {e^{ - i\omega t}}$ and $ {\rm Im} (\omega ) < 0$, we obtain
\begin{equation}
\left\{ \begin{array}{l}
R_{{\rm{in}}}^{\rm{H}} = {f_1}(r - {r_ + }){e^{ - ik{r_*}}},R_{{\rm{up}}}^{\rm{H}} = {f_2}(r - {r_ + }){e^{ik{r_*}}},\\
R_{{\rm{in}}}^\infty  = {g_1}({r^{ - 1}}){{\rm{e}}^{-i\omega {r_*}}},\quad \, R_{{\rm{up}}}^\infty  = {g_2}({r^{ - 1}}){{\rm{e}}^{ i\omega {r_*}}}.
\end{array} \right.
\end{equation}
When $\psi \sim {e^{ i\omega^* t}}$ and $ {\rm Im} (\omega^* ) > 0$, we obtain
\begin{equation}
\left\{ \begin{array}{l}
R_{{\rm{in}}}^{\rm{H}} ={f_2}(r - {r_ + }){e^{ik^*{r_*}}},R_{{\rm{up}}}^{\rm{H}} ={f_1}(r - {r_ + }){e^{ - ik^*{r_*}}},\\
R_{{\rm{in}}}^\infty  ={g_{2}}({r^{ - 1}}){{\rm{e}}^{ i\omega^* {r_*}}},\quad \,  R_{{\rm{up}}}^\infty  = {g_{1}}({r^{ - 1}}){{\rm{e}}^{-i\omega^* {r_*}}}.
\end{array} \right.
\end{equation}
In this work, the superscript asterisk denotes complex conjugation (e.g., $\omega^*$ represents the complex conjugate of $\omega$).
Note that when Iyer and Will first proposed the WKB method to solve the QNMs \cite{Iyer:1986np}, they discussed the boundary conditions corresponding to the  $\pm$ sign of the QNM's real part. However, they obtained an incorrect result:
\begin{equation}
{n} = \left\{ {\begin{array}{*{20}{l}}
{0,1,2,\ldots,} & {{\rm Re}(\omega) > 0,} \\
{-1,-2,\ldots,} & {{\rm Re}(\omega) < 0.}
\end{array}
} \right.
\end{equation}
Our analysis yields the correct result:
\begin{equation}
{n} = \left\{ {\begin{array}{*{20}{l}}
{0,1,2,\ldots,} & {{\rm Im}(\omega) < 0,} \\
{-1,-2,\ldots,} & {{\rm Im}(\omega^*) > 0.}
\end{array}
} \right.
\end{equation}
This can be numerically verified using the WKB computational code from Vritika \cite{QNM_Vritika}.
}

{
For the case $\psi \sim e^{i\omega^* t}$, we can similarly obtain the eigenvalue equation for the QNM frequencies with complex conjugation.
\begin{subequations}\label{eq:HeunC_QNM_Conj}
  \begin{align}
{\tilde B_{\ell m}^{{\rm{inc}}} = \mathop {\lim }\limits_{|{x_r}| \to \infty } \frac{{{\rm{HeunC}}({\alpha _r},\beta _r^- ,\gamma _r^+ ,{\delta _r},{\eta _r},{x_r})}}{{{{\rm{e}}^{ - {\alpha _r}{x_r}}}{x_r}^{ - \frac{{\beta _r^-  + \gamma _r^+  + 2}}{2} + \frac{{{\delta _r}}}{{{\alpha _r}}}}}} = 0}&, \nonumber\\
{{\rm{if}}{\kern 1pt} {\kern 1pt} {\alpha _r} = \alpha _r^+ }&;\\
{\tilde B_{\ell m}^{{\rm{inc}}} = \mathop {\lim }\limits_{|{x_r}| \to \infty } \frac{{{\rm{HeunC}}({\alpha _r},\beta _r^- ,\gamma _r^+ ,{\delta _r},{\eta _r},{x_r})}}{{{x_r}^{ - \frac{{\beta _r^-  + \gamma _r^+  + 2}}{2} - \frac{{{\delta _r}}}{{{\alpha _r}}}}}} = 0}&, \nonumber\\
{{\rm{if}}{\kern 1pt} {\kern 1pt} {\alpha _r} = \alpha _r^- }&.
  \end{align}
\end{subequations}
with a constraint condition
\begin{equation}\label{eq:QNM-Cond-MFD-Conj}
\arg(r) = 2{\rm{k}}\pi - \frac{\pi}{2}(1+\varepsilon) - \arg(\omega^* ), \quad{\rm{k}} \in \mathbb{Z},
\end{equation}
and a assumption
\begin{equation}\label{eq:QNM-assume-conj}
   \mathop {\lim }\limits_{r \to \infty } \frac{{R_{{\rm{up}}}^\infty }}{{R_{{\rm{in}}}^\infty }} = \mathop {\lim }\limits_{r \to \infty } \frac{{{g_1}({r^{ - 1}})}}{{{g_2}({r^{ - 1}})}}{{\rm{e}}^{-2i\omega^* {r_*}}} = 0.
\end{equation}
where the choice of $\varepsilon$ is similar to that in \cref{eq:QNM-Cond-MFD-new}.

In numerical calculations, the spatial variable $r$ is typically given a sufficiently large value to approximate spatial infinity.
To ensure that the assumed equation \eqref{eq:QNM-assume} approaches machine precision ($10^{-16}$) at infinity, we require
\begin{equation}\label{eq:tilde_g}
\tilde g = \frac{g_2(r^{-1})}{g_1(r^{-1})} = r^{\mathbf{k}}, \quad \mathbf{k} \in \mathbb{Z}^-,
\end{equation}
where $\mathbb{Z}^-$ is the set of negative integers.
Asymptotic analysis of the RTE \eqref{eq:GFoRTE} at infinity reveals that:
\begin{itemize}
    \item For $\psi \sim e^{-i\omega t}$, \cref{eq:tilde_g} holds when $s > 0$.
    \item For $\psi \sim e^{i\omega^* t}$, \cref{eq:tilde_g} holds when $s < 0$.
\end{itemize}
In other words:
\begin{itemize}
    \item When $s > 0$, using \cref{eq:HeunC_QNM} yields higher-precision and more efficient QNM spectra.
    \item When $s < 0$, using \cref{eq:HeunC_QNM_Conj} yields higher-precision and more efficient QNM spectra.
    \item When $s = 0$, both equations show equivalent accuracy.
\end{itemize}
}

{
There are some symmetries:
\begin{enumerate}
\item They are complex conjugate with each other,
\begin{equation}
  \quad {{\kern 1pt} _s}{A_{\ell m}}(a\omega^* ) = {{\kern 1pt} _s}A{_{\ell m}^*}(a\omega ).
\end{equation}
  \item Eigenvalues for $m$ and $-m$ are related,
\begin{subequations}\label{eq:symmetry_m}
  \begin{align}
{\mathop{\rm Re}\nolimits} \left[ {\omega (m)} \right] &=  - {\mathop{\rm Re}\nolimits} \left[ {\omega( - m)} \right],\nonumber \\
 {\rm Im} \left[ {\omega (m)} \right] &=   {\rm Im} \left[ {\omega( - m)} \right], \\
{\kern 1pt}_s{A_{\ell m}} (a\omega)&= {\kern 1pt} {{\kern 1pt} _s}A_{\ell , - m}^*(a\omega ).
\end{align}
\end{subequations}
  \item Eigenvalues for $s$ and $-s$ are related,
\begin{equation}\label{eq:symmetry_s}
  {{\kern 1pt} _{ - s}}{A_{\ell m}} = {\kern 1pt} {{\kern 1pt} _s}A_{\ell m}^* + 2s.
\end{equation}
 \item Eigenvalues for ${\omega}$ and ${\omega ^*}$ are related,
 \begin{equation}\label{}
   {{\kern 1pt} _s}{A_{\ell m}}(a\omega^*) = {\kern 1pt} {{\kern 1pt} _s}A_{\ell m}^*(a\omega ).
 \end{equation}
\item These two solutions are often referred to as mirror solutions since the relation,
 \begin{equation}{\omega _{{\text{II}}}}(\ell ,m) =  - { {{\omega^* _{\text{I}}}(\ell , - m)}}.\end{equation}
\end{enumerate}
}
{
\subsection{Norm Errors and Numerical Setup}\label{app:CompDetail_B}
In mathematics and numerical analysis, $L_2$-error and $L_{\infty}$-error measure the differences between functions or vectors. Their definitions are as follows:
\begin{itemize}
  \item The $L_2$-error (Mean Squared Error) of discrete data $\{x_i\}_{i=1}^{\rm n}$ and $\{y_i\}_{i=1}^{\rm n}$ is defined as
   \begin{equation}
     \lVert x - y\rVert_{L_2}=\sqrt{\frac{1}{\rm n}\sum_{i = 1}^{\rm n}(x_i - y_i)^{2}}.
   \end{equation}
  \item The $L_{\infty}$-error (Maximum or Chebyshev error) of discrete data $\{x_i\}_{i=1}^{\rm n}$ and $\{y_i\}_{i=1}^{\rm n}$ is defined as
   \begin{equation}
     \lVert x - y\rVert_{L_{\infty}}=\max_{1\leq i\leq {\rm n}}\vert x_i - y_i\vert.
   \end{equation}
\end{itemize}
The numerical setup is specified as follows:
\begin{enumerate}
    \item Processor: Intel Core i7-12700H 2.70 GHz.
    \item Default machine epsilon: $\epsilon=10^{-16}$ .
    \item Default radial coordinate cutoff: $|r|=30$ .
    \item Parallel computation (when used): 12 cores.
\end{enumerate}
}

\section{Asymptotic Formula of HeunC Function at Infinity}\label{app:AsymptoticFormula}

We provide the asymptotic expression of confluent Heun function $\mathbb{H} ={\rm{HeunC}}(\alpha_r ,{\beta_r} ,\gamma_r ,\delta_r ,\eta_r ;x_r)$ at infinity as follows,

\begin{widetext}
\begin{equation}
  \mathop {\lim }\limits_{|x| \to \infty }\mathbb{H}(x)= D_ \odot^{\beta_r} \;{x^{ - \frac{{{\beta_r}  + \gamma_r  + 2}}{2} - \frac{\delta_r }{\alpha_r }}} + D_\otimes ^{\beta_r} {{\rm{e}}^{ - \alpha_r x}}{x^{ - \frac{{{\beta_r}  + \gamma_r  + 2}}{2} + \frac{\delta_r }{\alpha_r }}} .
\end{equation}

Then, the connection coefficients $D_ \odot ^{\beta_r}$ and $D_ \otimes ^{\beta_r}$ are given by
\begin{align}
{}&\begin{array}{l}\label{eq:Dinc}
D_ \odot ^{\beta_r}  = \Xi _{{\rm n},\nu }^{\beta_r} D_{ \odot ,{\rm n},\nu }^{\beta_r} + {{\rm{e}}^{ - i\pi \left( {\nu  + \tfrac{1}{2}} \right)}}\frac{{\sin \pi \left( {\nu  + \frac{\delta_r }{\alpha_r }} \right)}}{{\sin \pi \left( {\nu  - \frac{\delta_r }{\alpha_r }} \right)}}\Xi _{ - {\rm n}, - \nu  - 1}^{\beta_r} D_{ \odot , - {\rm n}, - \nu  - 1}^{\beta_r} ,
\end{array}
\\
{}&{D_ \otimes ^{\beta_r}  = \Xi _{{\rm n},\nu }^{\beta_r} D_{ \otimes ,{\rm n},\nu }^{\beta_r}  + {{\rm{e}}^{i\pi \left( {\nu  + \tfrac{1}{2}} \right)}}\Xi _{ - {\rm n}, - \nu  - 1}^{\beta_r} D_{ \otimes , - {\rm n}, - \nu  - 1}^{\beta_r} ,}\label{eq:Dout}
\end{align}
with
\begin{align}
D_{ \odot ,{\rm n},\nu }^{\beta_r}  &= {\left( { - 1} \right)^{\frac{{\gamma_r  + {\beta_r}  + 2}}{2} + \frac{\delta_r }{\alpha_r }}}{\big( {\frac{\alpha_r }{2}} \big)^\tau } {\big( -{\frac{{i\alpha_r }}{2}} \big)^{ - \frac{{\gamma_r  + {\beta_r}  + 2}}{2} - \frac{\delta_r }{\alpha_r }}}{{\rm{e}}^{ - \frac{{i\pi \tau +\alpha_r}}{2}}}
\nonumber \\
&\times
{2^{ - 1 - \frac{\delta_r }{\alpha_r }}}{{\rm{e}}^{\frac{{i\pi }}{2}\big( {\nu  + 1 + \frac{\delta_r }{\alpha_r }} \big)}} \Xi _{{\rm n},\nu }^{\beta_r}\frac{{\Gamma \big( {\nu  + 1 + \frac{\delta_r }{\alpha_r }} \big)}}{{\Gamma \big( {\nu  + 1 - \frac{\delta_r }{\alpha_r }} \big)}},
\\
 D_{  \otimes,{\rm n},\nu }^{\beta_r}  &= {\big( { - 1} \big)^{\frac{{\gamma_r  + {\beta_r}  + 2}}{2} - \frac{\delta_r }{\alpha_r }}}{\left( {\frac{\alpha_r }{2}} \right)^\tau } {\left(- {\frac{{i\alpha_r }}{2}} \right)^{ - \frac{{\gamma_r  + {\beta_r}  + 2}}{2} + \frac{\delta_r }{\alpha_r }}} \nonumber \\
&\times {{\rm{e}}^{ - \frac{{i\pi \tau - \alpha_r}}{2}}}\Xi _{{\rm n},\nu }^{\beta_r}  \frac{ {2^{ - 1 + \frac{\delta_r }{\alpha_r }}}{{\rm{e}}^{ - \frac{{i\pi }}{2}\big( {\nu  + 1 - \frac{\delta_r }{\alpha_r }} \big)}}} {\sum\limits_{{\rm n} =  - \infty }^{ + \infty } {f_{\rm n}^\nu } }   \times\Big( \sum\limits_{{\rm n} =  - \infty }^{ + \infty } {{{( - 1)}^{\rm n}}} \frac{{{\big( {\nu  + 1 - \frac{\delta_r }{\alpha_r }} \big)}_{\rm n}}}{{{{\big( {\nu  + 1 + \frac{\delta_r }{\alpha_r }} \big)}_{\rm n}}}}f_{\rm n}^\nu  \Big),
\end{align}
and
\begin{align}\label{eq:Xinv}
&\begin{array}{*{20}{l}}
{\Xi _{{\rm n},\nu }^{\beta_r}  = \frac{{{2^{ - \nu }}{{\left( {\frac{\alpha_r }{2}} \right)}^{ - \hat \tau }}{{\rm{e}}^{\frac{{i\pi \hat \tau  + \alpha_r }}{2}}}\Gamma \left( {{\beta_r}  + 1} \right)\Gamma \left( {2\nu  + 2} \right)}}{{\Gamma \left( {\nu  + 1 + \frac{\delta_r }{\alpha_r }} \right)\Gamma \left( {\nu  + 1 - \frac{{{\beta_r}  + \gamma_r }}{2}} \right)\Gamma \left( {\nu  + 1 + \frac{{\gamma_r  - {\beta_r} }}{2}} \right)}}}\\
{ \times {{(\sum\limits_{{\rm n} =  - \infty }^0 {\frac{{{{( - 1)}^{\text{n}}}{{\left( {\nu  + 1 - \frac{{{\delta _r}}}{{{\alpha _r}}}} \right)}_{\text{n}}}}}{{( - {\text{n}})!{{\left( {2\nu  + 2} \right)}_{\text{n}}}{{\left( {\nu  + 1 + \frac{{{\delta _r}}}{{{\alpha _r}}}} \right)}_{\text{n}}}}}} f_{\text{n}}^\nu )}^{ - 1}}}
\end{array}
\\
&
\begin{array}{l}
 \times \Big( {\sum\limits_{{\rm n} = 0}^\infty  {{{\left( { - 1} \right)}^{\rm n}\frac{{\Gamma \left( {{\rm n} + 2\nu  + 1} \right)\Gamma \left( {{\rm n} + \nu  + 1 + \frac{{\gamma_r  - {\beta_r} }}{2}} \right)\Gamma \left( {{\rm n} + \nu  + 1 - \frac{{{\beta_r}  + \gamma_r }}{2}} \right)}}{{( {\rm n}!)\Gamma \left( {{\rm n} + \nu  + 1 + \frac{{{\beta_r}  - \gamma_r }}{2}} \right)\Gamma \left( {{\rm n} + \nu  + 1 + \frac{{{\beta_r}  + \gamma_r }}{2}} \right)}}f_{\rm n}^\nu } } \Big)},
\end{array}
\nonumber
\end{align}
where $\tau  = \frac{1}{4}\left( {3{\beta_r}  + \gamma_r  + \frac{{2\delta_r }}{\alpha_r }} \right)$ and $\hat \tau  = \frac{{\beta_r  - \gamma_r }}{4} + \nu  + \frac{\delta_r }{{2\alpha_r }}$.
\end{widetext}

\bibliography{mybibfile}

\end{document}